\documentclass[superscriptaddress,onecolumn,pre,nofootinbib]{revtex4}

\usepackage{amsmath}
\usepackage{graphicx,color}
\usepackage{amssymb}
\usepackage{amsfonts}
\usepackage{bm}

\newcommand{\be}{\begin{equation}}
\newcommand{\ee}{\end{equation}}
\newcommand{\bea}{\begin{eqnarray}}
\newcommand{\eea}{\end{eqnarray}}

\begin{document}
\sloppy


\title{Covariant theory of Bose-Einstein condensates in curved
spacetimes with electromagnetic interactions: the hydrodynamic approach}

\author{Pierre-Henri Chavanis}
\affiliation{Laboratoire de Physique
Th\'eorique, Universit\'e Paul
Sabatier, 118 route de Narbonne 31062 Toulouse, France}
\author{Tonatiuh Matos}
\affiliation{Departamento de F\'{\i}sica,
 Centro de Investigaci\'on y Estudios Avanzados del IPN\\
 A.P. 14-740,  07000, M\'exico, D.F., M\'exico}

\begin{abstract}

We develop a hydrodynamic representation of the
Klein-Gordon-Maxwell-Einstein equations. These equations combine quantum
mechanics, electromagnetism, and general relativity. We consider the case of an
arbitrary curved spacetime, the case of weak gravitational fields in a static
or expanding background, and the
nonrelativistic (Newtonian) limit. The Klein-Gordon-Maxwell-Einstein equations
govern the evolution of a
complex scalar field, possibly describing  self-gravitating
Bose-Einstein condensates, coupled to an electromagnetic field. They may find
applications in the context of dark matter, boson stars, and neutron stars with
a superfluid core.

\end{abstract}

\maketitle

\section{Introduction}
\label{sec_intro}

The fundamental equations of quantum mechanics (Schr\"odinger, Klein-Gordon,
Dirac) have an interesting history. A short historic was written by Dirac
\cite{dirachistoric} who retraced the early development of quantum mechanics.
According to Dirac, what we now call the Klein-Gordon (KG) equation [see Eq.
(\ref{fkg6})] was actually introduced by de
Broglie. He
proposed this wave
equation because
he noticed that there was an interesting connection between its solutions and
the relativistic motion of a particle. De Broglie postulated that these waves
were associated with the motion of the particle. In his thesis
\cite{brogliethese}, he introduced the correspondences $E=\hbar\omega$ and
${\bf p}=\hbar {\bf k}$ between the energy and impulse of a particle and the
pulsation and wave vector of a wave. He aimed at performing a real physical
synthesis, valid for all particles, of the notion of wave-corpuscle
duality that Einstein \cite{einstein} had
introduced for photons in
his theory of light quanta in 1905.\footnote{As reported by Bhaumik
\cite{bhaumik}, ``Langevin, who was de Broglie's
thesis advisor, was very skeptical about de Broglie's theory and contacted
Einstein to have his opinion. Einstein strongly supported de Broglie's work and
suggested to physicists to look for an evidence of the matter wave. A proof was
furnished soon by the accidental discovery of electron waves by Davisson and
Germer \cite{dg} in observing a diffraction pattern in a nickel crystal. Later, Einstein
mentioned to Rabi that he had thought about the equation for matter waves before
de Broglie but did not publish it because there was no evidence for it at that
time.''}

Schr\"odinger who was studying the motion of an electron
in an atom used the wave equation of de Broglie appropriately modified to take
into account the electromagnetic field in which the electron was moving. He
guessed what we now call the electromagnetic KG equation [see Eq. (\ref{fkg7})].
It is interesting to note that Schr\"odinger first
attempted to develop a {\it relativistic} wave theory of the hydrogen atom and
derived Eq. (\ref{fkg7}).
However, he abandoned it when he realized
that it was not giving the correct energy levels of the hydrogen
atom.  He was very depressed about it. He came back to the problem a
few months later and considered the nonrelativistic limit of Eq. (\ref{fkg7}).
He
obtained, in the absence of magnetic field, what is now called
the Schr\"odinger equation [see Eq. (\ref{fkg8})]. With this non-relativistic
approximation, he obtained results in agreement
with the observations apart from the fine structure
of the hydrogen spectrum which depends on the relativistic corrections.

If things really happened as Dirac describes in his
historic \cite{dirachistoric}, it is curious to
note that the manner Schr\"odinger introduced his equation in his
published papers \cite{schrodinger1,schrodinger2,schrodinger3,schrodinger4} is
totally different from his original approach retraced by Dirac. He
used only nonrelativistic arguments and did not rely on  Eq. (\ref{fkg7}). In
his first paper (27 January 1926)\footnote{Here and in the following, the
dates
correspond to the dates of submission.}
\cite{schrodinger1} he
obtained the eigenvalue equation (\ref{fsch5}) from a variational
principle. In his second paper (23 February
1926)
\cite{schrodinger2} he recovered this eigenvalue equation
 from an ingenious 
procedure obtained by combining the de Broglie relations and
the standard
wave equation (second order in time) with a
space dependent phase velocity (see Appendix \ref{sec_ori} for a brief
presentation of his
historical derivation). In his papers
\cite{schrodinger1,schrodinger2}, Schr\"odinger solved
the eigenvalue equation (\ref{fsch5}) for several potentials. In
addition to recovering the energy spectrum of the hydrogen atom
heuristically obtained by Bohr \cite{bohr1,bohr2}, he also derived the energy
spectrum of the harmonic oscillator and rotator, and found
agreement with the result obtained by  Heisenberg \cite{heisenberg}
from his more abstract matrix mechanics.\footnote{In this brief
review, we stick to what was originally called the ``wave (or undulatory)
mechanics'' initiated by de Broglie \cite{brogliethese} and
developed by Schr\"odinger
\cite{schrodinger1,schrodinger2,schrodinger3,schrodinger4}. We shall not discuss
in detail what was originally called the ``quantum mechanics'' (a term
introduced by Born)  based on the matrix
theory of Heisenberg \cite{heisenberg} further developed by Born and Jordan
\cite{bornjordan,bhj}. In Ref. \cite{schrodingerCORR} (18 March 1926),
Schr\"odinger showed the
equivalence between his wave mechanics and the Heisenberg-Born-Jordan quantum
mechanics.} In his third paper (10 May 1926)
\cite{schrodinger3}, Schr\"odinger applied his theory to the perturbation of the
hydrogen atom caused by an external homogeneous electric field (Stark effect).
In his fourth paper (21 June 1926) \cite{schrodinger4}, he obtained for the
first time the time-dependent equation (\ref{sch10}) that now bears his
name.
From this equation, he introduced a (charge) density
and a current (of charge) [see Eqs. (\ref{nr5}) and (\ref{nr7})] and derived a
local conservation equation
for  $\rho=|\psi|^2$ [see Eq. (\ref{nr6})]. This gives an interpretation to the
wave function in the sense that $|\psi|^2({\bf r},t)$ characterizes the
``presence'' of the particle at some point. Schr\"odinger thought that the wave
function represents a particle that is spread out, most of the particle being
where the modulus of the wave function $|\psi|^2$ is large. For example,
according to Schr\"odinger's view, the charge of the electron is not
concentrated in a point, but is spread out through the whole space, proportional
to the quantity $|\psi|^2$. On the other hand, Born (21 July 1926, 16 October
1926) \cite{bornf,born} developed a probabilistic interpretation of the wave
function.\footnote{As reported by
Bhaumik
\cite{bhaumik}, Born was strongly influenced by Einstein in his interpretation
as he stated in his Nobel lecture: ``Again an idea of Einstein's gave me the
lead. He had tried to make the duality of particles - light quanta or photons -
and waves comprehensible by interpreting the square of the optical wave
amplitudes as probability density for the occurrence of photons. This concept
could at once be carried over to the $\psi$-function:
$|\psi|^2$ ought to represent the probability
density for electrons (or other particles).''} Born proposed that the magnitude
of the wavefunction $\psi({\bf r},t)$ does not tell us how much of the particle
is at position ${\bf r}$ at time $t$, but rather the {\it probability} that the
particle is at ${\bf r}$ at time $t$. This gives an interpretation
to the wave function in the sense that $|\psi|^2$ represents the probability of presence of
the electron at some point. The correspondence principle [see Eq. (\ref{corrp})] was introduced
by Schr\"odinger (18 March
1926) \cite{schrodingerCORR} and de Broglie  (19 July
1926, 5 August 1926) \cite{broglie1,broglie2}. On the other hand, by making the
transformation of Eq. (\ref{hggp1}) introduced by  Wentzel (18
June 1926) \cite{wentzel}, Brillouin (5 July 1926)
\cite{brillouin}
and Kramers (9 September 1926) \cite{kramers}, it was found that
the Schr\"odinger equation reduces to the Hamilton-Jacobi equation in the
semi-classical limit $\hbar\rightarrow
0$.\footnote{This is the so-called WKB approximation which allows
one to study the semi-classical regime of a quantum system. One recovers the
classical mechanics
from the wave mechanics in the limit $\hbar\rightarrow 0$ in the same manner
that one recovers geometric optics from the theory of undulatory optics when the
wavelength $\lambda\rightarrow 0$.}  Although the fundamental
papers
of Schr\"odinger  \cite{schrodinger1,schrodinger2,schrodinger3,schrodinger4} are
written in German, he also wrote an english summary of his theory in Ref.
\cite{schrodingerPR} (3 September 1926).

The electromagnetic Klein-Gordon (KG) equation [see Eq. (\ref{fkg7})] appeared
in many
publications in a very short lapse of time: Klein  \cite{klein1} (28 April
1926),
Fock \cite{fock1} (11 June 1926), Schr\"odinger \cite{schrodinger4} (21 June
1926), De Donder and
Dungen \cite{ddd} (5 July
1926), de Broglie \cite{broglie1} (19 July 1926), Fock
\cite{fock2} (30 July 1926), de Broglie \cite{broglie2} (5 August 1926), Kudar
\cite{kudar} (30 August 1926), Ehrenfest and Uhlenbeck \cite{eu} (16 September
1926), Gamow and Iwanenko  \cite{gi} (19 September 1926),
Gordon \cite{gordon} (29 September 1926), De Donder  \cite{dd0} (11 october
1926), Schr\"odinger \cite{schrodingerKG12} (30 November 1926), Klein
\cite{klein2} (6 December
1926), and Schr\"odinger \cite{schrodingerKG2} (10 December
1926).\footnote{We note that Klein \cite{klein1} and
Fock \cite{fock2} derived their wave equation in a five dimensional spacetime
in an attempt to unify electromagnetism and gravitation. This is the so-called
Kaluza-Klein theory. The works of Klein \cite{klein1} and  Fock \cite{fock2}
were done independently to each other, and independently from the earlier work
of Kaluza \cite{kaluza}. Their approch was followed by  Ehrenfest and
Uhlenbeck
\cite{eu} and  Gamow and Iwanenko \cite{gi}.  We note that the
standard form of
the KG equation  formulated in a four dimensional
spacetime [see Eq. (\ref{fkg7})] first
appeared in Schr\"odinger's paper \cite{schrodinger4}.} These authors
obtained
the electromagnetic KG equation in
a flat spacetime (except \cite{klein1,fock2,eu,gi} who developed their theory in
a five dimensional curved spacetime). The KG equation taking
into account electromagnetism and general relativity in a four
dimensional curved spacetime [see Eq. (\ref{kg2})] was first written by De
Donder
\cite{ddd,dd1,dd2,donderp504,dd3}.  His work was further developed by Rosenfeld
\cite{rosenfeld1,rosenfeld2,rosenfeld3,rosenfeld4} (in a four or five
dimensional spacetime) and by de Broglie
\cite{broglie1927b}. The KG equation was
introduced by these authors in different
manners, either
from the correspondence
principle
\cite{schrodinger4,broglie1,broglie2,gordon,schrodingerKG12,klein2}
or from a variational principle
\cite{fock1,ddd,fock2,kudar,gordon,dd0,dd1,dd3,rosenfeld1,rosenfeld4}. The 
authors of Refs. \cite{gordon,klein2,schrodingerKG2,broglie1927b,broglie1927a,donderp504,
rosenfeld4} introduced a charge density [see Eq.
(\ref{charge4})] and a
current of charge [see Eq. (\ref{charge5})] and showed that the KG equation
conserves the charge [see
Eq. (\ref{charge6})].

This brief review shows that
there was an impressive activity during this
period. Clearly, 1926 is the year of quantum mechanics! We also note that
researchers were
interested since the start by finding a {\it relativistic} wave equation. This
is why the classical wave equation of Schr\"odinger (first order in time and
involving the complex number $i$ and the complex wavefunction $\psi$) was rather
original, and unexpected,  at that time. As we have seen, Schr\"odinger himself
was originally interested by finding a relativistic wave equation.
Klein
\cite{klein1,klein2} made the
transformation of Eq. (\ref{gpe1}) and showed the connection between the KG
equation and
the Schr\"odinger equation in the nonrelativistic limit $c\rightarrow +\infty$
(see also \cite{kudar,gordon,dd3}). On the other hand, by making the
WKB transformation [see Eq. (\ref{comp2})], it was found that the KG
equation reduces to the relativistic Hamilton-Jacobi equation in the
semi-classical limit $\hbar\rightarrow 0$.\footnote{The quantum Hamilton-Jacobi
equation
(\ref{comp3}) was written by Klein \cite{klein2} and De Donder
\cite{dd2} but they did not  write its complex
hydrodynamic
representation [see Eq. (\ref{comp7})]. It is interesting to note that De Donder
\cite{dd2} postulated the relation (\ref{comp3q}) 
to derive  the wave equation (\ref{kg2}). In his approach, Eq.
(\ref{comp3q}) is a fundamental equation connecting the classical mechanics to
the wave mechanics.} Therefore, the KG equation appeared to be a
physically relevant
relativistic wave equation since it returned the
correct results in the nonrelativistic limit $c\rightarrow
+\infty$ and in the semi-classical limit $\hbar \rightarrow 0$.

At that time, most physicists were satisfied with the KG equation.\footnote{In
his historic, Dirac \cite{dirachistoric} mentions that, although Schr\"odinger
discovered the KG equation, he was not ``bold enough to publish it'' because it
did not give results in agreement with observations for the hydrogen spectrum.
Actually, as we have previously indicated, Schr\"odinger {\it did} publish the
KG equation in Ref. \cite{schrodinger4}. This is even the first paper where the
KG equation appears under its standard form. However, at the end
of Ref. \cite{schrodingerPR}, Schr\"odinger points out difficulties with the
relativistic theory: (i) the difficulty to extend it to more than one
electron; (ii) its inability to account for the fine structure of the hydrogen
atom. Concerning point (ii), Schr\"odinger foresaw
that the deficiency  may be solved by taking the spin of the
electron into account.} However, Dirac was not happy with it
because the KG equation is second order in time and so one cannot apply to it
the transformation theory that Dirac cherished so much (it was his
``darling''\cite{dirachistoric}). In the Introduction of his paper on the
quantum theory of the
electron, Dirac \cite{dirac1} pointed out several problems with the KG equation.
From the KG equation, one can introduce a density that satisfies a local
conservation equation. However, this density is {\it not} definite positive so
it cannot be interpreted as a density probability.\footnote{As
indicated previously, the conserved density was interpreted as a
charge density
\cite{gordon,klein2,schrodingerKG2,broglie1927b,broglie1927a,donderp504,
rosenfeld4} (de Broglie \cite{broglie1927b,broglie1927a} interpreted it as a
density of corpuscles, apparently not realizing the problem of sign).} As a
result, the interpretation of the scalar
field (SF) $\varphi$ governed by the KG equation is unclear. Another difficulty
with
the KG
equation is that it allows negative kinetic energies as solution. These negative
kinetic energies cannot have a physical reality. Finally, the wave theory based
on the KG equation,
when applied to the hydrogen atom, does not give results in agreement with
experiments because the observed number of stationary states for an electron in
an atom is twice the number given by the theory. To account for this
observation, Uhlenbeck and Goudsmit  \cite{spin} introduced the idea that
the
electron has a
spin.\footnote{The idea of the spin of the electron goes back to Compton
\cite{compton}. It was also imagined by Kronig but Pauli discouraged him to
publish his results (see the very interesting
anecdotes of Dirac \cite{dirachistoric} on this subject).} This idea has
been incorporated in the nonrelativistic wave theory by
Pauli \cite{pauli} and Darwin \cite{darwin}, leading to the Pauli equation. This
equation gives results in agreement with the experiment for hydrogen-like
spectra to the first order of accuracy. However, this approach remains
heuristic. Furthermore it is not fully relativistic.

In order to solve the first difficulty (the indeterminate sign of the density),
Dirac \cite{dirac1} proposed another relativistic
extension of the Schr\"odinger equation. He obtained an equation, the Dirac
equation, that satisfies the requirements of relativity and that is first order
in time (his equation involves matrices
that generalize the matrices that Pauli \cite{pauli} introduced in his
equation). Therefore,
one can apply the transformation theory to it. In Dirac's
theory, one can introduce a density that is positive definite and that satisfies
a local conservation equation. This density can be
interpreted as the probability density of presence of the electron. Therefore,
the Dirac wave function $\psi$
has a clear physical interpretation. Furthermore, when including the
electromagnetic field, the Dirac equation  gave the electron a spin
$s=1/2$ and the correct magnetic moment without {\it ad hoc} assumption. That
was for Dirac a surprise and an unexpected bonus \cite{dirachistoric}.
Finally, the Dirac equation gave results in total agreement with observation of
the spectrum of hydrogen.

There remained a difficulty with the Dirac equation. It was possible for the
electron of charge $-e$ to have states of positive and negative kinetic energy,
this last case having no physical meaning. One cannot arbitrarily exclude
solutions with negative energy (as we do in classical relativity theory) since,
in general, a perturbation will cause transitions from states with positive
energy to states with negative energy.
Dirac solved the problem of negative energies via
the ``hole'' theory \cite{dirac2}. He assumed that all the negative
energy states are occupied by electrons (Dirac sea) on account of Pauli's
exclusion principle, except for a few states
with a low velocity. These vacant states are ``holes''. These holes behave as
particles with positive energy and charge $+e$. Dirac did not dare to postulate
a new particle. He published his work as a theory of electrons and protons,
hoping that in some unexplained way the Coulomb interaction between the
particles would account for the big difference of mass between electrons and
protons. This consideration was criticized by Weyl \cite{weyl} who published a
categorical
statement that
the new particle should have the same mass as the electron. This led to the
concept of antimatter. The first antiparticle, the
positron, was experimentally discovered by Anderson \cite{anderson} in
1932. Antimatter was unsuspected before Dirac's work.

Although the KG equation is not a
successful relativistic
generalization of the Schr\"odinger equation to describe the
electron and other spin-$1/2$ particles (fermions), this equation was
resurrected in
the context of quantum
field theory where it was shown to describe spin-$0$
particles (bosons) such as $\pi$-mesons, pions, or the Higgs boson.\footnote{The
application of the KG equation to spin-$0$ particles 
was first proposed by Pauli and Weisskopf \cite{pw}. They worked in terms of the
energy density [see Eq. (\ref{wkg2simp})] (different from the charge
density [see
Eq. (\ref{charge4})]) that is definite positive and satisfies a local
conservation equation [see Eq. (\ref{emt3})].}
There are therefore two possible theories for
particles, both relativistic, one for the
particles of zero spin satisfying the Bose-Einstein statistics
\cite{bose,einsteinb}, the other for
particles of spin $s=1/2$ satisfying the Fermi-Dirac
statistics \cite{fermi,dirac}. The Dirac equation
applies to
electrons and to other particles of spin $1/2$
like protons and quarks. The KG equation applies to certain kinds of mesons with
zero spin. These particles are neutral for a real SF and charged for a complex
SF.

In the 1960s, the equations of quantum mechanics found
new applications in the context of
Bose-Einstein condensates (BECs). The subject started
with the work of Bogoliubov \cite{bogoliubov} who attempted to explain the
phenomenon of superfluidity with a model of imperfect Bose gas with weak
repulsion between atoms. At $T=0$, all the bosons
are in the same
quantum state described by a single wavefunction $\psi({\bf
r},t)$.\footnote{For a system of $N$ bosons
in interaction
one has to consider in principle the $N$-body Schr\"odinger equation. It reduces
to a self-consistent one-body Schr\"odinger  equation when we neglect
fluctuations and implement a mean field approximation. In that case, the
$N$-body wave function is equal to a product of $N$ one-body wave functions
(Hartree approximation). This approximation
is exact in the case of BECs when $N\rightarrow +\infty$. This is equivalent to
starting from the Heisenberg equation of motion for the wave function operator
in the formalism of second quantification, as in the theory of Bogoliubov
\cite{bogoliubov},
and neglecting excitations.}  When the
particles interact through a pair contact potential \cite{hy,lhy},
the
evolution of
the wavefunction $\psi({\bf r},t)$ is described by a nonlinear version of the
Schr\"odinger equation called the Gross-Pitaevskii (GP) equation
\cite{gross1,gross2,gross3,pitaevskii1,pitaevskii2}
which involves a cubic nonlinearity [see Eq.
(\ref{nr4})]. A relativistic extension of the GP equation is provided by the KG
equation incorporating a potential $V(|\varphi|^2)$ that
takes self-interaction into account [see Eq. (\ref{fkg1})]. A pair contact
interaction between the bosons corresponds to a quartic potential in the KG
equation [see Eq. (\ref{lag4})].

The coupling between the KG equation and gravity through the Einstein
equations, leading to the Klein-Gordon-Einstein (KGE) equations, was
considered by Kaup (1968) \cite{kaup} and Ruffini and Bonazzola (1969)
\cite{rb} in the context of boson stars. In a sense, boson stars are
the descendent of the so-called {\it geons}
of Wheeler (1955) \cite{wheeler} except that they are built from scalar
particles (spin-$0$) instead of electromagnetic fields, i.e., spin-$1$ bosons.
These authors determined the maximum mass of boson stars in general relativity.
Ruffini and Bonazzola
\cite{rb} also considered the nonrelativistic limit of
their theory, leading to the Schr\"odinger-Poisson equation, describing
Newtonian self-gravitating BEC stars. In the works of Kaup
\cite{kaup} and Ruffini and Bonazzola
\cite{rb}, it was assumed that the bosons have no self-interaction.
Self-interacting boson stars were studied later by  Colpi {\it et al.}
(1986) \cite{colpi}. They  considered the case of a self-interacting
SF described by the KGE equations with a quartic potential and showed
that the self-interaction of the bosons can considerably increase the maximum
mass of boson stars.

More recently, it was proposed that dark matter (DM) halos could be made of a SF
described by the KGE
equations. Actually, at the galactic scale, the Newtonian limit is
valid so that
DM halos can be described by the Schr\"odinger-Poisson (SP) equations or by the
Gross-Pitaevskii-Poisson (GPP) equations.
Therefore, DM halos could be gigantic quantum objects made of BECs. The wave
properties of bosonic DM may
stabilize the system against gravitational collapse, providing halo cores
and sharply suppressing small-scale linear power. This may
solve the
problems of the cold dark matter (CDM) model such as the cusp problem and the
missing
satellite
problem. The  scalar
field dark matter (SFDM) model and the BEC dark matter (BECDM) model, also
called $\Psi$DM models,
have received much attention in the last years.\footnote{The bibliography on
boson stars and on SF/BEC dark matter is extensive. We refer the interested
reader to the Introduction of \cite{prd1}  for a short historic  and to the
reviews \cite{revueabril,revueshapiro,bookspringer} for an exhaustive
list of references.}

The Schr\"odinger equation has always been regarded as an abstract equation
because it describes the evolution of a  complex wavefunction
$\psi({\bf r},t)$  whose interpretation is unclear and still makes debate. For
that reason, other representations of the Schr\"odinger equation  have been
proposed. In a very intriguing early work, Madelung (25 October
1926)
\cite{madelung} made
the transformation of Eqs. (\ref{hggp5}) and
(\ref{whydro2})  and showed that
the
Schr\"odinger
equation is mathematically equivalent to hydrodynamic
equations, namely the equation of
continuity  [see Eq. (\ref{class1})] and the Euler equation [see Eq.
(\ref{class3})] for a pressureless irrotational
perfect fluid
with an additional quantum potential [see Eq. (\ref{qpc})] arising from the
finite value of $\hbar$. The
quantum potential, or quantum force,
accounts for the Heisenberg uncertainty principle.  Because
of this quantum term, the particle's motion
does not follow the laws of classical mechanics. The paper of
Madelung was welcome with
skepticism.\footnote{His  hydrodynamic approach
was further developed by Kennard \cite{kennard} in 1927. However, apart from
that work, the paper of Madelung was very little quoted.} For
example, in his review of quantum mechanics, Pauli expresses the opinion that
the hydrodynamic approach of Madelung  is not very
interesting (see the comment in Ref. \cite{spiegel}).

At about the same period, de
Broglie \cite{broglie1927a,broglie1927b,broglie1927c}
developed a relativistic hydrodynamic representation of the
KG equation (he was apparently not aware of the work of Madelung).
He made the transformation of Eqs. (\ref{hydro1}) and
(\ref{hydro4}) and derived the relativistic quantum Euler
equations (\ref{hydro7}) and (\ref{hydro8}) that contain the Lorentz invariant
quantum potential (\ref{hydro9}).\footnote{Actually,
these
relativistic
hydrodynamic equations were derived by London (25 February 1927) \cite{london} a
little before de Broglie's first paper  on the subject (1st April
1927) \cite{broglie1927a}.} This
is the relativistic version of the
classical Madelung quantum potential (\ref{qpc}). De Broglie
\cite{broglie1927c} interpreted the
quantum force as a force of internal
tensions existing around the corpuscles (see also
Rosenfeld \cite{rosenfeld3,rosenfeld4}). He showed that everything
happens as if the particles had an effective mass  depending on the
quantum potential [see Eq. (\ref{db2})]. He also interpreted
the continuity equation (\ref{db8}) as a conservation equation for a
density transported by a velocity [see Eq. (\ref{db6})]. The aim of de Broglie
was to provide a
causal and objective interpretation of
wave mechanics, in accordance with the wish expressed many times by Einstein,
and in contrast to the purely probabilistic interpretation of
quantum mechanics
put forward by Born, Bohr, and Heisenberg. This is what he called the pilot wave
theory
because, in
virtue of Eqs. (\ref{hydro1c}) and (\ref{hydro4}),  the particle is guided by
the wave $\psi$. The pilot wave theory of de Broglie
was criticized by Pauli during the October
1927 Solvay Physics Conference (see the comment in
\cite{solvay}), and de Broglie abandoned
it.

The  results of Madelung \cite{madelung} and de
Broglie \cite{broglie1927a,broglie1927b,broglie1927c}  were
rediscovered by Bohm \cite{bohm1,bohm2} in 1952 in relation to his
interpretation of the quantum theory in terms of ``hidden'' variables. For that
reason, the quantum potential is sometimes called the Bohm
potential. Developing the works of Madelung  \cite{madelung}, de Broglie
\cite{broglie1927a,broglie1927b,broglie1927c} and Bohm \cite{bohm1,bohm2},
Takabayasi \cite{takabayasi1,takabayasi2} proposed a formulation of classical
and relativistic quantum mechanics in terms of hydrodynamic equations based on
the Schr\"odinger and KG equations. He emphasized the role of the quantum force
that is responsible for the ``blurring'' of the classical trajectory. He
interpreted the diffusion of wave packets, interference effects and tunnel
effects in terms of this quantum force.  He also criticized certain
aspects of Bohm's interpretation. This renewal of
interest for a causal interpretation of quantum mechanics stimulated de Broglie
to return to the problem again and undertake a fresh examination of his old
ideas \cite{ndb1,ndb2,ndb3,ndb4}.

The hydrodynamic representation of Madelung \cite{madelung} and de Broglie
 \cite{broglie1927a,broglie1927b,broglie1927c}  may lack a clear physical
interpretation in the case where the Schr\"odinger
equation or the KG equation describes just {\it one} particle. However, it
takes more sense when the Schr\"odinger equation, the
GP equation, or the KG equation describes a BEC made of many particles in the
same quantum state. In that case, the BEC can be interpreted as
a real fluid
described by the quantum Euler equation.\footnote{One interesting aspect of
BECs is related to
their superfluid properties. Their velocity field ${\bf
v}=\nabla S/m$ (where $S$ is the action) is irrotational
 but there may exist vortical motion due to
singular point vortices with circulation quantized in units of $h/m$ as
argued by Onsager in a footnote \cite{onsager} and Feynman \cite{feynman}
(this fact was actually discovered by Dirac
\cite{diracmm} in a more general context in which an electromagnetic field is
present). In contrast to classical hydrodynamics, the cores of
vortices are completely determined by the de Broglie length and all energies
are finite.} For
self-interacting BECs, there is an additional
pressure force coming from the
potential of self-interaction. This hydrodynamic representation was
developed by Gross \cite{gross2,gross3}.\footnote{A
quantum hydrodynamics for a many body system was previously 
developed by Landau \cite{landauS} and London \cite{londonS} in connection to
superfluidity
and superconductivity. It seems that neither Landau, London and Gross were
aware of the works of Madelung and de Broglie.} He
writes \cite{gross3}: ``We have a model
of a fluid which is more satisfactory than that of classical fluid dynamics''.

More recently, the Madelung hydrodynamic representation of the (nonrelativistic)
Schr\"odinger, or GP, equation has been used by B\"ohmer and Harko
\cite{bohmer}, Chavanis \cite{prd1,prd2,chavaniscosmo,chavaniskpz} and 
Rindler-Daller and Shapiro \cite{tanja,revueshapiro}  among
others to describe BEC dark matter (in a
static or expanding background), and by Chavanis and Harko
\cite{chavharko} to describe BEC stars such as boson stars or neutron stars with
a superfluid core. On the other hand, the de Broglie hydrodynamic representation
of the (relativistic) KG
equation has been used by Su\'arez
and Matos \cite{abrilMNRAS,smz} to
study the formation of structures in the universe, assuming that dark matter is
in the form of a fundamental SF with a quartic potential.
In the works of Su\'arez
and Matos \cite{abrilMNRAS,smz}, the SF is taken to be real and
the
gravitational
potential is
introduced by hand in the KG equation, and assumed to be determined by the
classical
Poisson equation where the source is the rest-mass density $\rho$. This leads to
the Klein-Gordon-Poisson (KGP) equations. However, this treatment is not
self-consistent since it combines
relativistic and nonrelativistic equations. A self-consistent relativistic
treatment was developed by Su\'arez and Chavanis \cite{abrilph1,abrilph2} who
derived the hydrodynamic
representation of a complex SF evolving through the KG equation and coupled to
gravity through
the Einstein
equations in the weak field approximation. This corresponds to the
Klein-Gordon-Einstein (KGE) equations. They used the
conformal Newtonian gauge which takes into account metric perturbations up to
first order and considered only scalar perturbations. This is sufficient to
calculate observational cosmological consequences of the SF dynamics in the
linear regime.

On the other hand, the case where the SF interacts
with an electromagnetic field in a flat spacetime has been considered recently by Matos and
collaborators \cite{mrm,mc}. They derived the hydrodynamic representation of the
Klein-Gordon-Maxwell (KGM) equations. We note that the electromagnetic field was already taken into account in the pioneering works of de Broglie \cite{broglie1927a,broglie1927b,broglie1927c} and Takabayasi \cite{takabayasi1,takabayasi2}.

In this paper, we combine these different approaches (electromagnetism and
gravity) and consider the general case where the SF interacts with
an electromagnetic field in a curved spacetime. Therefore, we derive the  
hydrodynamic representation of the Klein-Gordon-Maxwell-Einstein (KGME)
equations. To our knowledge, this general case has not been treated previously.
We consider the
fully nonlinear Einstein equations and the weak field approximation in a static
and expanding background. We recover the
nonrelativistic (Newtonian) results when $c\rightarrow +\infty$. This is probably the most
complete model that we can consider. The KGME equations describe self-gravitating
Bose-Einstein condensates coupled to an electromagnetic field. They may find
applications in the context of dark matter, boson stars, and neutron stars with
a superfluid core.

The paper is organized as follows. In Sec.
\ref{sec_kgme}, we consider the KGME equations in an
arbitrary curved spacetime. In Sec. \ref{sec_gpme}, we make the Klein
transformation and derive the Gross-Pitaevskii-Maxwell-Einstein (GPME)
equations. In Sec. \ref{sec_wfa}, we consider the KGME and GPME 
equations in the weak field approximation $\Phi/c^2\ll 1$. In Sec.
\ref{sec_rnl}, we consider the
nonrelativistic limit $c\rightarrow +\infty$ of the GPME equations leading to
the Gross-Pitaevskii-Maxwell-Poisson (GPMP) equations. In each case, we derive
a complex and real hydrodynamic representation of the field equations. The
Appendices regroup complementary results. In Appendix \ref{sec_id},
we list various identities that are useful in our calculations. In Appendices
\ref{sec_kgp} and \ref{sec_kgps}, we consider the generalized
Klein-Gordon-Maxwell-Poisson (KGMP) equations and the generalized GPMP equations
in an
expanding and static background. In
Appendix \ref{sec_kgm}, we derive the relativistic and nonrelativistic
eigenvalues equations associated with the KG and Schr\"odinger equations. In
Appendix \ref{sec_pilot} we discuss the relation between our approach and the
pilot wave theory of de Broglie. In Appendix \ref{sec_ori}, we recall the
historical derivation of the Schr\"odinger equation in complement to the
discussion given in our Introduction.

\section{The Klein-Gordon-Maxwell-Einstein equations}
\label{sec_kgme}

\subsection{The Lagrangian of the scalar field coupled to the electromagnetic field}
\label{sec_lag}

We consider a  complex SF which is a continuous
function of space and time defined at each point by
$\varphi(x^\mu)=\varphi(x,y,z,t)$. The action of the relativistic SF
is
\begin{equation}
S_\varphi=\int \mathcal{L}_\varphi \sqrt{-g}\, d^4x,
\label{lag1}
\end{equation}
where $\mathcal{L}_\varphi=\mathcal{L}_\varphi(\varphi,
\varphi^*,\partial_\mu\varphi,\partial_\mu\varphi^*)$
is the Lagrangian density and $g={\rm det}(g_{\mu\nu})$ is the determinant of
the metric tensor. In the absence of electromagnetic field, the 
Lagrangian of the SF is
\begin{eqnarray}
{\cal
L}^{(0)}_\varphi=\frac{1}{2}g^{\mu\nu}\partial_{\mu}\varphi^*\partial_{\nu}
\varphi-\frac{m^2c^2}{2\hbar^2}|\varphi|^2
-V(|\varphi|^2).
\label{lag2}
\end{eqnarray}
The
first term is the
kinetic energy, the second term (quadratic) is the rest-mass energy, and
the third term is the self-interaction energy. In some applications, we
shall consider a
quartic SF potential of the form $V(|\varphi|^2)=\frac{m^2}{2\hbar^4}
\lambda_s|\varphi|^4$.
If the SF describes a
BEC at $T=0$, the quartic potential
can be rewritten as
\begin{equation}
V(|\varphi|^2)=\frac{2\pi a_s
m}{\hbar^2}|\varphi|^4,
\label{lag4}
\end{equation}
where $a_s$ denotes the s-scattering length of the bosons (in that case
$\lambda_s=4\pi a_s\hbar^2/m$).

We assume that the SF
interacts with an  electromagnetic field described by the  quadripotential
$A^{\mu}$. We introduce the  Faraday tensor\footnote{In general relativity,
the Faraday tensor should be written as
$F_{\mu\nu}=D_{\mu}A_{\nu}-D_{\nu}A_{\mu}$ where $D$ is the covariant
derivative defined by Eq. (\ref{id0}). Using Eqs. (\ref{id1}) and
(\ref{id2}), we have
$D_{\mu}A_{\nu}-D_{\nu}A_{\mu}=\partial_{\mu}A_{\nu}-\partial_{\nu}A_{\mu}$
leading to Eq. (\ref{lag5}).}
\begin{equation}
F_{\mu\nu}=\partial_{\mu}A_{\nu}-\partial_{\nu}A_{\mu}.
\label{lag5}
\end{equation}
The Lagrangian of the SF coupled to an electromagnetic field is
obtained from the Lagrangian of the SF in the absence of electromagnetic field
by making the substitution
$\partial_{\mu}\rightarrow \partial_{\mu}+\frac{i
e}{\hbar}A_{\mu}$, where $e$ is the elementary charge, and adding the Lagrangian of the electromagnetic field
$-\frac{1}{4\mu_0}F^{\mu\nu}F_{\mu\nu}$, where $\mu_0$ is the permeability of
free space (for future reference we recall that the vacuum permeability
$\mu_0$ and the vacuum permittivity $\epsilon_0$ satisfy the relation
$\epsilon_0\mu_0=1/c^2$). This yields
\begin{eqnarray}
{\cal L}_\varphi=\frac{1}{2}g^{\mu\nu}\left (\partial_{\mu}\varphi^*-i
\frac{e}{\hbar}
A_{\mu}\varphi^*\right )\left (\partial_{\nu}\varphi+i \frac{e}{\hbar}
A_{\nu}\varphi\right )-\frac{m^2c^2}{2\hbar^2}|\varphi|^2
-V(|\varphi|^2)-\frac{1}{4\mu_0}F^{\mu\nu}F_{\mu\nu}.
\label{lag6}
\end{eqnarray}
The Lagrangian of the SF can be expanded as
\begin{eqnarray}
{\cal
L}_\varphi=\frac{1}{2}g^{\mu\nu}\partial_{\mu}\varphi^*\partial_{\nu}
\varphi+\frac { 1 } { 2 } g^ {
\mu\nu}i \frac{e}{\hbar} A_{\nu}\varphi\partial_{\mu}\varphi^*
-\frac{1}{2}g^{\mu\nu}i \frac{e}{\hbar}
A_{\mu}\varphi^*\partial_{\nu}\varphi+\frac{1}{2}g^{\mu\nu}\frac{e^2}{\hbar^2}
A_{\mu}A_{\nu}|\varphi|^2\nonumber\\
-\frac{m^2c^2}{2\hbar^2}|\varphi|^2
-V(|\varphi|^2)-\frac{1}{4\mu_0}F^{\mu\nu}F_{\mu\nu}.
\label{lag7}
\end{eqnarray}

\subsection{The energy-momentum tensor}
\label{sec_emt}

Taking the variation of the SF action (\ref{lag1}) with respect
to the metric $g_{\mu\nu}$, we get
\begin{equation}
\delta S_\varphi=\frac{1}{2}\int T_{\mu\nu}\, \delta g^{\mu\nu} \sqrt{-g}\, 
d^4x, 
\label{tb10}
\end{equation}
where
\begin{eqnarray}
T_{\mu\nu}=2\frac{\partial\mathcal{L}_\varphi}{\partial
g^{\mu\nu}}-g_{\mu\nu}\mathcal{L}_{\varphi}
\label{emt1}
\end{eqnarray}
is the energy-momentum tensor of the SF. For the Lagrangian (\ref{lag7}), it
takes the form
\begin{eqnarray}
T_{\mu\nu}=\frac{1}{2}(\partial_{\mu}\varphi^*
\partial_{\nu}\varphi+\partial_{\nu}\varphi^* \partial_{\mu}\varphi)
-g_{\mu\nu}\left
\lbrack\frac{1}{2}g^{\rho\sigma}\partial_{\rho}\varphi^*\partial_{\sigma}
\varphi-\frac{m^2c^2}{2\hbar^2}|\varphi|^2-V(|\varphi|^2)\right
\rbrack\nonumber\\
+\frac{1}{2} \biggl
(\frac{ie}{\hbar}A_{\nu}\varphi
\partial_{\mu}\varphi^*-\frac{ie}{\hbar}A_{\mu}\varphi^*
\partial_{\nu}\varphi
+2\frac{e^2}{\hbar^2}A_{\mu}A_{\nu}\varphi
\varphi^*+\frac{ie}{\hbar}A_{\mu}\varphi
\partial_{\nu}\varphi^*-\frac{ie}{\hbar}A_{\nu}\varphi^* \partial_{\mu}\varphi
\biggr )\nonumber\\
-\frac{1}{2}g_{\mu\nu}g^{\rho\sigma}\biggl
(i\frac{e}{\hbar}A_{\sigma}\varphi\partial_{\rho}\varphi^*-i\frac{e}{\hbar}A_{
\rho}
\varphi^*\partial_{\sigma}\varphi+\frac{e^2}{\hbar^2}A_{\rho}A_{\sigma}
\varphi\varphi^*\biggr
)+\frac{1}{\mu_0}\left
(F_{\mu\alpha}F_{\,\,\,\nu}^{\alpha}+\frac{1}{4}g_{\mu\nu} F^ { \alpha\beta }
F_{\alpha\beta}\right ).
\label{emt2}
\end{eqnarray}
We can also obtain the energy-momentum tensor of a SF
coupled to an electromagnetic field from the  energy-momentum tensor
$T_{\mu\nu}^{(0)}$ of
a SF in the absence of electromagnetic field by making the
substitution
$\partial_{\mu}\rightarrow
\partial_{\mu}+\frac{i
e}{\hbar}A_{\mu}$ and adding the energy-momentum tensor of
the electromagnetic
field
$\frac{1}{\mu_0}(F_{\mu\alpha}F_{\,\,\,\nu}^{\alpha}+\frac{1}{4}g_{\mu\nu}F^{
\alpha\beta}
F_{\alpha\beta})$. 

The conservation of the energy-momentum tensor, which
results via the  Noether theorem from the invariance of the Lagrangian density
under continuous translations in space and time, writes
\begin{eqnarray}
D_{\nu}T^{\mu\nu}=0.
\label{emt3}
\end{eqnarray}
By analogy with the energy-momentum tensor of a perfect fluid, the energy
density and the pressure tensor of the SF are defined by
\begin{eqnarray}
\epsilon=T_0^0,\qquad 
P_i^j=-T_i^j.
\end{eqnarray}
On the other hand, using Eq. (\ref{emt2}), one can show
that the
quantity
$(m^2/\hbar^2)|\varphi|^2$ tends to the rest-mass density $\rho$ in the
non-relativistic limit $c\rightarrow +\infty$ (see Appendix A of
\cite{abrilph1}). Therefore, we shall interpret
\begin{eqnarray}
\rho=\frac{m^2}{\hbar^2}|\varphi|^2
\label{emt4}
\end{eqnarray} 
as a pseudo rest-mass density. In the relativistic regime, the density $\rho$
defined by Eq. (\ref{emt4}) has not a clear physical interpretation (it is
not the rest-mass density and $\rho c^2$ is not the energy density).
However,  it can
always be introduced as a convenient notation \cite{abrilph1}. It is
only in the
non-relativistic limit $c\rightarrow +\infty$ that $\rho$ can be identified with
the
rest-mass density.

\subsection{The Klein-Gordon equation}
\label{sec_kg}

The equation of motion for the SF can be derived from the principle
of least action. Imposing $\delta
S_{\varphi}=0$ for arbitrary variations
$\delta\varphi$
and $\delta\varphi^*$ of the field, we obtain the Euler-Lagrange
equation
\begin{eqnarray}
D_{\mu}\left\lbrack\frac{\partial {\cal L}_\varphi}{\partial
(\partial_{\mu}\varphi)^*}\right\rbrack-\frac{\partial {\cal
L}_\varphi}{\partial\varphi^*}=0.
\label{kg1}
\end{eqnarray}
For the Lagrangian (\ref{lag7}),
 this leads to the electromagnetic KG
equation
\begin{equation}
\square_{\rm
e}\varphi+\frac{m^2c^2}{\hbar^2}\varphi+2\frac{dV}{d|\varphi|^2}\varphi=0,
\label{kg2}
\end{equation}
where $\square_e$ is the electromagnetic d'Alembertian operator in a curved
spacetime
\begin{equation}
\square_e\varphi=\left (D_{\mu}+i\frac{e}{\hbar}A_{\mu}\right
)\left (\partial^{\mu}+i\frac{e}{\hbar}A^{\mu}\right
)\varphi.
\label{kg3}
\end{equation}
The KG equation of a SF
coupled to an electromagnetic field can also be obtained from the KG equation of
a SF in the absence of electromagnetic field by making
the substitution $D_{\mu}\rightarrow
D_{\mu}+\frac{i
e}{\hbar}A_{\mu}$.
For the quartic potential (\ref{lag4}), the KG equation
(\ref{kg2}) becomes
\begin{equation}
\square_e\varphi+\frac{m^2c^2}{\hbar^2}\varphi+\frac{8\pi a_s
m}{\hbar^2}|\varphi|^2\varphi=0.
\label{kg5}
\end{equation}
Using the identities of Appendix \ref{sec_id}, the electromagnetic
d'Alembertian
operator (\ref{kg3}) can be written in the equivalent forms 
\begin{equation}
\square_e\varphi=D_{\mu}\partial^{\mu}\varphi+i\frac{e}{\hbar}
(D_{\mu}A^{ \mu })\varphi+2i\frac{e} {
\hbar}A_{\mu}\partial^{\mu}\varphi-\frac{e^2}{\hbar^2}A_{\mu}A^{\mu}
\varphi,
\label{kg7}
\end{equation}
\begin{equation}
\square_e\varphi=g^{\mu\nu}\partial_{\mu}\partial_{\nu}\varphi-g^{\mu\nu}\Gamma_
{\mu\nu}^{\sigma}\partial_{\sigma}\varphi+i \frac{e}{\hbar}
g^{\mu\nu}\left (
\partial_{\mu}A_{\nu}-\Gamma_{\mu\nu}^{
\sigma} A_{\sigma}\right )\varphi
+2i \frac{e}{\hbar} g^{\mu\nu}
A_{\mu}\partial_{\nu}\varphi-\frac{e^2}{\hbar^2}
g^{\mu\nu}A_{\mu}A_{\nu}\varphi,
\label{kg8}
\end{equation}
\begin{eqnarray}
\square_e\varphi=g^{\mu\nu}\left
(\partial_{\mu}+i\frac{e}{\hbar}A_{\mu}\right )\left
(\partial_{\nu}+i\frac{e}{\hbar}A_{\nu}\right
)\varphi-g^{\mu\nu}\Gamma_{\mu\nu}^{\sigma}\left
(\partial_{\sigma}+i\frac{e}{\hbar}A_{\sigma}\right
)\varphi.
\label{kg9}
\end{eqnarray}
We note that the first term in Eq. (\ref{kg7}) corresponds 
to the d'Alembertian operator in a curved spacetime in the absence of
electromagnetic field
\begin{equation}
\square\varphi=D_{\mu}\partial^{\mu}\varphi=g^{\mu\nu}D_{\mu}\partial_{\nu}
\varphi=g^{ \mu\nu} \left
(\partial_{\mu}\partial_{\nu}\varphi-\Gamma_{\mu\nu}^{\sigma}\partial_{\sigma}
\varphi \right ).
\label{kg3b}
\end{equation}
We also note that the second term in Eq. (\ref{kg7}) disappears if we make the
choice
of the Lorentz gauge 
\begin{equation}
D_{\mu}A^{\mu}=0.
\label{comp1qw}
\end{equation}

\subsection{The charge density and the current of charge}
\label{sec_charge}

We define the quadricurrent of the SF by 
\begin{eqnarray}
J_{\mu}=-\frac{m}{ 2i\hbar }
(\varphi^*\partial_{\mu}\varphi-\varphi\partial_{\mu}\varphi^*)-\frac{e
m}{\hbar^2}
|\varphi|^2A_{\mu}.
\label{charge1}
\end{eqnarray}
As recalled in the Introduction, there is
a
difficulty with the
interpretation of the KG equation because the density $J_0$ is not
definite positive. As a result, it cannot
be interpreted as a density probability, or as a mass density. However, it can
be interpreted as a charge density that can take positive or negative values.
Therefore, we introduce the quadricurrent of charge
\begin{eqnarray}
(J_e)_{\mu}\equiv\frac{e}{m}J_{\mu}=-\frac{e}{2i\hbar}
(\varphi^*\partial_{\mu}\varphi-\varphi\partial_{\mu}\varphi^*)-\frac{e^2}{\hbar
^2} |\varphi|^2A_{\mu}.
\label{charge3}
\end{eqnarray}
We note that the current vanishes for a real scalar field.
Therefore, only complex scalar fields are charged. We define the
charge density
by
$\rho_e=({J_e})_0/c$ and the current of charge by ${\bf
J}_e=((J_e)_x,(J_e)_y,(J_e)_z)=(-(J_e)_1,-(J_e)_2,-(J_e)_3)$ with the {\it
lower} indices.\footnote{This
is
not the usual
convention but we find that the equations are simpler when we use this
convention. Note that our convention (with the minus sign) reduces to
the usual one in the case of a flat spacetime.} Similarly, we
define the electric potential by $U/c=A_0$ and the potential vector by ${\bf
A}=(A_x,A_y,A_z)=(-A_1,-A_2,-A_3)$. The charge density and the current of
charge are then
given
by
\begin{eqnarray}
\rho_e =-\frac {e} { 2i\hbar c^2 }
\left
(\varphi^*\frac{\partial\varphi}{\partial
t}-\varphi\frac{\partial\varphi^*}{\partial t}\right )    -\frac { e^2 } {
\hbar^2c^2 }
|\varphi|^2 U,
\label{charge4}
\end{eqnarray}
\begin{eqnarray}
{\bf J}_e=\frac{e}{2i\hbar}
(\varphi^*\nabla\varphi-\varphi\nabla\varphi^*)-\frac{e^2}{\hbar^2}
|\varphi|^2 {\bf A}.
\label{charge5}
\end{eqnarray}

Taking the divergence of Eq. (\ref{charge3})
and using the KG
equation
(\ref{kg2}), one can show that
\begin{equation}
D_{\mu}J_e^{\mu}=0.
\label{charge6}
\end{equation}
This equation expresses the local conservation of the charge of a complex SF.
The global
conservation of charge can be
obtained as follows. Integrating Eq. (\ref{charge6}) over the whole
physical space and using Eq.
(\ref{id0}), we get
\begin{eqnarray}
0&=&\int D_{\mu}J_e^{\mu} \sqrt{-g}\, d^3x=\int \partial_{\mu}\left
(\sqrt{-g} J_e^{\mu}\right ) \, d^3x\nonumber\\
&=&\int \partial_{0}\left
(\sqrt{-g} J_e^{0}\right ) \, d^3x+\int \partial_i\left
(\sqrt{-g} {J}_e^i\right ) \, d^3x=\frac{1}{c}\int \partial_{t}\left
(\sqrt{-g} J_e^{0}\right ) \, d^3x,
\label{charge7}
\end{eqnarray}
where the last equality is obtained because the second integral in the second
line can be converted into a surface term that vanishes at
infinity. If we define the total charge by 
\begin{eqnarray}
Q=\frac{1}{c}\int J_e^{0}\sqrt{-g}\, d^3x,
\label{charge8}
\end{eqnarray}
we find that Eq. (\ref{charge7}) takes the form
\begin{eqnarray}
\frac{dQ}{dt}=0.
\label{charge9}
\end{eqnarray}
This equation expresses the global conservation of the charge of a complex
SF. The conserved quadricurrent of charge and the conserved charge
result via the Noether theorem from the invariance of the Lagrangian density
under a global phase transformation $\varphi\rightarrow\varphi e^{-i\theta}$ of
the complex SF.

\subsection{The Maxwell equations}

The Maxwell equations can be obtained from the
principle
of least action. Imposing $\delta
S_{\varphi}=0$ for arbitrary variations
$\delta A^{\nu}$ of the quadripotential, we obtain the Euler-Lagrange
equations
\begin{eqnarray}
D_{\mu}\left\lbrack\frac{\partial {\cal L}_\varphi}{\partial
(\partial_{\mu}A_{\nu})}\right\rbrack-\frac{\partial {\cal
L}_\varphi}{\partial A_{\nu}}=0.
\label{max0}
\end{eqnarray}
For the Lagrangian (\ref{lag6}), we have
\begin{eqnarray}
\frac{\partial {\cal
L}_\varphi}{\partial A_{\nu}}=-J_e^{\nu},\qquad \frac{\partial {\cal
L}_\varphi}{\partial
(\partial_{\mu}A_{\nu})}=-\frac{1}{\mu_0}F^{\mu\nu},
\label{max0a}
\end{eqnarray}
leading to the Maxwell equations
\begin{eqnarray}
D_{\mu}F^{\mu\nu}=\mu_0 J_e^{\nu}.
\label{max1}
\end{eqnarray}
For a complex SF, the quadricurrent of
charge $J_e^{\mu}$  is given by Eq. (\ref{charge3}). The conservation of
charge expressed
by Eq.
(\ref{charge6}) is included in the Maxwell equations (it results from the
anti-symmetry of the Faraday tensor). Substituting Eq. (\ref{lag5}) into the
Maxwell equations (\ref{max1}), we obtain the field equations satisfied by the
quadripotential
\begin{eqnarray}
D^{\mu}D_{\mu}A_{\nu}-D^{\mu}D_{\nu}A_{\mu}=\mu_0 (J_e)_{\nu}.
\label{lw1}
\end{eqnarray}
Using the identities (\ref{comp10b}) and (\ref{comp10c}), we can
rewrite Eq. (\ref{lw1}) as
\begin{eqnarray}
\square A_{\nu}-D_{\nu}D_{\mu}A^{\mu}-R_{\mu\nu}A^{\mu}=\mu_0 (J_e)_{\nu},
\label{lw1new}
\end{eqnarray}
where $R_{\mu\nu}$ is the Ricci tensor. With the choice of the Lorentz gauge
(\ref{comp1qw}), Eq.
(\ref{lw1new}) reduces to
\begin{eqnarray}
\square A_{\nu}-R_{\mu\nu}A^{\mu}=\mu_0 (J_e)_{\nu}.
\label{lw1new1}
\end{eqnarray}

\subsection{The Einstein equations}

The Einstein-Hilbert action in general relativity is defined by
\begin{equation}
S_g=\frac{c^4}{16\pi G}\int R \sqrt{-g}\, d^4x,
\label{kge7}
\end{equation}
where  $R$ is the Ricci scalar and $G$ is Newton's
gravitational
constant.  Its variation with respect to the metric $g_{\mu\nu}$
is given by \cite{weinberg}:
\begin{equation}
\delta S_g=-\frac{c^4}{16\pi
G}\int \left(R_{\mu\nu}-\frac{1}{2}g_{\mu\nu}R\right)\delta
g^{\mu\nu} \sqrt{-g} \, d^4x.
\label{kge8}
\end{equation}
The total action (SF $+$
gravity) is $S=S_{\varphi}+S_g$. The field equations can be obtained from the
principle of least action. Imposing 
$\delta S=0$ for arbitrary variations  $\delta g_{\mu\nu}$ of the metric, and
using Eqs.
(\ref{tb10}) and
(\ref{kge8}), we obtain the Einstein equations
\begin{equation}
R_{\mu\nu}-\frac{1}{2}g_{\mu\nu}R=\frac{8\pi G}{c^4}T_{\mu\nu}.
\label{ein1}
\end{equation}
These are a set of $10$ equations that describe the fundamental
interaction
between gravity and matter as a result of the curvature of spacetime. The
energy-momentum tensor  $T^{\mu\nu}$ is the source of the
gravitational field in the Einstein field equations of general relativity
in the sense that it determines the metric $g_{\mu\nu}$. The conservation of
the energy-momentum tensor expressed by Eq. (\ref{emt3}) is included in
the Einstein equations. Eqs. (\ref{kg2}), (\ref{max1}) and (\ref{ein1}) form
the KGME equations.

\subsection{The complex hydrodynamic equations}
\label{sec_che}

In this section, we restrict ourselves to the  electromagnetic KG equation 
(\ref{kg2}) 
without self-interaction ($V=0$). We
write the SF in the WKB form
\begin{equation}
\varphi\propto e^{i {\cal S}_{\rm tot}/\hbar},
\label{comp2}
\end{equation}
where ${\cal S}_{\rm tot}$ is a complex action. Substituting Eq. (\ref{comp2})
into Eq. (\ref{kg2}), we obtain the complex quantum
relativistic Hamilton-Jacobi equation
\begin{eqnarray}
(\partial_{\mu}{\cal S}_{\rm tot}+eA_{\mu})(\partial^{\mu}{\cal S}_{\rm
tot}+eA^{\mu})-m^2c^2=i\hbar D_{\mu}(\partial^{\mu}{\cal S}_{\rm tot}+e
A^{\mu})
=i\hbar \square {\cal S}_{\rm tot}+ie\hbar D_{\mu}A^{\mu}.
\label{comp3}
\end{eqnarray}
For $\hbar=0$, we recover the relativistic Hamilton-Jacobi equation (in that
case ${S}_{\rm tot}$ is real):
\begin{equation}
\left\lbrack {S}_{\rm tot} \right\rbrack\equiv (\partial_{\mu}{S}_{\rm
tot}+eA_{\mu})(\partial^{\mu}{S}_{\rm
tot}+eA^{\mu})-m^2c^2=0.
\label{comp4qw}
\end{equation}
With the choice of the Lorentz gauge (\ref{comp1qw}), the
complex quantum relativistic Hamilton-Jacobi equation
(\ref{comp3}) can be written as
\begin{equation}
\left\lbrack {\cal S}_{\rm tot} \right\rbrack=i\hbar \square {\cal S}_{\rm
tot}.
\label{comp3q}
\end{equation}

We introduce a
complex quadrivelocity
\begin{equation}
{U}_{\mu}=-\frac{\partial_{\mu}{\cal S}_{\rm
tot}+eA_{\mu}}{m}
\label{comp5}
\end{equation}
whose components are
\begin{equation}
{U}_{0}=-\frac{1}{mc}\left (\frac{\partial {\cal S}_{\rm
tot}}{\partial t}+e U\right ),\qquad {\bf {U}}=\frac{\nabla {\cal
S}_{\rm
tot}-e{\bf A}}{m}.
\label{comp6}
\end{equation}
In consistency with our previous conventions, we have defined ${\bf
U}=(U_x,U_y,U_z)=(-U_1,-U_2,-U_3)$. We also introduce a
complex energy ${\cal E}_{\rm
tot}$ such that $U_0={\cal E}_{\rm tot}/mc$. According to Eq. (\ref{comp6}), we
have
\begin{eqnarray}
{\cal E}_{\rm tot}=-\frac { \partial {\cal S}_{\rm tot}}{\partial
t}-eU.
\label{etot}
\end{eqnarray}
Substituting Eq. (\ref{comp5}) into Eq. (\ref{comp3}), we find that the complex
quantum relativistic Hamilton-Jacobi equation takes the form
\begin{equation}
{U}_{\mu}{U}^{\mu}-c^2=-i\frac{\hbar}{m}D_{\mu}{U}^{\mu}.
\label{comp7}
\end{equation}
For $\hbar=0$, it reduces to (in that case ${u}_{\mu}$ is real):
\begin{equation}
{u}_{\mu}{u}^{\mu}=c^2.
\label{comp7wq}
\end{equation}
This corresponds to the relativistic equation of mechanics
$p_{\mu}p^{\mu}=m^2c^2$ where $m u^{\mu}$ represents the impulse $p^{\mu}$.
Taking the gradient of Eq. (\ref{comp7}) and using the
identities (\ref{comp10}), (\ref{comp10b}) and (\ref{comp10c}), we obtain the
equation
\begin{equation}
{U}^{\mu}D_{\nu}{U}_{\mu}=-i\frac{\hbar}{2m}(D_{\mu}D_{\nu}U^{\mu}-R_{\mu\nu}U^{
\mu}).
\label{comp11}
\end{equation}
Using the relation
\begin{equation}
D_{\mu}{U}_{\nu}-D_{\nu}{U}_{\mu}=-\frac{e}{m} F_{\mu\nu
}
\label{comp12}
\end{equation}
obtained from Eqs. (\ref{lag5}), (\ref{comp5}) and (\ref{comp10d}), we can
rewrite Eq. (\ref{comp11}) as
\begin{equation}
\frac{d{U}_{\nu}}{d\tau}\equiv
{U}^{\mu}D_{\mu}{U}_{\nu}=-i\frac{\hbar}{2m } \square
{U}_{\nu}-\frac{e}{m}{U}^{\mu}F_{\mu\nu}-i\frac{\hbar
e}{2m^2}D^{\mu}F_{\mu\nu}+i\frac{\hbar}{2m}R_{\mu\nu}U^{\mu}.
\label{comp14}
\end{equation}
Equation (\ref{comp14})  can be interpreted as a complex quantum relativistic
Euler-Lorentz
equation. The first term on the r.h.s. is a
relativistic viscous term with a complex viscosity
\begin{equation}
\nu=\frac{i\hbar}{2m},
\label{comp14vis}
\end{equation}
the second
term is a complex Lorentz force, the third term is a peculiar complex
electromagnetic
quantum force and the fourth term arises from the curvature of spacetime.
The KG equation (\ref{kg2}) is equivalent to the complex hydrodynamic
equation
(\ref{comp14}). For $\hbar=0$, we recover the
relativistic Euler-Lorentz
equation (in that case ${u}_{\mu}$ is real):
\begin{equation}
\frac{d{u}_{\nu}}{d\tau}\equiv
{u}^{\mu}D_{\mu}{u}_{\nu}=-\frac{e}{m}{u}^{\mu}F_{\mu\nu
}.
\label{comp14bb}
\end{equation}

\subsection{The real hydrodynamic equations}
\label{sec_rhe}

We write the SF in the de Broglie
form
\begin{equation}
\varphi=\frac{\hbar}{m}\sqrt{\rho}e^{i S_{\rm tot}/\hbar},
\label{hydro1}
\end{equation}
where $\rho$ is the pseudo rest-mass density  (\ref{emt4}) and
\begin{eqnarray}
S_{\rm tot}=\frac{\hbar}{2i}\ln \left (\frac{\varphi}{\varphi^*}\right )
\label{hydro1c}
\end{eqnarray}
is a real  action. Making the de Broglie transformation (\ref{hydro1}) in the
electromagnetic
KG equation
(\ref{kg2}),
and separating real and imaginary parts, we
obtain the pair of equations
\begin{eqnarray}
D_{\mu}\left\lbrack \rho\left (\partial^{\mu}S_{\rm tot}+e A^{\mu}\right
)\right\rbrack=0,
\label{hydro2}
\end{eqnarray}
\begin{eqnarray}
\left (\partial_{\mu}S_{\rm tot}+e A_{\mu}\right
)\left (\partial^{\mu} S_{\rm tot}+e A^{\mu}\right
)-\hbar^2\frac{\square\sqrt{\rho}}{\sqrt{\rho}}-m^2c^2-2 m^2 V'(\rho)=0.
\label{hydro3}
\end{eqnarray}
Equation (\ref{hydro2}) can be interpreted as a continuity equation and Eq. 
(\ref{hydro3}) can be interpreted as a  quantum relativistic Hamilton-Jacobi
equation with a relativistic covariant quantum potential
\begin{eqnarray}
Q=-\frac{\hbar^2}{2m}\frac{\square\sqrt{\rho}}{\sqrt{\rho}}.
\label{hydro9}
\end{eqnarray}
For $\hbar=0$, we recover the relativistic Hamilton-Jacobi equation
(\ref{comp4qw}) with the additional term
$-2m^2V'(\rho)$.

Following de Broglie, we introduce the quadrivelocity
\begin{eqnarray}
u_{\mu}=-\frac{\partial_{\mu}S_{\rm tot}+eA_{\mu}}{m}
\label{hydro4}
\end{eqnarray}
whose components are
\begin{eqnarray}
u_0=-\frac{1}{mc}\left (\frac { \partial S_{\rm tot}}{\partial
t}+eU\right ), \qquad {\bf u}=\frac{\nabla S_{\rm tot}-e{\bf A}}{m},
\label{hydro5}
\end{eqnarray}
where we have defined ${\bf u}=(u_x,u_y,u_z)=(-u_1,-u_2,-u_3)$. We
also introduce an energy $E_{\rm tot}$ such
that $u_0=E_{\rm tot}/mc$. According to Eq. (\ref{hydro5}), we have
\begin{eqnarray}
E_{\rm tot}=-\frac { \partial S_{\rm tot}}{\partial
t}-eU.
\label{hydro5c}
\end{eqnarray}
Using Eq. (\ref{hydro4}), we can rewrite Eqs.
(\ref{hydro2}) and (\ref{hydro3}) as
\begin{eqnarray}
D_{\mu}\left (\rho u^{\mu}\right )=0,
\label{hydro7}
\end{eqnarray}
\begin{eqnarray}
u_{\mu}u^{\mu}=\frac{\hbar^2}{m^2}\frac{\square\sqrt{\rho}}{\sqrt{\rho}}
+c^2+2V'(\rho).
\label{hydro8}
\end{eqnarray}
Under this form, it is clear that Eq. (\ref{hydro7}) can be
interpreted as a continuity equation (we show in Sec. \ref{sec_hrc} that it is
equivalent to the local charge conservation). On the other hand, Eq.
(\ref{hydro8}) can be interpreted as a quantum relativistic
Hamilton-Jacobi or Bernoulli equation.
For $\hbar=0$, we recover  the relativistic Hamilton-Jacobi equation
(\ref{comp7wq}) with the additional term
$2V'(\rho)$. Taking the gradient of
Eq. (\ref{hydro8}) and using the identity (\ref{comp10}) and the relation
\begin{eqnarray}
D_{\mu}u_{\nu}-D_{\nu}u_{\mu}=-\frac{e}{m} F_{\mu\nu
}
\label{hydro11}
\end{eqnarray}
obtained from Eqs. (\ref{lag5}), (\ref{hydro4}) and (\ref{comp10d}),
we get
\begin{eqnarray}
\frac{du_{\nu}}{d\tau}\equiv u^{\mu}D_{\mu}u_{\nu}=\frac{\hbar^2}{2m^2}
D_ { \nu } \left (\frac{
\square\sqrt {\rho} } { \sqrt {\rho} }\right
)-\frac{e}{m}u^{\mu}F_{ \mu\nu }+D_{\nu}V'(\rho).
\label{hydro12b}
\end{eqnarray}
Equation (\ref{hydro12b}) can be interpreted as a quantum relativistic
Euler-Lorentz equation. 
The first term on the r.h.s. is the relativistic quantum force, the second
term is the Lorentz force and the third term is a pressure force arising from
the self-interaction of the SF.
The KG equation (\ref{kg2}) is equivalent to the hydrodynamic
equations (\ref{hydro7}), (\ref{hydro8}) and (\ref{hydro12b}). For $\hbar=0$, we
recover the
relativistic Euler-Lorentz
equation (\ref{comp14bb}) with the additional term $D_{\nu}V'(\rho)$.

\subsection{The hydrodynamic representation of the current}
\label{sec_hrc}

According to Eq. (\ref{hydro1c}), we have
\begin{eqnarray}
\partial_{\mu} S_{\rm
tot}=\frac{\hbar}{2i|\varphi|^2}
(\varphi^*\partial_{\mu}\varphi-\varphi\partial_ {\mu}\varphi^*),
\label{hydro13}
\end{eqnarray}
\begin{eqnarray}
\frac{\partial S_{\rm tot}}{\partial t}=\frac{\hbar}{2i|\varphi|^2}\left
(\varphi^*\frac{\partial\varphi}{\partial
t}-\varphi\frac{\partial\varphi^*}{\partial
t}\right ),\qquad \nabla S_{\rm tot}= \frac{\hbar}{2i|\varphi|^2}\left
(\varphi^*\nabla\varphi-\varphi\nabla\varphi^*\right ).
\label{compw}
\end{eqnarray}
As a result, the quadricurrent of charge, the charge density
and the current of charge defined by Eqs. (\ref{charge3})-(\ref{charge5}) can be
rewritten as
\begin{eqnarray}
(J_e)_{\mu}=-\frac{e}{m}\rho\frac{\partial_{\mu}S_{\rm
tot}+eA_{\mu}}{m},\qquad \rho_e=-\frac{e}{mc}\rho\frac{
\frac { \partial S_{\rm tot}}{\partial
t}+eU}{mc}, \qquad {\bf J}_e= \frac{e}{m}\rho \frac{\nabla
S_{\rm tot}-e{\bf
A}}{m}.
\label{hydro14b}
\end{eqnarray}
Comparing these expressions with Eqs. (\ref{hydro4})-(\ref{hydro5c}), we
obtain
\begin{eqnarray}
(J_e)_{\mu}=\frac{e}{m}\rho u_{\mu},\qquad \rho_e=\frac{e}{mc}\rho
u_{0}=\frac{e}{m^2c^2}\rho E_{\rm tot},\qquad {\bf J}_e=\frac{e}{m}\rho {\bf u}.
\label{hydro17}
\end{eqnarray}
Using the relation (\ref{hydro17}) between the quadricurrent of charge
and the quadrivelocity, we find that the  continuity equation (\ref{hydro7})
is equivalent to the local charge conservation equation
(\ref{charge6}). On the other hand, the Maxwell equations (\ref{max1}) can be
rewritten as
\begin{eqnarray}
D_{\mu}F^{\mu\nu}=\mu_0 \frac{e}{m} \rho u^{\nu}.
\label{hydro20}
\end{eqnarray}

\subsection{The general relativistic London equation}
\label{sec_ggle}

In the case where the term
$\partial_{\mu}S_{\rm tot}$ can be neglected as compared to $eA_{\mu}$ in Eq.
(\ref{hydro14b}), the quadricurrent of charge, the charge density and the
current of charge reduce
to
\begin{eqnarray}
(J_e)_{\mu}=-\frac{e^2}{m^2}\rho A_{\mu},\qquad \rho_e=-\frac{e^2}{m^2c^2}\rho
U, \qquad {\bf J}_e=- \frac{e^2}{m^2}\rho {\bf
A}.
\label{hydro21}
\end{eqnarray}
These equations are very similar to the London
equations that were introduced phenomenologically by the brothers London
\cite{brother} in 1934 in their theory of superconductivity.\footnote{In their
Concluding Remarks, they discuss the link between their phenomenological
equations and the formulae for the charge density and the current of charge
associated with the KG equation. This exactly corresponds to the presentation
that we have given here (we arrived at these results independently). It is often
said that the London equations are purely phenomenological. Actually, they can
be {\it derived} from the KG equation. This gives a precise  mathematical
meaning to the density $\rho$ that appears in the London equations.  In their
phenomenological approach, $\rho$ is identified with the particle density. In
our approach (and in their Concluding Remarks), $\rho$ corresponds to the {\it
pseudo} rest-mass density (\ref{emt4}) that has
not a straightforward interpretation. However, in many cases, it is expected to
be of the same order of magnitude as the particle density.} Substituting Eq.
(\ref{hydro21})
into  Eq. (\ref{lw1new}), we obtain the general
relativistic London equation
\begin{eqnarray}
\square A_{\nu}-D_{\nu}D_{\mu}A^{\mu}-R_{\mu\nu}A^{\mu}=-\mu_0\frac{e^2}{m^2}
\rho A_{\nu}.
\label{lw1b}
\end{eqnarray}
With the choice of the Lorentz gauge (\ref{comp1qw}), Eq.
(\ref{lw1b}) reduces
to
\begin{eqnarray}
\square A_{\nu}-R_{\mu\nu}A^{\mu}=-\mu_0\frac{e^2}{m^2}
\rho A_{\nu}.
\label{lw1c}
\end{eqnarray}

\section{The Gross-Pitaevskii-Maxwell-Einstein equations}
\label{sec_gpme}

\subsection{The general relativistic Gross-Pitaevskii equation}
\label{sec_ggp}

The KG equation (\ref{kg2}) expressed in terms of the SF
$\varphi$ does not have a well-defined limit when $c\rightarrow +\infty$. In
order to recover the GP equation in the nonrelativistic limit
$c\rightarrow +\infty$ we have to make the Klein transformation
\begin{eqnarray}
\varphi({\bf r},t)=\frac{\hbar}{m}e^{-i m c^2 t/\hbar}\psi({\bf r},t).
\label{gpe1}
\end{eqnarray}
The new field $\psi$ will be called the pseudo wavefunction. It is related to
the pseudo
rest-mass density (\ref{emt4}) by 
\begin{eqnarray}
\rho=|\psi|^2.
\label{dwp}
\end{eqnarray}
It is only in the nonrelativistic limit $c\rightarrow +\infty$ that $\psi$
can be identified with the wavefunction and that $\rho$ can be identified with
the rest-mass density. However, we can always make the Klein transformation
(\ref{gpe1}) in the KG equation (\ref{kg2}) even if we are not in the
nonrelativistic limit \cite{abrilph1}. In that case, we obtain the
general relativistic GP equation (see Appendix
\ref{sec_det}):
\begin{eqnarray}
i\hbar c\partial^0\psi-\frac{\hbar^2}{2m}\square_e\psi
+\frac{1}{2}mc^2\left (g^{00}-1\right )\psi-e c  A^0
\psi-m \frac{dV}{d|\psi|^2}\psi-\frac{1}{2}i\hbar c
g^{\mu\nu}\Gamma_{\mu\nu}^0\psi=0.
\label{ggp1}
\end{eqnarray}
It can also be written as
\begin{eqnarray}
i\hbar g^{00}\frac{\partial\psi}{\partial t}-\frac{\hbar^2}{2m}\left
(D_{\mu}+i\frac{e}{\hbar}A_{\mu}\right
)\left (\partial^{\mu}+i\frac{e}{\hbar}A^{\mu}\right
)\psi
+\frac{1}{2}mc^2\left (g^{00}-1\right )\psi-e c g^{00} A_0
\psi-m \frac{dV}{d|\psi|^2}\psi\nonumber\\
+i\hbar c g^{0j} \left (\partial_j+i\frac{e}{\hbar}A_j\right
)\psi-\frac{1}{2}i\hbar c g^{\mu\nu}\Gamma_{\mu\nu}^0\psi=0.
\label{ggp2}
\end{eqnarray}
The GP equation (\ref{ggp1}) coupled to the Maxwell equations (\ref{max1}) and
to
the Einstein equations (\ref{ein1}) written in terms of
$\psi$ form the GPME equations.

\subsection{The charge density and the current of charge}

Using the Klein transformation (\ref{gpe1}), the quadricurrent of charge, the
charge density and the current of charge defined by Eqs.
(\ref{charge3})-(\ref{charge5}) can be expressed in terms of the pseudo
wavefunction $\psi$ as
\begin{eqnarray}
(J_e)_{\mu}=-\frac{e\hbar}{2im^2}\left (\psi^*\partial_{\mu}\psi-\psi\partial_{\mu}\psi^*
-\frac{2imc}{\hbar}|\psi|^2\delta_{\mu}^{0}\right )
-\frac{e^2}{m^2}|\psi|^2A_{\mu},
\label{mal2}
\end{eqnarray}
\begin{eqnarray}
\rho_e =-\frac {e \hbar } { 2im^2c^2 }
\left
(\psi^*\frac{\partial\psi}{\partial
t}-\psi\frac{\partial\psi^*}{\partial t}-\frac{2imc^2}{\hbar}|\psi|^2\right )
-\frac { e^2 } { m^2 c^2 }
|\psi|^2 U,
\label{gpe7}
\end{eqnarray}
\begin{eqnarray}
{\bf J}_e=\frac{e\hbar}{2im^2}
(\psi^*\nabla\psi-\psi\nabla\psi^*)-\frac{e^2}{m^2}
|\psi|^2 {\bf A}.
\label{gpe8}
\end{eqnarray}

\subsection{The complex hydrodynamic equations}
\label{sec_wa}

In this section, we restrict ourselves to the  electromagnetic KG equation 
(\ref{kg2}) without self-interaction ($V=0$). We write the pseudo wave
function under the WKB form 
\begin{eqnarray}
\psi\propto e^{i{\cal S}/\hbar},
\label{hggp1}
\end{eqnarray}
where ${\cal S}$ is a complex action. We introduce a complex quadrivelocity
\begin{equation}
{V}_{\mu}=-\frac{\partial_{\mu}{\cal S}+eA_{\mu}}{m},
\qquad {V}_{0}=-\frac{1}{mc}\left (\frac{\partial {\cal S}}{\partial t}+eU\right
),
\qquad {\bf V}=\frac{\nabla {\cal S}-e{\bf A}}{m}.
\label{hggp2}
\end{equation}
We also introduce a complex energy ${\cal E}$ such that $V_0={\cal E}/mc$.
According to Eq. (\ref{hggp2}), we have
\begin{eqnarray}
{\cal E}=-\frac { \partial {\cal S}}{\partial
t}-e U.
\label{hydro5d}
\end{eqnarray}
Combining Eqs. (\ref{gpe1}) and (\ref{hggp1}), we obtain
\begin{eqnarray}
\varphi\propto e^{i({\cal S}-mc^2t)/\hbar}.
\label{suna}
\end{eqnarray}
Comparing this equation with Eq. (\ref{comp2}), we get
\begin{eqnarray}
{\cal S}_{\rm tot}={\cal S}-mc^2 t.
\label{hggp3}
\end{eqnarray}
Therefore, we find the relations
\begin{eqnarray}
U_{\mu}=V_{\mu}+c \delta_{\mu}^{0},\qquad  U^{\mu}=V^{\mu}+c g^{\mu 0},
\label{hggp4}
\end{eqnarray}
\begin{eqnarray}
U_{0}=V_{0}+c,\qquad  {\bf U}={\bf V},\qquad {\cal E}_{\rm tot}={\cal E}+mc^2.
\label{hggp8c}
\end{eqnarray}
In order to obtain the complex hydrodynamic representation of
the general relativistic GP equation (\ref{ggp1}), we can perform the WKB
transformation (\ref{hggp1}), introduce the quadrivelocity (\ref{hggp2}) and
proceed as in Sec. \ref{sec_che}. Alternatively, using the relations
(\ref{hggp3})-(\ref{hggp8c}), we can directly rewrite the complex
hydrodynamic equations
of Sec. \ref{sec_che} in terms of ${\cal S}$ and $V_{\mu}$ instead of  ${\cal
S}_{\rm tot}$ and $U_{\mu}$. The complex quantum relativistic
Hamilton-Jacobi equation (\ref{comp7}) and the complex quantum relativistic
Euler-Lorentz equation (\ref{comp14}) become 
\begin{eqnarray}
V_{\mu}V^{\mu}=c^2(1-g^{00})-2cV^0-i\frac{\hbar}{m}D_{\mu}V^{\mu}+i\frac{\hbar
c}{m}g^{\mu\nu}\Gamma_{\mu\nu}^0,
\label{hggp4b}
\end{eqnarray}
\begin{eqnarray}
\frac{d{V}_{\nu}}{d\tau}\equiv
{V}^{\mu}D_{\mu}{V}_{\nu}=-c D^0V_{\nu}-i\frac{\hbar}{2m } \square
{V}_{\nu}-\frac{e}{m}{V}^{\mu}F_{\mu\nu}-\frac{e}{m}cF_{\nu}^0-i\frac{\hbar
e}{2m^2}D^{\mu}F_{\mu\nu}\nonumber\\
+i\frac{\hbar}{2m}R_{\mu\nu}V^{\mu}+i\frac{\hbar
c}{2m}R_
{\nu}^{0}+cV^{\mu}\Gamma_{\mu\nu}^0+c^2g^{\mu0}\Gamma_{\mu\nu}^{0}
+i\frac { \hbar c}{2m}D^{\mu}\Gamma_{\mu\nu}^0.
\label{comp14as}
\end{eqnarray}
These complex hydrodynamic
equations are equivalent to the general relativistic GP equation (\ref{ggp1}).

\subsection{The real hydrodynamic equations}
\label{sec_rheb}

We write the pseudo wave function under the
Madelung form
\begin{eqnarray}
\psi({\bf r},t)=\sqrt{\rho({\bf r},t)} e^{i{S({\bf r},t)}/\hbar},
\label{hggp5}
\end{eqnarray}
where $\rho$ is the  pseudo rest-mass density (\ref{dwp})  and
\begin{eqnarray}
S=\frac{1}{2}i\hbar\ln\left (\frac{\psi^*}{\psi}\right )
\label{whydro1c}
\end{eqnarray}
is a real action. We introduce a quadrivelocity
\begin{equation}
{v}_{\mu}=-\frac{\partial_{\mu}{S}+eA_{\mu}}{m},\qquad
{v}_{0}=-\frac{1}{mc}\left (\frac{\partial S}{\partial t}+eU\right ),\qquad {\bf
v}=\frac{\nabla S-e{\bf A}}{m}.
\label{hggp6b}
\end{equation}
We also introduce an energy $E$ such that $v_0=E/mc$. According to Eq.
(\ref{hggp6b}), we have
\begin{eqnarray}
E=-\frac { \partial S}{\partial
t}-e U.
\label{hydro5e}
\end{eqnarray}
Combining Eqs.  (\ref{gpe1}) and (\ref{hggp5}), we obtain
\begin{eqnarray}
\varphi=\frac{\hbar}{m}\sqrt{\rho}e^{i(S-mc^2t)/\hbar}.
\label{chc3}
\end{eqnarray}
Comparing this equation with Eq. (\ref{hydro1}), we get 
\begin{eqnarray}
{S}_{\rm tot}={S}-mc^2 t.
\label{hggp7}
\end{eqnarray}
Therefore, we find the relations
\begin{eqnarray}
u_{\mu}=v_{\mu}+c \delta_{\mu}^{0},\qquad  u^{\mu}=v^{\mu}+c g^{\mu 0},
\label{hggp8}
\end{eqnarray}
\begin{eqnarray}
u_{0}=v_{0}+c,\qquad  {\bf u}={\bf v},\qquad E_{\rm tot}=E+mc^2.
\label{hggp8b}
\end{eqnarray}
In order to obtain the real hydrodynamic representation of
the general relativistic GP equation (\ref{ggp1}), we can perform the
Madelung
transformation (\ref{hggp5}), introduce the quadrivelocity (\ref{hggp6b}) and
proceed as in Sec. \ref{sec_rhe}. Alternatively, using the relations
(\ref{hggp7})-(\ref{hggp8b}), we can directly rewrite the complex
hydrodynamic equations
of Sec. \ref{sec_rhe}  in terms of $S$ and $v_{\mu}$ instead of  $S_{\rm tot}$
and $u_{\mu}$. The continuity equation (\ref{hydro7}), the
quantum relativistic 
Hamilton-Jacobi equation (\ref{hydro8}), and the quantum relativistic
 Euler-Lorentz equation
(\ref{hydro12b}) become
\begin{eqnarray}
D_{\mu}\left (\rho v^{\mu}\right )+c\partial^{0}\rho=0,
\label{ot4}
\end{eqnarray}
\begin{eqnarray}
v_{\mu}v^{\mu}=-2 c
v^0+c^2(1-g^{00})+\frac{\hbar^2}{m^2}\frac{\square\sqrt{\rho}}{\sqrt{ \rho} }+2
\frac{dV}{d\rho},
\label{ot5}
\end{eqnarray}
\begin{eqnarray}
\frac{dv_{\nu}}{d\tau}\equiv
v^{\mu}D_{\mu}v_{\nu}=-c D^0v_{\nu}+\frac{\hbar^2}{2m^2}
D_ { \nu } \left (\frac{
\square\sqrt {\rho} } { \sqrt {\rho} }\right
)-\frac{e}{m}v^{\mu}F_{ \mu\nu }-\frac{e}{m}c F_{\nu
}^0+D_{\nu}V'(\rho)+cv^{\mu}\Gamma_{\mu\nu}^0+c^2g^{\mu 0}\Gamma_{\mu\nu}^0.
\label{hydro12as}
\end{eqnarray}
These hydrodynamic equations are equivalent to the general
relativistic GP equation (\ref{ggp1}).

\subsection{The hydrodynamic representation of the current}

According to Eq. (\ref{whydro1c}), we have
\begin{eqnarray}
\partial_{\mu}S=\frac{1}{2}i\hbar \frac{1}{|\psi|^2}
(\psi \partial_{\mu}\psi^*-\psi^*\partial_{\mu}\psi),
\label{chc1}
\end{eqnarray}
\begin{eqnarray}
\frac{\partial S}{\partial t}=\frac{1}{2}i\hbar \frac{1}{|\psi|^2}\left
(\psi\frac{\partial\psi^*}{\partial t}-\psi^*\frac{\partial\psi}{\partial
t}\right ),\qquad \nabla S=\frac{1}{2}i\hbar
\frac{1}{|\psi|^2}\left
(\psi\nabla\psi^*-\psi^*\nabla\psi\right ).
\label{chc1b}
\end{eqnarray}
As a result, the quadricurrent of charge, the charge density
and the current of charge defined by Eqs. (\ref{mal2})-(\ref{gpe8}) can be
rewritten as
\begin{eqnarray}
(J_e)_{\mu}=-\frac{e}{m^2}\rho(\partial_{\mu}S-mc\delta_{\mu}^0+eA_{\mu}),
\qquad \rho_ { e} =-\frac {e} {m^2c^2 } \rho\left
(\frac{\partial S}{\partial
t}-mc^2+eU\right ),\qquad {\bf J}_e=\frac{e}{m^2}\rho (\nabla S-e{\bf A}).
\label{mal3}
\end{eqnarray}
Comparing these expressions with Eq. (\ref{hggp6b}) we obtain
\begin{eqnarray}
(J_e)_{\mu}=\frac{e}{m}\rho (v_{\mu}+\delta_{\mu}^0 c),\qquad \rho_e=\frac{e}{mc}\rho
(v_{0}+c)=\frac{e}{m}\rho \left (1+\frac{E}{mc^2}\right ),\qquad {\bf
J}_e=\frac{e}{m}\rho {\bf v}.
\label{hydro17b}
\end{eqnarray}
These equations can be directly obtained from the results of Sec.
\ref{sec_hrc} by using Eqs. (\ref{hggp7})-(\ref{hggp8b}).

\section{The weak field approximation}
\label{sec_wfa}

In this section, we consider the  KGME equations
in the weak field approximation $\Phi/c^2\ll 1$. The equations that we derive
are valid
at the order $O(\Phi/c^2)$.

\subsection{The  conformal Newtonian Gauge}

We shall work with the conformal Newtonian gauge which is a perturbed form
of the Friedmann-Lema\^itre-Robertson-Walker  (FLRW) line element \cite{ma}. We
consider the simplest form of 
Newtonian gauge, only taking
into account scalar perturbations which are the ones that contribute to
the formation of structures in cosmology. We neglect anisotropic
stresses. We assume that the Universe is flat in
agreement with the observations of the cosmic microwave background (CMB).
Under these conditions, the line element is given by
\begin{equation}
ds^2=c^2\left(1+2\frac{\Phi}{c^2}\right)dt^2-a(t)^2\left(1-2\frac{\Phi}{c^2}
\right)\delta_{ij}dx^idx^j,
\label{conf1}
\end{equation}
where ${\Phi}/{c^2}\ll 1$. In this metric, $\Phi({\bf r},t)$ represents the
gravitational potential of
classical Newtonian gravity and $a(t)$ is the scale factor. The expression of
the Christoffel symbols necessary to derive the equations of the following
subsections can be found, e.g., in Appendix A of \cite{li}.

\subsection{The electromagnetic field and the first couple of Maxwell equations}

We define the electromagnetic field as a function of the
potentials by
\begin{eqnarray}
{\bf E}=-\frac{\partial {\bf A}}{\partial t}-\nabla U,\qquad {\bf
B}=\frac{1}{a}\nabla\times {\bf A}.
\label{em1}
\end{eqnarray}
Taking the curl of the electric field and the divergence of the magnetic
field, we obtain the first couple of Maxwell equations
\begin{eqnarray}
\frac{1}{a}\nabla\times {\bf E}=-\frac{\partial {\bf B}}{\partial t}-H{\bf
B},\qquad \nabla\cdot {\bf B}=0,
\label{em2}
\end{eqnarray}
where $H=\dot a/a$ is the Hubble parameter. The Faraday
tensor (\ref{lag5}) can be expressed in terms of the
electromagnetic field as
\begin{eqnarray}
F_{\mu\nu}=\left(\begin{array} {cccc}
   0    &  E_x/c  &  E_y/c  &  E_z/c  \\
-E_x/c  &   0    & -aB_z   &  aB_y  \\
-E_y/c  & aB_z &    0    &  -aB_x  \\
-E_z/c  & -aB_y   & aB_x &   0
    \end{array}
 \right).
\end{eqnarray}

\subsection{The Klein-Gordon equation}

In the weak field approximation, using the Newtonian gauge, the electromagnetic
d'Alembertian operator can be
written as
\begin{equation}
\left (1+\frac{2\Phi}{c^2}\right
)\square_e\varphi=\frac{1}{c^2}\left (\frac{\partial}{\partial
t}+i\frac{e}{\hbar}U\right
)^2\varphi-\frac{1}{a^2}\left (1+\frac{4\Phi}{c^2}\right
)\left
(\nabla-i\frac{e}{\hbar}{\bf A}\right
)^2\varphi+\left (3\frac{H}{c^2}-\frac{4}{c^4}\frac{\partial\Phi}{\partial
t}\right )\left (\frac{\partial}{\partial
t}+i\frac{e}{\hbar}U\right
)\varphi
\label{fkg4}
\end{equation}
or, equivalently, as
\begin{eqnarray}
\left (1+\frac{2\Phi}{c^2}\right
)\square_e\varphi=\frac{1}{c^2}\frac{\partial^2\varphi}{\partial
t^2}+\frac{3H}{c^2}\frac{\partial\varphi}{\partial
t}-\frac{1}{a^2}\left(1+\frac{4\Phi}{c^2}\right)\Delta\varphi
-\frac{4}{c^4}\frac{\partial\Phi}{
\partial t}\frac{\partial\varphi}{\partial
t}+i \frac{e}{\hbar} \left \lbrack\frac{1}{a^2}\left
(1+\frac{4\Phi}{c^2}\right )\nabla\cdot
{\bf A}+\frac{1}{c^2}\frac{\partial
U}{\partial t}\right \rbrack\varphi\nonumber\\
-i \frac{e}{\hbar} \biggl ( \frac{4}{c^4}U\frac{\partial\Phi}{\partial
t}-\frac{3}{c^2}H U\biggr )
\varphi+\frac{e^2}{\hbar^2}\left\lbrack \frac{1}{a^2}\left
(1+\frac{4\Phi}{c^2}\right ){\bf A}^2-\frac{1}{c^2}U^2\right\rbrack
\varphi
+2i \frac{e}{\hbar} \left \lbrack\frac{1}{a^2}\left (1+\frac{4\Phi}{c^2}\right
){\bf A}\cdot\nabla\varphi+\frac{U}{c^2}\frac{\partial \varphi}{\partial
t}\right \rbrack.
\label{mal1}
\end{eqnarray}
We note that the four first terms in Eq. (\ref{mal1}) correspond to the
d'Alemberian operator in the
absence of electromagnetic field
\begin{eqnarray}
\left (1+\frac{2\Phi}{c^2}\right
)\square=\frac{1}{c^2}\frac{\partial^2}{\partial
t^2}+\frac{3H}{c^2}\frac{\partial}{\partial
t}-\frac{1}{a^2}\left(1+\frac{4\Phi}{c^2}\right)\Delta
-\frac{4}{c^4}\frac{\partial\Phi}{
\partial t}\frac{\partial}{\partial
t}.
\label{conf2}
\end{eqnarray}
The KG equation (\ref{kg2}) takes the form
\begin{eqnarray}
\frac{1}{c^2}\frac{\partial^2\varphi}{\partial
t^2}+\frac{3H}{c^2}\frac{\partial\varphi}{\partial
t}-\frac{1}{a^2}\left(1+\frac{4\Phi}{c^2}\right)\Delta\varphi
-\frac{4}{c^4}\frac{\partial\Phi}{
\partial t}\frac{\partial\varphi}{\partial
t}+i \frac{e}{\hbar} \left \lbrack\frac{1}{a^2}\left
(1+\frac{4\Phi}{c^2}\right )\nabla\cdot
{\bf A}+\frac{1}{c^2}\frac{\partial
U}{\partial t}\right \rbrack\varphi\nonumber\\
-i \frac{e}{\hbar} \biggl ( \frac{4}{c^4}U\frac{\partial\Phi}{\partial
t}-\frac{3}{c^2}H U\biggr )
\varphi+\frac{e^2}{\hbar^2}\left\lbrack \frac{1}{a^2}\left
(1+\frac{4\Phi}{c^2}\right ){\bf A}^2-\frac{1}{c^2}U^2\right\rbrack
\varphi
+2i \frac{e}{\hbar} \left \lbrack\frac{1}{a^2}\left (1+\frac{4\Phi}{c^2}\right
){\bf A}\cdot\nabla\varphi+\frac{U}{c^2}\frac{\partial \varphi}{\partial
t}\right \rbrack\nonumber\\
+\left (1+\frac{2\Phi}{c^2}\right) \frac{m^2
c^2}{\hbar^2}\varphi
+2\left(1+2\frac{\Phi}{c^2}\right)\frac{dV}{d|\varphi|^2}\varphi =0.
\label{wkg1}
\end{eqnarray}
Using the expression (\ref{emt2}) of
the energy-momentum tensor, we find that the energy
density and the pressure are given by
\begin{eqnarray}
\epsilon=T_0^0=\frac{1}{2c^2}\left (1-\frac{2\Phi}{c^2}\right
)\left
|\frac{\partial\varphi}{\partial t}\right |^2
+\frac{1}{2a^2}\left (1+\frac{2\Phi}{c^2}\right
)|\nabla\varphi|^2+i\frac{e}{2\hbar}\frac{U}{c^2}\left
(1-\frac{2\Phi}{c^2}\right )\left
(\varphi\frac{\partial\varphi^*}{\partial
t}-\varphi^*\frac{\partial\varphi}{\partial
t}\right )\nonumber\\
+\frac{e^2}{2\hbar^2}\frac{U^2}{c^2}\left (1-\frac{2\Phi}{c^2}\right
)|\varphi|^2
-\frac{1}{2a^2}\left (1+\frac{2\Phi}{c^2}\right
)\frac{ie}{\hbar}{\bf A}\cdot
(\varphi\nabla\varphi^*-\varphi^*\nabla\varphi)
+\frac{1}{2a^2}\left (1+\frac{2\Phi}{c^2}\right
)\frac{e^2}{\hbar^2}{\bf
A}^2|\varphi|^2\nonumber\\
+\frac{m^2c^2}{2\hbar^2}|\varphi|^2+V(|\varphi|^2)+\frac{\epsilon_0}{2a^2}{\bf
E}^2+\frac{1}{2\mu_0a^2}\left (1+\frac{4\Phi}{c^2}\right ){\bf
B}^2,
\label{wkg2}
\end{eqnarray}
\begin{eqnarray}
P=-\frac{1}{3}(T_1^1+T_2^2+T_3^3)=\frac{1}{2c^2}\left (1-\frac{2\Phi}{c^2}\right
)\left
|\frac{\partial\varphi}{\partial t}\right |^2
-\frac{1}{6a^2}\left (1+\frac{2\Phi}{c^2}\right
)|\nabla\varphi|^2+i\frac{e}{2\hbar}\frac{U}{c^2}\left
(1-\frac{2\Phi}{c^2}\right )\left
(\varphi\frac{\partial\varphi^*}{\partial
t}-\varphi^*\frac{\partial\varphi}{\partial
t}\right )\nonumber\\
+\frac{e^2}{2\hbar^2}\frac{U^2}{c^2}\left (1-\frac{2\Phi}{c^2}\right
)|\varphi|^2
+\frac{1}{6a^2}\left (1+\frac{2\Phi}{c^2}\right
)\frac{ie}{\hbar}{\bf A}\cdot
(\varphi\nabla\varphi^*-\varphi^*\nabla\varphi)
-\frac{1}{6a^2}\left (1+\frac{2\Phi}{c^2}\right
)\frac{e^2}{\hbar^2}{\bf
A}^2|\varphi|^2\nonumber\\
-\frac{m^2c^2}{2\hbar^2}|\varphi|^2-V(|\varphi|^2)+\frac{\epsilon_0}{6a^2}{\bf
E}^2+\frac{1}{6\mu_0a^2}\left (1+\frac{4\Phi}{c^2}\right ){\bf
B}^2.\qquad 
\label{wkg3}
\end{eqnarray}
They can be written in the simpler form
\begin{eqnarray}
\epsilon=\frac{1}{2c^2}\left (1-\frac{2\Phi}{c^2}\right
)\left
|\frac{\partial\varphi}{\partial t}+i\frac{e}{\hbar}U\varphi\right |^2
+\frac{1}{2a^2}\left (1+\frac{2\Phi}{c^2}\right
)\left |\nabla\varphi-i\frac{e}{\hbar}{\bf
A}\varphi\right |^2\nonumber\\
+\frac{m^2c^2}{2\hbar^2}|\varphi|^2+V(|\varphi|^2)+\frac{\epsilon_0}{2a^2}{\bf
E}^2+\frac{1}{2\mu_0a^2}\left (1+\frac{4\Phi}{c^2}\right ){\bf
B}^2,
\label{wkg2simp}
\end{eqnarray}
\begin{eqnarray}
P=\frac{1}{2c^2}\left (1-\frac{2\Phi}{c^2}\right
)\left
|\frac{\partial\varphi}{\partial t}+i\frac{e}{\hbar}U\varphi\right |^2
-\frac{1}{6a^2}\left (1+\frac{2\Phi}{c^2}\right
)\left
|\nabla\varphi-i\frac{e}{\hbar}{\bf
A}\varphi\right |^2\nonumber\\
-\frac{m^2c^2}{2\hbar^2}|\varphi|^2-V(|\varphi|^2)+\frac{\epsilon_0}{6a^2}{\bf
E}^2+\frac{1}{6\mu_0a^2}\left (1+\frac{4\Phi}{c^2}\right ){\bf
B}^2.\qquad 
\label{wkg3simp}
\end{eqnarray} 
They can also be expressed in
terms of the charge
density (\ref{charge4}) and the  current of charge (\ref{charge5}) as
\begin{eqnarray}
\epsilon=\frac{1}{2c^2}\left (1-\frac{2\Phi}{c^2}\right
)\left
|\frac{\partial\varphi}{\partial t}\right |^2
+\frac{1}{2a^2}\left (1+\frac{2\Phi}{c^2}\right
)|\nabla\varphi|^2-\left (1-\frac{2\Phi}{c^2}\right
) \rho_e U\nonumber\\
-\frac{e^2}{2\hbar^2}\frac{U^2}{c^2}\left (1-\frac{2\Phi}{c^2}\right
)|\varphi|^2
-\frac{1}{a^2} \left
(1+\frac{2\Phi}{c^2}\right
){\bf A}\cdot {\bf J}_e   -\frac{1}{2a^2}\left
(1+\frac{2\Phi}{c^2}\right
)\frac{e^2}{\hbar^2}{\bf
A}^2|\varphi|^2\nonumber\\
+\frac{m^2c^2}{2\hbar^2}|\varphi|^2+V(|\varphi|^2)+\frac{\epsilon_0}{2a^2}
{\bf
E}^2+\frac{1}{2\mu_0a^2}\left (1+\frac{4\Phi}{c^2}\right ){\bf
B}^2,
\label{wkg4}
\end{eqnarray}
\begin{eqnarray}
P=\frac{1}{2c^2}\left (1-\frac{2\Phi}{c^2}\right
)\left
|\frac{\partial\varphi}{\partial t}\right |^2
-\frac{1}{6a^2}\left (1+\frac{2\Phi}{c^2}\right
)|\nabla\varphi|^2-\left (1-\frac{2\Phi}{c^2}\right
) \rho_e U\nonumber\\
-\frac{e^2}{2\hbar^2}\frac{U^2}{c^2}\left (1-\frac{2\Phi}{c^2}\right
)|\varphi|^2
+\frac{1}{3 a^2} \left
(1+\frac{2\Phi}{c^2}\right
){\bf A}\cdot {\bf J}_e   +\frac{1}{6a^2}\left
(1+\frac{2\Phi}{c^2}\right
)\frac{e^2}{\hbar^2}{\bf
A}^2|\varphi|^2\nonumber\\
-\frac{m^2c^2}{2\hbar^2}|\varphi|^2-V(|\varphi|^2)+\frac{\epsilon_0}{6a^2}
{\bf E}^2+\frac{1}{6\mu_0a^2}\left (1+\frac{4\Phi}{c^2}\right ){\bf
B}^2.
\label{wkg5}
\end{eqnarray}

\subsection{The local conservation of charge}
\label{sec_cc}

The local charge conservation equation (\ref{charge6}) can
be written as
\begin{eqnarray}
D_{\mu}J_e^{\mu}=g^ { \mu\nu }
\partial_{\mu}(J_e)_{\nu}-g^{\mu\nu}\Gamma_{\mu\nu}^{\sigma}(J_e)_{\sigma}=0.
\label{cc1}
\end{eqnarray}
In the weak field approximation, using the conformal Newtonian gauge, it takes
the form
\begin{eqnarray}
\frac{\partial \rho_e}{\partial t}+\frac{1}{a^2}\left
(1+\frac{4\Phi}{c^2}\right )\nabla\cdot {\bf J}_e-\frac{4}{c^2}\frac{\partial
\Phi}{\partial t} \rho_e+3H\rho_e=0.
\label{cc2}
\end{eqnarray}
The total charge is given by 
\begin{eqnarray}
Q=a^3\int\rho_e\left (1-\frac{4\Phi}{c^2}\right )\, d{\bf r}.
\label{cc3}
\end{eqnarray}

\subsection{The second couple of Maxwell equations}

The Maxwell equations (\ref{max1}) can be written as
\begin{eqnarray}
D_{\mu}F^{\mu}_{\,\,\,\,\nu}=D_{\mu}(g^{\mu\sigma}F_{\sigma\nu})=\mu_0
(J_e)_{\nu}.
\label{max2}
\end{eqnarray}
Using Eqs.
(\ref{id5}) and (\ref{id6}), we obtain
\begin{eqnarray}
g^{\mu\sigma}\left
(\partial_{\mu}F_{\sigma\nu}-\Gamma_{\sigma\mu}^{\rho}F_{\rho\nu}
-\Gamma_{\nu\mu}^{\rho}F_{\sigma\rho}\right )=\mu_0(J_e)_{\nu}.
\label{max5}
\end{eqnarray}
In the weak field approximation, using the conformal Newtonian gauge, the
second couple of Maxwell equations writes
\begin{eqnarray}
\frac{1}{a^2}\left (1+\frac{2\Phi}{c^2}\right
)\nabla\cdot {\bf E}-\frac{2}{a^2c^2}{\bf E}\cdot \nabla
\Phi=\frac{\rho_e}{\epsilon_0},
\label{wmax1}
\end{eqnarray}
\begin{eqnarray}
-\left (1-\frac{2\Phi}{c^2}\right )\frac{1}{c^2}\frac{\partial
{\bf
E}}{\partial t}+\frac{2}{c^4}\frac{\partial\Phi}{\partial t}{\bf
E}-\left (1-\frac{2\Phi}{c^2}\right )\frac{H}{c^2}{\bf E}+\frac{2}{c^2
a}\nabla\Phi\times {\bf B}+\frac{1}{a} \left (1+\frac{2\Phi}{c^2}\right
)\nabla\times {\bf B}=\mu_0{\bf J}_e.
\label{wmax2}
\end{eqnarray}
On the other hand, the
Lorentz gauge (\ref{comp1qw}) takes the form
\begin{eqnarray}
\frac{1}{c^2}\frac{\partial U}{\partial t}+\frac{1}{a^2}\left
(1+\frac{4\Phi}{c^2}\right )\nabla\cdot {\bf
A}-\frac{4U}{c^4}\frac{\partial\Phi}{\partial t}+\frac{3H}{c^2}U=0.
\label{wmax3}
\end{eqnarray}
Combining Eqs. (\ref{em1}), (\ref{conf2}), (\ref{wmax1}), and
(\ref{wmax2}), and making the choice of  the Lorentz gauge (\ref{wmax3}), we
find that the field
equations satisfied by the potentials $U$ and ${\bf
A}$ are
\begin{eqnarray}
\square U-\frac{2}{a^2}H\left (1+\frac{2\Phi}{c^2}\right )\nabla\cdot {\bf
A}+\frac{4}{a^2c^2}\frac{\partial\Phi}{\partial t}\nabla\cdot {\bf
A}-\frac{4U}{c^4}\frac{\partial^2\Phi}{\partial t^2}+\left
(1-\frac{2\Phi}{c^2}\right )\frac{3\dot
H}{c^2}U-\frac{2}{a^2c^2}{\bf E}\cdot \nabla\Phi=\frac{\rho_e}{\epsilon_0},
\label{wmax4}
\end{eqnarray}
\begin{eqnarray}
\square {\bf A}+\left (1-\frac{2\Phi}{c^2}\right )\frac{2H}{c^2}{\bf
E}-\frac{2}{c^2}\frac{\partial\Phi}{\partial t}{\bf
E}+\frac{2}{c^2a^2}\nabla\Phi\times (\nabla\times {\bf
A})-\frac{4}{a^2c^2}\nabla\Phi\cdot \nabla {\bf
A}+\frac{4U}{c^4}\frac{\partial\nabla\Phi}{\partial t}=\mu_0 {\bf
J}_{e}.
\label{wmax5}
\end{eqnarray}

\subsection{The Einstein equations}

The time-time component of the Einstein equations (\ref{ein1}) is
\begin{eqnarray}
R_0^0-\frac{1}{2}R=\frac{8\pi G}{c^4}T_0^0,
\label{wein1}
\end{eqnarray}
where the time-time component of the energy-momentum tensor $T_0^0$ is equal to
the
energy density $\epsilon$. In the weak field approximation, using  the Newtonian
conformal gauge, we find
\begin{eqnarray}
R_0^0-\frac{1}{2}R=\frac{3H^2}{c^2}+\frac{2}{a^2c^2}\Delta\Phi-\frac{6}{c^4}
H\left
(\frac{\partial\Phi}{\partial t}+H\Phi\right ).
\label{wein2}
\end{eqnarray}
Therefore, Eq. (\ref{wein1}) can be written as
\begin{eqnarray}
\frac{\Delta\Phi}{4\pi Ga^2}=\frac{\epsilon}{c^2}-\frac{3H^2}{8\pi
G}+\frac{3H}{4\pi G
c^2}\left (\frac{\partial\Phi}{\partial t}+H\Phi\right ).
\label{wein3}
\end{eqnarray}
Substituting Eq.
(\ref{wkg2}) into Eq. (\ref{wein3}), we obtain
\begin{eqnarray}
\frac{\Delta\Phi}{4\pi G a^2}=\frac{1}{2c^4}\left (1-\frac{2\Phi}{c^2}\right
)\left |\frac{\partial\varphi}{\partial t}\right |^2
+\frac{1}{2a^2c^2}\left (1+\frac{2\Phi}{c^2}\right
)|\nabla\varphi|^2+\frac{m^2}{2\hbar^2}|\varphi|^2+\frac{1}{c^2}
V(|\varphi|^2)-\frac{3H^2}{8\pi G}+\frac{3H}{4\pi G
c^2}\left (\frac{\partial\Phi}{\partial t}+H\Phi\right )\nonumber\\
+i\frac{e}{2\hbar}\frac{U}{c^4}\left (1-\frac{2\Phi}{c^2}\right
)\left
(\varphi\frac{\partial\varphi^*}{\partial
t}-\varphi^*\frac{\partial\varphi}{\partial
t}\right )
+\frac{e^2}{2\hbar^2}\frac{U^2}{c^4}\left (1-\frac{2\Phi}{c^2}\right
)|\varphi|^2
-\frac{1}{2a^2}\left (1+\frac{2\Phi}{c^2}\right
)\frac{ie}{\hbar c^2}{\bf A}\cdot
(\varphi\nabla\varphi^*-\varphi^*\nabla\varphi)\nonumber\\
+\frac{1}{2a^2}\left (1+\frac{2\Phi}{c^2}\right
)\frac{e^2}{\hbar^2c^2}{\bf A}^2|\varphi|^2
+\frac{\epsilon_0}{2a^2}\frac{{\bf
E}^2}{c^2}+\frac{1}{2\mu_0a^2}\left (1+\frac{4\Phi}{c^2}\right
)\frac{{\bf
B}^2}{c^2}.\nonumber\\
\label{wein4}
\end{eqnarray}
Equations (\ref{wkg1}), (\ref{wmax1}), (\ref{wmax2}) and (\ref{wein4}) form the
KGME equations in the weak field approximation. The charge density and the
current of charge appearing in the Maxwell equations (\ref{wmax1}) and
(\ref{wmax2}) are given in terms of $\varphi$ by Eqs.
(\ref{charge4}) and (\ref{charge5}).

\subsection{The Gross-Pitaevskii-Maxwell-Einstein equations}

Making the Klein transformation (\ref{gpe1})
in the KGME equations (\ref{wkg1}) and
(\ref{wein4}), we
obtain
\begin{eqnarray}
i\hbar\frac{\partial\psi}{\partial t}-\frac{\hbar^2}{2m
c^2}\frac{\partial^2\psi}{\partial t^2}-\frac{3}{2}H\frac{\hbar^2}{m
c^2}\frac{\partial\psi}{\partial t}
+\frac{\hbar^2}{2 m a^2}\left
(1+\frac{4\Phi}{c^2}\right )\Delta\psi-m\Phi \psi
-\left
(1+\frac{2\Phi}{c^2}\right )m \frac{dV}{d|\psi|^2}\psi+\frac{3}{2}i\hbar
H\psi\nonumber\\
+\frac{2\hbar^2}{m c^4}\frac{\partial\Phi}{\partial t}\left
(\frac{\partial \psi}{\partial t}-\frac{i m c^2}{\hbar}\psi\right )
-i
\frac{e\hbar}{2m} \left \lbrack\frac{1}{a^2}\left
(1+\frac{4\Phi}{c^2}\right )\nabla\cdot
{\bf A}+\frac{1}{c^2}\frac{\partial
U}{\partial t}\right \rbrack\psi
+i \frac{e\hbar}{2m} \biggl ( 4\frac{U}{c^4}
\frac{\partial\Phi}{\partial
t}
-3\frac{1}{c^2}H U\biggr )
\psi\nonumber\\
-\frac{e^2}{2m}\left\lbrack \frac{1}{a^2}\left
(1+\frac{4\Phi}{c^2}\right ){\bf A}^2-\frac{1}{c^2}U^2\right\rbrack
\psi
-i \frac{e\hbar}{m} \biggl \lbrack\frac{1}{a^2}\left
(1+\frac{4\Phi}{c^2}\right
){\bf A}\cdot\nabla\psi
+\frac{U}{c^2}\left
(\frac{\partial\psi}{\partial t}-\frac{imc^2}{\hbar}\psi\right )\biggr
\rbrack=0,
\label{gpe2}
\end{eqnarray}
\begin{eqnarray}
\frac{\Delta\Phi}{4\pi G a^2}=\left (1-\frac{\Phi}{c^2}\right
)|\psi|^2
+\frac{\hbar^2}{2m^2c^4}\left (1-\frac{2\Phi}{c^2}\right )\left
|\frac{\partial\psi}{\partial t}\right |^2+\frac{\hbar^2}{2a^2c^2m^2}\left
(1+\frac{2\Phi}{c^2}\right )|\nabla\psi|^2
+\frac{1}{c^2}V(|\psi|^2)\nonumber\\
-\frac{\hbar}{m c^2}\left
(1-\frac{2\Phi}{c^2}\right ){\rm Im} \left (\frac{\partial\psi}{\partial
t}\psi^*\right ) -\frac{3H^2}{8\pi G}+\frac{3H}{4\pi G c^2}\left
(\frac{\partial\Phi}{\partial t}+H\Phi\right )
+i\frac{e\hbar}{2m^2}\frac{U}{c^4}\left (1-\frac{2\Phi}{c^2}\right )\left
(\psi\frac{\partial\psi^*}{\partial
t}-\psi^*\frac{\partial\psi}{\partial
t}\right )\nonumber\\
-\frac{e}{mc^2}U\left
(1-\frac{2\Phi}{c^2}\right
)|\psi|^2+\frac{e^2}{2m^2}\frac{U^2}{c^4}\left
(1-\frac{2\Phi}{c^2}\right
)|\psi|^2
-\frac{1}{2a^2}\left (1+\frac{2\Phi}{c^2}\right
)\frac{ie\hbar}{m^2c^2}{\bf A}\cdot
(\psi\nabla\psi^*-\psi^*\nabla\psi)\nonumber\\
+\frac{1}{2a^2}\left (1+\frac{2\Phi}{c^2}\right
)\frac{e^2}{m^2c^2}{\bf A}^2|\psi|^2
+\frac{\epsilon_0}{2a^2}\frac{{\bf
E}^2}{c^2}+\frac{1}{2\mu_0a^2}\left (1+\frac{4\Phi}{c^2}\right
)\frac{{\bf
B}^2}{c^2}.
\label{gpe3}
\end{eqnarray}
For the quartic potential (\ref{lag4}), we get
\begin{equation}
V(|\psi|^2)=\frac{2\pi a_s\hbar^2}{m^3}|\psi|^4.
\label{gpe4}
\end{equation}
The energy density and the
pressure can be
written as
\begin{eqnarray}
\epsilon=\left (1-\frac{\Phi}{c^2}\right
)c^2|\psi|^2
+\frac{\hbar^2}{2m^2c^2}\left (1-\frac{2\Phi}{c^2}\right )\left
|\frac{\partial\psi}{\partial t}\right |^2+\frac{\hbar^2}{2a^2m^2}\left
(1+\frac{2\Phi}{c^2}\right )|\nabla\psi|^2
+V(|\psi|^2)\nonumber\\
-\frac{\hbar}{m}\left
(1-\frac{2\Phi}{c^2}\right ){\rm Im} \left (\frac{\partial\psi}{\partial
t}\psi^*\right )
+i\frac{e\hbar}{2m^2}\frac{U}{c^2}\left (1-\frac{2\Phi}{c^2}\right )\left
(\psi\frac{\partial\psi^*}{\partial
t}-\psi^*\frac{\partial\psi}{\partial
t}\right )\nonumber\\
-\frac{e}{m}U\left
(1-\frac{2\Phi}{c^2}\right
)|\psi|^2+\frac{e^2}{2m^2}\frac{U^2}{c^2}\left
(1-\frac{2\Phi}{c^2}\right
)|\psi|^2
-\frac{1}{2a^2}\left (1+\frac{2\Phi}{c^2}\right
)\frac{ie\hbar}{m^2}{\bf A}\cdot
(\psi\nabla\psi^*-\psi^*\nabla\psi)\nonumber\\
+\frac{1}{2a^2}\left (1+\frac{2\Phi}{c^2}\right
)\frac{e^2}{m^2}{\bf A}^2|\psi|^2
+\frac{\epsilon_0}{2a^2}{\bf
E}^2+\frac{1}{2\mu_0a^2}\left (1+\frac{4\Phi}{c^2}\right
){\bf
B}^2,
\label{gpe5}
\end{eqnarray}
\begin{eqnarray}
P=-\Phi |\psi|^2
+\frac{\hbar^2}{2m^2c^2}\left (1-\frac{2\Phi}{c^2}\right )\left
|\frac{\partial\psi}{\partial t}\right |^2-\frac{\hbar^2}{6a^2m^2}\left
(1+\frac{2\Phi}{c^2}\right )|\nabla\psi|^2
-V(|\psi|^2)\nonumber\\
-\frac{\hbar}{m}\left
(1-\frac{2\Phi}{c^2}\right ){\rm Im} \left (\frac{\partial\psi}{\partial
t}\psi^*\right )
+i\frac{e\hbar}{2m^2}\frac{U}{c^2}\left (1-\frac{2\Phi}{c^2}\right )\left
(\psi\frac{\partial\psi^*}{\partial
t}-\psi^*\frac{\partial\psi}{\partial
t}\right )\nonumber\\
-\frac{e}{m}U\left
(1-\frac{2\Phi}{c^2}\right
)|\psi|^2+\frac{e^2}{2m^2}\frac{U^2}{c^2}\left
(1-\frac{2\Phi}{c^2}\right
)|\psi|^2
+\frac{1}{6a^2}\left (1+\frac{2\Phi}{c^2}\right
)\frac{ie\hbar}{m^2}{\bf A}\cdot
(\psi\nabla\psi^*-\psi^*\nabla\psi)\nonumber\\
-\frac{1}{6a^2}\left (1+\frac{2\Phi}{c^2}\right
)\frac{e^2}{m^2}{\bf A}^2|\psi|^2
+\frac{\epsilon_0}{6a^2}{\bf
E}^2+\frac{1}{6\mu_0a^2}\left (1+\frac{4\Phi}{c^2}\right
){\bf
B}^2.
\label{gpe6}
\end{eqnarray}
Equations (\ref{wmax1}), (\ref{wmax2}), (\ref{gpe2}) and
(\ref{gpe3})
form
the GPME equations in the weak field
approximation.
The charge density and
the current of charge appearing in the Maxwell equations (\ref{wmax1}) and
(\ref{wmax2}) are given  in terms of the pseudo wavefunction $\psi$
by Eqs. (\ref{gpe7}) and (\ref{gpe8}).

{\it Remark:} By
introducing the electromagnetic d'Alembertian (\ref{mal1}), the GP equation
(\ref{gpe2})
can be written in the more
compact form 
\begin{eqnarray}
i\hbar\frac{\partial\psi}{\partial
t}+\frac{3}{2}i\hbar
\left (H-\frac{4}{3c^2}\frac{\partial\Phi}{\partial
t}\right )\psi-\frac{\hbar^2}{2m}\left (1+\frac{2\Phi}{c^2}\right
)\square_e\psi-eU\psi-m\Phi \psi
-\left
(1+\frac{2\Phi}{c^2}\right )m \frac{dV}{d|\psi|^2}\psi=0.
\label{gpe2wb}
\end{eqnarray}
This equation can also  be directly derived from Eq. (\ref{ggp1}) in the weak
field approximation.

\subsection{Hydrodynamic representation of the Gross-Pitaevskii equation}
\label{sec_eau}

We can obtain a hydrodynamic representation
of the GPME equations (\ref{gpe2}) and (\ref{gpe3})  by making
the Madelung transformation (\ref{hggp5}). We
introduce the velocity field\footnote{We
must be careful that this definition differs from that of Eq. (\ref{hggp6b})
because of the presence of the scale factor $a$. However, in order to avoid the
proliferation of notations, we have named the
velocity in Sec. \ref{sec_rheb} and in the present section by the same
symbol
${\bf v}$.}
\begin{eqnarray}
{\bf v}=\frac{\nabla S-e {\bf A}}{ma}.
\label{whydro2}
\end{eqnarray}
We note that
\begin{eqnarray}
\nabla\times {\bf v}=-\frac{e}{m a}\nabla\times {\bf A}=-\frac{e}{m} {\bf B}.
\label{whydro3}
\end{eqnarray}
This relation shows that the magnetic field creates a vorticity field. This
vorticity is equal to twice the Larmor pulsation $\omega_L=-(e/2m)B$. In the
absence of magnetic field, the velocity field is irrotational. We also
introduce the energy
\begin{eqnarray}
E=-\frac{\partial
S}{\partial t}-e U.
\label{whydro4}
\end{eqnarray}
We note that
\begin{eqnarray}
\nabla E+ma\frac{\partial {\bf v}}{\partial t}=-maH{\bf v}+e{\bf E}.
\label{whydro4qq}
\end{eqnarray}
Substituting Eq. (\ref{hggp5}) into the
GPME equations (\ref{gpe2}) and (\ref{gpe3}), and
separating real and imaginary
parts, we obtain the hydrodynamic equations
\begin{eqnarray}
\frac{\partial\rho}{\partial t}+3H\left
(1+\frac{E}{mc^2}\right )\rho+\frac{1}{a}\left
(1+\frac{4\Phi}{c^2}\right )\nabla\cdot (\rho {\bf
v})=-\frac{1}{mc^2}\frac{\partial}{\partial
t}(\rho E)+\frac{4}{c^2}\left (1+\frac{E}{mc^2}\right
)\rho\frac{\partial\Phi}{\partial t},
\label{whydro5}
\end{eqnarray}
\begin{eqnarray}
\frac{\partial S}{\partial t}+\frac{1}{2 m a^2}\left
(1+\frac{4\Phi}{c^2}\right )(\nabla S-e{\bf A})^2=-\frac{\hbar^2}{2 m
c^2}\frac{\frac{\partial^2\sqrt{\rho}}{\partial
t^2}}{\sqrt{\rho}}
+\left (1+\frac{4\Phi}{c^2}\right )\frac{\hbar^2}{2
m a^2}\frac{\Delta\sqrt{\rho}}{\sqrt{\rho}}\nonumber\\
-m\Phi-eU-\left (1+\frac{2\Phi}{c^2}\right )m h(\rho)
+\frac{E^2}{2mc^2}-\left
(3H-\frac{4}{c^2}\frac{\partial\Phi}{\partial
t}\right )\frac{\hbar^2}{4 m c^2 \rho}\frac{\partial\rho}{\partial t},
\label{whydro6}
\end{eqnarray}
\begin{eqnarray}
\frac{\partial {\bf v}}{\partial t}+H{\bf v}+\frac{1}{a}({\bf v}\cdot
\nabla){\bf v}=-\frac{\hbar^2}{2m^2ac^2}\nabla \left
(\frac{\frac{\partial^2\sqrt{\rho}}{\partial t^2}}{\sqrt{\rho}}\right
)
+\frac{\hbar^2}{2m^2a^3}\nabla\left\lbrack \left (1+\frac{4\Phi}{c^2}\right
)\frac{\Delta\sqrt{\rho}}{\sqrt{\rho}}\right\rbrack-\frac{1}{a}\nabla\Phi-\frac{
1}{\rho a}\nabla p\nonumber\\
-\frac{2}{a c^2}\nabla (h\Phi)-\frac{2}{a c^2}\nabla (\Phi
{\bf v}^2)
+\frac{e}{m a}\left ({\bf E}+{\bf v}\times {\bf B}\right
)+\frac{1}{2am^2c^2}\nabla
(E^2)
-\frac{3}{4}\frac{\hbar^2}{a m^2 c^2} H\nabla \left
(\frac{1}{\rho}\frac{\partial\rho}{\partial t}\right )+\frac{\hbar^2}{a m^2
c^4}\nabla \left (\frac{\partial\Phi}{\partial
t}\frac{1}{\rho}\frac{\partial\rho}{\partial t}\right ),\nonumber\\
\label{whydro7}
\end{eqnarray}
\begin{eqnarray}
\frac{\Delta\Phi}{4\pi G a^2}=\left (1-\frac{\Phi}{c^2}\right )\rho
+\frac{\hbar^2}{2m^2c^4}\left (1-\frac{2\Phi}{c^2}\right )\left\lbrack
\frac{1}{4\rho}\left (\frac{\partial\rho}{\partial t}\right
)^2+\frac{\rho}{\hbar^2}\left (\frac{\partial S}{\partial t}\right
)^2\right\rbrack
+\frac{\hbar^2}{2a^2c^2m^2}\left (1+\frac{2\Phi}{c^2}\right )\left\lbrack
\frac{1}{4\rho}(\nabla\rho)^2+\frac{\rho}{\hbar^2}(\nabla S)^2\right\rbrack
\nonumber\\
+\frac{1}{c^2}V(\rho)-\frac{1}{m c^2}\left
(1-\frac{2\Phi}{c^2}\right )\rho\frac{\partial S}{\partial t}
-\frac{3H^2}{8\pi G}+\frac{3H}{4\pi G c^2}\left (\frac{\partial\Phi}{\partial
t}+H\Phi\right )
+\frac{e}{m^2}\frac{U}{c^4}\left (1-\frac{2\Phi}{c^2}\right
)\rho\frac{\partial S}{\partial t}
-\frac{e}{mc^2}U\left
(1-\frac{2\Phi}{c^2}\right
)\rho\nonumber\\
+\frac{e^2}{2m^2}\frac{U^2}{c^4}\left
(1-\frac{2\Phi}{c^2}\right
)\rho
-\frac{1}{a^2}\left (1+\frac{2\Phi}{c^2}\right
)\frac{e}{m^2c^2}\rho {\bf A}\cdot \nabla S
+\frac{1}{2a^2}\left (1+\frac{2\Phi}{c^2}\right
)\frac{e^2}{m^2c^2}{\bf A}^2\rho
+\frac{\epsilon_0}{2a^2}\frac{{\bf
E}^2}{c^2}+\frac{1}{2\mu_0a^2}\left (1+\frac{4\Phi}{c^2}\right
)\frac{{\bf
B}^2}{c^2},\nonumber\\
\label{whydro8}
\end{eqnarray}
where $h(\rho)=V'(\rho)$ is a pseudo  enthalpy and $p(\rho)$ is a pseudo
pressure defined by the relation $h'(\rho)=p'(\rho)/\rho$ \cite{prd1}. The
pseudo pressure is
explicitly given by $p(\rho)=\rho h(\rho)-\int h(\rho)\, d\rho$, i.e.,
\begin{eqnarray}
p(\rho)=\rho V'(\rho)-V(\rho).
\label{whydro8b}
\end{eqnarray}
The pseudo velocity of sound is $c_s^2=p'(\rho)=\rho V''(\rho)$. In order to
obtain Eq. (\ref{whydro7}), we have taken the gradient of Eq.
(\ref{whydro6}) and
we have used the identity $({\bf v}\cdot\nabla){\bf
v}=\nabla
({{\bf v}^2}/{2})-{\bf v}\times
(\nabla\times {\bf v})$
together with Eq. (\ref{whydro3}). For the
quartic potential (\ref{gpe4}), we get
\begin{eqnarray}
V(\rho)=\frac{2\pi a_s\hbar^2}{m^3}\rho^2, \qquad h(\rho)=\frac{4\pi
a_s\hbar^2}{m^3}\rho,\qquad p(\rho)=\frac{2\pi a_s\hbar^2}{m^3}\rho^2,\qquad
c_s^2=\frac{4\pi a_s\hbar^2}{m^3}\rho.
\label{whydro8c}
\end{eqnarray}
The  pseudo pressure is
given by a polytropic  equation of state of index $\gamma=2$. We
note that this quadratic equation of state coincides with the equation
of state of a nonrelativistic self-interacting BEC \cite{bogoliubov,revuebec}.
This coincidence is not obvious since Eqs.
(\ref{whydro5})-(\ref{whydro8}) are
valid in the relativistic regime. The
interpretation of this equation of state is, however, not direct because $\rho$
and $p$ are a pseudo rest-mass density and a pseudo pressure that coincide with
the real rest-mass
density
and the real pressure of a BEC only in the nonrelativistic limit $c\rightarrow
+\infty$.

The energy density and the
pressure can be written in terms of hydrodynamic
variables as
\begin{eqnarray}
\epsilon=\left (1-\frac{\Phi}{c^2}\right )\rho c^2
+\frac{\hbar^2}{2m^2c^2}\left (1-\frac{2\Phi}{c^2}\right )\left\lbrack
\frac{1}{4\rho}\left (\frac{\partial\rho}{\partial t}\right
)^2+\frac{\rho}{\hbar^2}\left (\frac{\partial S}{\partial t}\right
)^2\right\rbrack
+\frac{\hbar^2}{2a^2m^2}\left (1+\frac{2\Phi}{c^2}\right )\left\lbrack
\frac{1}{4\rho}(\nabla\rho)^2+\frac{\rho}{\hbar^2}(\nabla S)^2\right\rbrack
\nonumber\\
+V(\rho)-\frac{1}{m}\left
(1-\frac{2\Phi}{c^2}\right )\rho\frac{\partial S}{\partial t}
+\frac{e}{m^2}\frac{U}{c^2}\left
(1-\frac{2\Phi}{c^2}\right
)\rho\frac{\partial S}{\partial t}
-\frac{e}{m}U\left
(1-\frac{2\Phi}{c^2}\right
)\rho\nonumber\\
+\frac{e^2}{2m^2}\frac{U^2}{c^2}\left
(1-\frac{2\Phi}{c^2}\right
)\rho
-\frac{1}{a^2}\left (1+\frac{2\Phi}{c^2}\right
)\frac{e}{m^2}\rho {\bf A}\cdot \nabla S
+\frac{1}{2a^2}\left (1+\frac{2\Phi}{c^2}\right
)\frac{e^2}{m^2}{\bf A}^2\rho
+\frac{\epsilon_0}{2a^2}{\bf
E}^2+\frac{1}{2\mu_0a^2}\left (1+\frac{4\Phi}{c^2}\right
){\bf
B}^2,\nonumber\\
\label{whydro10}
\end{eqnarray}
\begin{eqnarray}
P=-\rho\Phi
+\frac{\hbar^2}{2m^2c^2}\left (1-\frac{2\Phi}{c^2}\right )\left\lbrack
\frac{1}{4\rho}\left (\frac{\partial\rho}{\partial t}\right
)^2+\frac{\rho}{\hbar^2}\left (\frac{\partial S}{\partial t}\right
)^2\right\rbrack
-\frac{\hbar^2}{6a^2m^2}\left (1+\frac{2\Phi}{c^2}\right )\left\lbrack
\frac{1}{4\rho}(\nabla\rho)^2+\frac{\rho}{\hbar^2}(\nabla S)^2\right\rbrack
\nonumber\\
-V(\rho)-\frac{1}{m}\left
(1-\frac{2\Phi}{c^2}\right )\rho\frac{\partial S}{\partial t}
+\frac{e}{m^2}\frac{U}{c^2}\left
(1-\frac{2\Phi}{c^2}\right
)\rho\frac{\partial S}{\partial t}
-\frac{e}{m}U\left
(1-\frac{2\Phi}{c^2}\right
)\rho\nonumber\\
+\frac{e^2}{2m^2}\frac{U^2}{c^2}\left
(1-\frac{2\Phi}{c^2}\right
)\rho
+\frac{1}{3a^2}\left (1+\frac{2\Phi}{c^2}\right
)\frac{e}{m^2}\rho {\bf A}\cdot \nabla S
-\frac{1}{6a^2}\left (1+\frac{2\Phi}{c^2}\right
)\frac{e^2}{m^2}{\bf A}^2\rho
+\frac{\epsilon_0}{6a^2}{\bf
E}^2+\frac{1}{6\mu_0a^2}\left (1+\frac{4\Phi}{c^2}\right
){\bf
B}^2.\nonumber\\
\label{whydro11}
\end{eqnarray}
We note that, in general, the pressure $P$ defined by Eq. (\ref{whydro11})
differs from the
pressure $p$ defined by Eq. (\ref{whydro8b}). However, it can be shown
that they
coincide for a spatially homogeneous SF \cite{abrilph1,abrilphprep} in certain
approximations.

The hydrodynamic equations  (\ref{whydro5})-(\ref{whydro8}) have a
clear physical
interpretation. Equation (\ref{whydro5}), corresponding to the imaginary
part of the relativistic GP equation, is the  continuity equation.
We note that $\int\rho\, d{\bf r}$ is not conserved in the relativistic regime.
However, one can show  (see Sec. \ref{sec_cchr}) that the
continuity equation (\ref{whydro5}) is equivalent to the charge conservation
equation (\ref{cc2}).  Equation
(\ref{whydro6}),
corresponding to the real part of the relativistic GP equation,
is the  quantum relativistic  Hamilton-Jacobi or Bernoulli equation.
Equation (\ref{whydro7}), obtained by
taking the gradient of  Eq. (\ref{whydro6}), is the quantum
relativistic Euler-Lorentz
equation. It includes the Lorentz force $(e/m)({\bf E}+{\bf v}\times
{\bf B})$. Equation
(\ref{whydro8}) is the Einstein equation. These equations are coupled to the
Maxwell equations (\ref{wmax1}) and 
(\ref{wmax2}) in which the charge density and the current of charge are
expressed in terms of hydrodynamic variables according to Eq. (\ref{chc6}).
We stress
that the hydrodynamic equations (\ref{whydro5})-(\ref{whydro8})  are equivalent
to the GPME equations (\ref{gpe2}) and (\ref{gpe3}) which
are themselves equivalent
to the
KGME equations (\ref{wkg1}) and (\ref{wein4}).  Equations
(\ref{em2}), (\ref{wmax1}), (\ref{wmax2}) and (\ref{whydro5})-(\ref{whydro8})
form the
quantum barotropic 
Euler-Lorentz-Maxwell-Einstein (ELME) equations in the weak field
approximation.

{\it Remark:} By
introducing the d'Alembertian (\ref{conf2}), the hydrodynamic equations
(\ref{whydro5})-(\ref{whydro7}) can be written in the more
compact form
\begin{eqnarray}
\frac{\partial}{\partial t}\left \lbrack \left
(1+\frac{E}{mc^2}\right )\rho\right\rbrack+3\left
(H-\frac{4}{3c^2}\frac{\partial\Phi}{\partial
t}\right )\left
(1+\frac{E}{mc^2}\right )\rho+\frac{1}{a}\left
(1+\frac{4\Phi}{c^2}\right )\nabla\cdot (\rho {\bf
v})=0,
\label{kr1}
\end{eqnarray}
\begin{eqnarray}
\frac{\partial S}{\partial t}+\frac{1}{2 m a^2}\left
(1+\frac{4\Phi}{c^2}\right )(\nabla S-e{\bf A})^2=-\frac{\hbar^2}{2
m}\left (1+\frac{2\Phi}{c^2}\right
)\frac{\square\sqrt{\rho}}{\sqrt{\rho}}
-m\Phi-eU-\left (1+\frac{2\Phi}{c^2}\right )m h(\rho)
+\frac{E^2}{2mc^2},
\label{kr2}
\end{eqnarray}
\begin{eqnarray}
\frac{\partial {\bf v}}{\partial t}+H{\bf v}+\frac{1}{a}({\bf v}\cdot
\nabla){\bf v}=-\frac{\hbar^2}{2m^2a}\nabla
\left\lbrack \left (1+\frac{2\Phi}{c^2}\right
)\frac{\square\sqrt{\rho}}{\sqrt{\rho}}\right\rbrack-\frac{1}{a}\nabla\Phi-\frac
{
1}{\rho a}\nabla p\nonumber\\
-\frac{2}{a c^2}\nabla (h\Phi)-\frac{2}{a c^2}\nabla (\Phi
{\bf v}^2)
+\frac{e}{m a}\left ({\bf E}+{\bf v}\times {\bf B}\right
)+\frac{1}{2am^2c^2}\nabla
(E^2).
\label{kr3}
\end{eqnarray}
Introducing the energy (\ref{whydro4}), we can write the quantum relativistic
Hamilton-Jacobi equation (\ref{kr2}) as
\begin{eqnarray}
E+\frac{E^2}{2mc^2}=\frac{1}{2}m\left
(1+\frac{4\Phi}{c^2}\right ){\bf v}^2+\frac{\hbar^2}{2
m}\left (1+\frac{2\Phi}{c^2}\right
)\frac{\square\sqrt{\rho}}{\sqrt{\rho}}
+m\Phi+\left (1+\frac{2\Phi}{c^2}\right )m h(\rho).
\label{kr4}
\end{eqnarray}
This is a second degree equation for $E$ whose solutions are
\begin{eqnarray}
E=mc^2\left\lbrack -1\pm \sqrt{1+\left
(1+\frac{4\Phi}{c^2}\right )\frac{{\bf v}^2}{c^2}+\frac{\hbar^2}{
m^2c^2}\left (1+\frac{2\Phi}{c^2}\right
)\frac{\square\sqrt{\rho}}{\sqrt{\rho}}
+2\frac{\Phi}{c^2}+2\left (1+\frac{2\Phi}{c^2}\right )\frac{1}{c^2}
h(\rho)}\right\rbrack.
\label{kr5d}
\end{eqnarray}
This expression can be substituted into  Eqs.  (\ref{kr1}) and (\ref{kr3}) to
obtain a closed system of hydrodynamic equations. 

\subsection{Vortices and magnetic monopoles}

The velocity field corresponding to the Madelung transformation is defined by
Eq.
(\ref{whydro2}). The vorticity is then given by Eq.
(\ref{whydro3}). When ${\bf A}={\bf 0}$, the flow is irrotational
($\nabla\times {\bf v}={\bf 0}$). More
generally, the equation ${\bf v}+({e}/{ma}){\bf A}={\nabla S}/{ma}$ implies
the  relation
\begin{equation}
\label{vorticity0}
\nabla
\times \left ({\bf v}+\frac{e}{ma}{\bf A}\right )={\bf 0}\qquad \forall {\bf r}
\quad {\rm where}\quad \rho({\bf r})\neq 0.
\end{equation}
This relation is valid only at the points where $\nabla S/ma$
is well defined, i.e., at the points where the wavefunction (or the density)
does not vanish. When the wavefunction vanishes, its phase does not have any
meaning and neither $S$ nor $\nabla S$ is well defined (the velocity is
singular). At such points, known as nodal points, $\nabla \times [{\bf
v}+(e/ma){\bf A}]$ does not vanish in general, leading to the appearance of
singularities. If we consider the circulation of
${\bf v}+({e}/{ma}){\bf A}$ around a nodal point, we have
\begin{equation}
\label{vorticity1}
\Gamma=\oint \left ({\bf v}+\frac{e}{ma}{\bf A}\right ) \cdot d{\bf
l}=\frac{1}{ma}\oint \nabla S\cdot d{\bf l}=\frac{1}{ma}\oint dS=2\pi
n\frac{\hbar}{ma}\qquad n=0,\pm 1,\pm 2,...
\end{equation}
since the phase $S/\hbar$, when it exists, is defined up to a multiple of
$2\pi$. This relation shows that the circulation around a nodal point is
quantized in units of $h/ma$. The integer $n$ is the circulation number. Using
the Stokes theorem, we get
\begin{equation}
\label{vorticity2}
\Gamma=\int \left (\nabla \times {\bf v}+\frac{e}{m}{\bf B}\right )\cdot d{\bf
S}=2\pi n\frac{\hbar}{ma}.
\end{equation}
Therefore, $\nabla \times {\bf v}+(e/m){\bf B}$
vanishes everywhere except on certain singular points (nodal points) where it
has
singularities of the $\delta$-type.
We note that
the derivation of Eqs. (\ref{vorticity0})-(\ref{vorticity2}) does not
depend upon the dynamical equations so the results of this section are valid for
many types of waves and flows.

The previous arguments were developed by Onsager \cite{onsager}
and Feynman \cite{feynman} in the case of a quantum fluid without
electromagnetic field. In that case, the singularities correspond to vortices.
When ${\bf B}={\bf 0}$, Eqs. (\ref{vorticity1}) and (\ref{vorticity2})
reduce to (from now on, we take $a=1$):
\begin{equation}
\label{vorticity3}
\oint {\bf v} \cdot d{\bf
l}=\int \left (\nabla \times {\bf v}\right )\cdot d{\bf
S}=2\pi n\frac{\hbar}{m},
\end{equation}
so that the circulation of the velocity is quantized in units
of $h/m$. Dirac \cite{diracmm} previously developed these arguments in a
more general
context in which an electromagnetic field can be present. He focused on the
situation where $\oint {\bf v}\cdot d{\bf l}=\int (\nabla\times {\bf v})\cdot
d{\bf S}=0$. In that case, Eqs.  (\ref{vorticity1}) and (\ref{vorticity2})
reduce to
\begin{equation}
\label{vorticity4}
\oint {\bf A} \cdot d{\bf l}=\int {\bf B}\cdot d{\bf
S}=2\pi n\frac{\hbar}{e}.
\end{equation}
Following Dirac \cite{diracmm}, if  the magnetic field is created by  a magnetic
monopole, Eq. (\ref{vorticity4}) implies that its magnetic charge $\mu$
satisfies the relation
\begin{equation}
\label{vorticity5}
\frac{e\mu}{4\pi\epsilon_0\hbar c} =\frac{1}{2} n.
\end{equation}
This is known as the Dirac quantization condition. The hypothetical existence of
a magnetic monopole would imply that the electric charges must be quantized in
certain units. Inversely, the existence of the electric charges implies that the
magnetic charges of the hypothetical magnetic monopoles, if they exist, must be
quantized in units inversely proportional to the elementary electric charge.
Equation (\ref{vorticity5}) can be compared with the famous relation of quantum
electrodynamics 
\begin{equation}
\label{vorticity6}
\frac{e^2}{4\pi\epsilon_0\hbar c}=\alpha\simeq 137,
\end{equation}
where $\alpha$ is Sommerfeld's fine structure constant. The initial goal of
Dirac \cite{diracmm} was to unravel the meaning of Eq. (\ref{vorticity6}). At
the end of his
seminal paper on magnetic monopoles, he expressed his
disappointment that his theory leads to the reciprocity relation
(\ref{vorticity5}) between electricity and magnetism, instead of the purely
electronic quantum condition (\ref{vorticity6}).

\subsection{The charge and the current in the hydrodynamic representation}
\label{sec_cchr}

The charge density and the current of charge are given by Eq. (\ref{mal3}).
Comparing Eq. (\ref{mal3}) with Eqs. (\ref{whydro2}) and (\ref{whydro4}), they
can be expressed in terms of hydrodynamic variables as
\begin{eqnarray}
\rho_{e} =\frac {e}{m} \rho \left (1+\frac{E}{mc^2}\right ),\qquad {\bf
J}_e=a\frac{e}{m}\rho {\bf v}.
\label{chc6}
\end{eqnarray}
Using Eq. (\ref{chc6}), we can check that the equation of
continuity (\ref{kr1}) is equivalent to the local charge conservation equation
(\ref{cc2}).

\subsection{The London equations}
\label{sec_lonwc}

In the case where the charge density and the current of charge can be
approximated by Eq. (\ref{hydro21}), we obtain from Eqs. (\ref{wmax4}) and
(\ref{wmax5}) the London equations in the weak field approximation
\begin{eqnarray}
\square U-\frac{2}{a^2}H\left (1+\frac{2\Phi}{c^2}\right )\nabla\cdot {\bf
A}+\frac{4}{a^2c^2}\frac{\partial\Phi}{\partial t}\nabla\cdot {\bf
A}-\frac{4U}{c^4}\frac{\partial^2\Phi}{\partial t^2}+\left
(1-\frac{2\Phi}{c^2}\right )\frac{3\dot
H}{c^2}U-\frac{2}{a^2c^2}{\bf E}\cdot
\nabla\Phi=-\frac{e^2}{m^2c^2\epsilon_0}\rho U,
\label{wmax4l}
\end{eqnarray}
\begin{eqnarray}
\square {\bf A}+\left (1-\frac{2\Phi}{c^2}\right )\frac{2H}{c^2}{\bf
E}-\frac{2}{c^2}\frac{\partial\Phi}{\partial t}{\bf
E}+\frac{2}{c^2a^2}\nabla\Phi\times (\nabla\times {\bf
A})-\frac{4}{a^2c^2}\nabla\Phi\cdot \nabla {\bf
A}+\frac{4U}{c^4}\frac{\partial\nabla\Phi}{\partial
t}=-\frac{e^2\mu_0}{m^2}\rho {\bf A}.
\label{wmax5l}
\end{eqnarray}

\subsection{Hydrodynamic representation of the Klein-Gordon equation}

We can obtain a hydrodynamic representation
of the KGME equations (\ref{wkg1}) and (\ref{wein4})  by making
the Broglie transformation (\ref{hydro1}) and introducing the
velocity field
\begin{equation}
{\bf u}=\frac{\nabla S_{\rm
tot}-e{\bf A}}{ma}
\label{hykg1}
\end{equation}
and the energy 
\begin{equation}
E_{\rm tot}=-\frac{\partial
S_{\rm tot}}{\partial t}-eU.
\label{hykg2}
\end{equation}
A first possibility to obtain the hydrodynamic equations is to
substitute Eq. (\ref{hydro1}) into Eqs.  (\ref{wkg1}) and (\ref{wein4}), and
proceed as in Sec. \ref{sec_eau}. Alternatively, using the
relations (\ref{hggp7})-(\ref{hggp8b}), we can directly rewrite the hydrodynamic
equations of Sec. \ref{sec_eau} in terms of $S_{\rm tot}$, $E_{\rm tot}$
and ${\bf u}$ instead of $S$, $E$ and ${\bf v}$. The continuity equation
(\ref{kr1}), the quantum Hamilton-Jacobi equation (\ref{kr2}), and the
Euler-Lorentz equation (\ref{kr3}) become
\begin{eqnarray}
\frac{\partial}{\partial t}\left ( \frac{E_{\rm
tot}}{mc^2}\rho\right )+3\left
(H-\frac{4}{3c^2}\frac{\partial\Phi}{\partial
t}\right )\frac{E_{\rm tot}}{mc^2}\rho+\frac{1}{a}\left
(1+\frac{4\Phi}{c^2}\right )\nabla\cdot (\rho {\bf
v})=0,
\label{qkr1}
\end{eqnarray}
\begin{eqnarray}
\frac{\partial S_{\rm tot}}{\partial t}+\frac{1}{2 m a^2}\left
(1+\frac{4\Phi}{c^2}\right )(\nabla S_{\rm tot}-e{\bf A})^2=-\frac{\hbar^2}{2
m}\left (1+\frac{2\Phi}{c^2}\right
)\frac{\square\sqrt{\rho}}{\sqrt{\rho}}
-m\Phi-eU\nonumber\\
-\left (1+\frac{2\Phi}{c^2}\right )m h(\rho)
+\frac{E_{\rm tot}^2}{2mc^2}-E_{\rm tot}-\frac{1}{2}mc^2,
\label{qkr2}
\end{eqnarray}
\begin{eqnarray}
\frac{\partial {\bf u}}{\partial t}+H{\bf u}+\frac{1}{a}({\bf u}\cdot
\nabla){\bf u}=-\frac{\hbar^2}{2m^2a}\nabla
\left\lbrack \left (1+\frac{2\Phi}{c^2}\right
)\frac{\square\sqrt{\rho}}{\sqrt{\rho}}\right\rbrack-\frac{1}{a}\nabla\Phi-\frac
{
1}{\rho a}\nabla p\nonumber\\
-\frac{2}{a c^2}\nabla (h\Phi)-\frac{2}{a c^2}\nabla (\Phi
{\bf u}^2)
+\frac{e}{m a}\left ({\bf E}+{\bf u}\times {\bf B}\right
)+\frac{1}{2am^2c^2}\nabla
(E_{\rm tot}^2)-\frac{1}{a m}\nabla E_{\rm tot}.
\label{qkr3}
\end{eqnarray}
Introducing the energy (\ref{hykg2}), we can write the quantum relativistic
Hamilton-Jacobi equation (\ref{qkr2}) as
\begin{eqnarray}
E_{\rm tot}^2=m^2c^4+m^2c^2\left
(1+\frac{4\Phi}{c^2}\right ){\bf u}^2+\hbar^2c^2\left (1+\frac{2\Phi}{c^2}\right
)\frac{\square\sqrt{\rho}}{\sqrt{\rho}}
+2m^2c^2\Phi+2m^2c^2\left (1+\frac{2\Phi}{c^2}\right ) h(\rho).
\label{kr6}
\end{eqnarray}
This equation can be interpreted as the quantum generalization (also taking
into account general relativity) of the relativistic
equation
of mechanics $E_{\rm tot}^2=p^2c^2+m^2c^4$ where  $m{\bf u}$
represents the impulse ${\bf p}$. Finally, the charge
density and the current of charge (\ref{chc6}) can be expressed in terms of
hydrodynamic variables as
\begin{eqnarray}
\rho_{e} =\frac{e\rho E_{\rm tot}}{m^2c^2},\qquad {\bf
J}_e=a\frac{e}{m}\rho {\bf u}.
\label{qchc6}
\end{eqnarray}

\section{The nonrelativistic limit}
\label{sec_rnl}

In this section, we consider the nonrelativistic limit  $c\rightarrow +\infty$ 
of the KGME equations. As explained at the begining of Sec. \ref{sec_gpme}, in
order to derive the nonrelativistic equations from the KGME equations, we
first have to make the Klein transformation (\ref{gpe1}) and take the limit
$c\rightarrow
+\infty$ in the GPME equations.

\subsection{The Gross-Pitaevskii-Maxwell-Poisson equations}

For $c\rightarrow +\infty$, the GPME equations  (\ref{gpe2}) and (\ref{gpe3})
reduce to
\begin{eqnarray}
i\hbar\frac{\partial\psi}{\partial
t}+\frac{3}{2}i\hbar
H\psi=-\frac{\hbar^2}{2 m a^2}\left (\nabla-i\frac{e}{\hbar}{\bf
A}\right )^2\psi+m\Phi \psi+eU\psi
+m\frac{dV}{d|\psi|^2}\psi,
\label{nr1}
\end{eqnarray}
\begin{eqnarray}
\frac{\Delta\Phi}{4\pi G a^2}=|\psi|^2 -\frac{3H^2}{8\pi G}.
\label{nr2}
\end{eqnarray}

\subsection{The mass density and the current of mass}

One can associate to the GP equation (\ref{nr1}) a density and a current
defined by
\begin{eqnarray}
\rho =|\psi|^2, \qquad {\bf J}=\frac{\hbar}{2im}
(\psi^*\nabla\psi-\psi\nabla\psi^*)-\frac{e}{m}
|\psi|^2 {\bf A}.
\label{nr5}
\end{eqnarray}
They satisfy the local conservation equation
\begin{eqnarray}
\frac{\partial \rho}{\partial t}+\frac{1}{a^2}\nabla\cdot {\bf
J}+3H\rho=0.
\label{nr6}
\end{eqnarray}
Since the density $\rho =|\psi|^2$ is definite positive, it can be interpreted
as a probability density, or as a mass
density. Accordingly, the current ${\bf J}$ can be interpreted as a current of
probability, or as a current of mass. In that case, Eq. (\ref{nr6}) expresses
the local conservation of the normalization condition for the probability
density, or the local conservation of mass. The total mass is $M=a^3\int\rho\,
d{\bf r}$.

\subsection{The charge density and the current of charge}

For $c\rightarrow +\infty$, the
charge density and the current of charge defined by Eqs. (\ref{gpe7}) and
(\ref{gpe8}) reduce to
\begin{eqnarray}
\rho_e =\frac {e}{m}|\psi|^2, \qquad {\bf J}_e=\frac{e\hbar}{2im^2}
(\psi^*\nabla\psi-\psi\nabla\psi^*)-\frac{e^2}{m^2}
|\psi|^2 {\bf A}.
\label{nr7}
\end{eqnarray}
Comparing Eqs. (\ref{nr5}) and (\ref{nr7}), we obtain the relations
\begin{eqnarray}
\rho_e =\frac {e}{m}\rho, \qquad {\bf J}_e=\frac{e}{m} {\bf J}
\label{nr8}
\end{eqnarray}
between the density and current of charge and the density and current of mass.
The local charge conservation equation (\ref{cc2}) reduces to
\begin{eqnarray}
\frac{\partial \rho_e}{\partial t}+\frac{1}{a^2}\nabla\cdot {\bf
J}_e+3H\rho_e=0.
\label{nr9c}
\end{eqnarray}
It is equivalent to the local mass conservation equation (\ref{nr6}). The total
charge is $Q=a^3\int\rho_e\, d{\bf r}$. The
Maxwell equations are given by Eqs. (\ref{s20}) and (\ref{s21}), the
Lorentz gauge by Eq. (\ref{sa1}), and the field equations for $U$ and ${\bf A}$
by Eqs. (\ref{sa2}) and (\ref{sa3}). Of course, we cannot take the
limit $c\rightarrow +\infty$ in the Maxwell equations since they have a
relativistic origin. Equations
(\ref{nr1}), (\ref{nr2}), (\ref{s20}) and (\ref{s21})
form the GPMP equations.

\subsection{The complex hydrodynamic equations}

In this section, we restrict ourselves to the electromagnetic Schr\"odinger
equation corresponding to the GP equation (\ref{nr1}) without
self-interaction ($V=0$). It can be written as
\begin{eqnarray}
i\hbar\frac{\partial\psi}{\partial
t}+\frac{3}{2}i\hbar
H\psi=-\frac{\hbar^2}{2 m a^2}\Delta\psi+\frac{i\hbar e}{m a^2}{\bf A}\cdot
\nabla\psi+\frac{i\hbar e}{2m a^2}(\nabla\cdot {\bf A})\psi+\frac{e^2}{2m
a^2}{\bf A}^2\psi+m\Phi\psi+eU\psi.
\label{ch1}
\end{eqnarray}
We note that the third term on the r.h.s. disappears if we make the
choice of the Coulomb gauge
\begin{eqnarray}
\nabla\cdot {\bf A}=0.
\label{ch1g}
\end{eqnarray}
Making the
WKB transformation (\ref{hggp1}) in the Schr\"odinger equation (\ref{ch1}), we
obtain the complex quantum Hamilton-Jacobi equation
\begin{eqnarray}
\frac{\partial {\cal S}}{\partial t}+\frac{1}{2m a^2}(\nabla{\cal S}-e{\bf
A})^2+eU+m\Phi=\frac{i\hbar}{2ma^2}\nabla\cdot (\nabla {\cal S}-e{\bf
A})+\frac{3}{2}i\hbar H.
\label{ch3}
\end{eqnarray}
We introduce the complex velocity
\begin{eqnarray}
{\bf V}=\frac{\nabla{\cal S}-e{\bf A}}{ma}.
\label{ch4}
\end{eqnarray}
We note that
\begin{eqnarray}
\nabla\times {\bf V}=-\frac{e}{ma}\nabla\times {\bf A}=-\frac{e}{m}{\bf B}.
\label{ch5}
\end{eqnarray}
Using Eq. (\ref{ch5}), the general identities $({\bf V}\cdot\nabla){\bf
V}=\nabla
({{\bf V}^2}/{2})-{\bf
V}\times (\nabla\times {\bf V})$ and $\Delta{\bf V}=\nabla (\nabla\cdot {\bf
V})-\nabla\times(\nabla\times {\bf V})$ reduce to
\begin{eqnarray}
({\bf V}\cdot\nabla){\bf V}=\nabla \left (\frac{{\bf V}^2}{2}\right
)+\frac{e}{m}{\bf
V}\times {\bf B},\qquad \Delta{\bf V}=\nabla (\nabla\cdot {\bf
V})+\frac{e}{m}\nabla\times {\bf B}.
\label{ch6}
\end{eqnarray}
Taking the gradient of Eq. (\ref{ch3}), and using Eqs. (\ref{em1}), (\ref{ch4})
and (\ref{ch6}), we obtain the complex hydrodynamic equation
\begin{eqnarray}
\frac{\partial {\bf V}}{\partial t}+\frac{1}{a}({\bf V}\cdot\nabla){\bf
V}+H{\bf V}=\frac{i\hbar}{2ma^2}\nabla(\nabla\cdot {\bf V})+\frac{e}{ma}({\bf
E}+{\bf V}\times {\bf
B})-\frac{1}{a}\nabla\Phi.
\label{ch8in}
\end{eqnarray}
It can also be written as
\begin{eqnarray}
\frac{\partial {\bf V}}{\partial t}+\frac{1}{a}({\bf V}\cdot\nabla){\bf
V}+H{\bf V}=\frac{i\hbar}{2ma^2}\Delta{\bf
V}+\frac{e}{ma}({\bf E}+{\bf V}\times {\bf
B})-\frac{ie\hbar}{2m^2a^2}\nabla\times {\bf B}-\frac{1}{a}\nabla\Phi.
\label{ch8}
\end{eqnarray}
Equation (\ref{ch8}) can be interpreted as a complex quantum Euler-Lorentz
equation. The first term on the r.h.s. is similar to a viscous term
with a complex viscosity
\begin{eqnarray}
\nu=\frac{i\hbar}{2ma^2},
\label{ch9}
\end{eqnarray}
the second term is a complex Lorentz force
$(e/ma)({\bf E}+{\bf V}\times {\bf B})$, the third term in a peculiar complex
electromagnetic quantum force   $-(ie\hbar/2m^2a^2)\nabla\times {\bf B}$, and
the last term is the gravitational force.

{\it Remark:} Introducing the complex energy 
\begin{eqnarray}
{\cal E}=-\frac { \partial {\cal S}}{\partial
t}-e U,
\label{paris}
\end{eqnarray}
the
quantum Hamilton-Jacobi equation (\ref{ch3}) takes the form
\begin{eqnarray}
{\cal E}=\frac{1}{2}m{\bf V}^2+m\Phi-\frac{i\hbar}{2a}\nabla\cdot {\bf V}-\frac{3}{2}i\hbar H.
\end{eqnarray}

\subsection{The real hydrodynamic equations}

Making the
Madelung transformation defined by Eqs. (\ref{hggp5}) and (\ref{whydro2}) in
the GPMP equations (\ref{nr1})
and (\ref{nr2}), or taking the limit
$c\rightarrow +\infty$ in the quantum barotropic ELME equations
(\ref{whydro5})-(\ref{whydro8}),
 we obtain the hydrodynamic equations
\begin{eqnarray}
\frac{\partial\rho}{\partial t}+3H\rho+\frac{1}{a}\nabla\cdot (\rho {\bf
v})=0,
\label{class1}
\end{eqnarray}
\begin{eqnarray}
\frac{\partial S}{\partial t}+\frac{1}{2 m a^2}(\nabla S-e{\bf
A})^2=\frac{\hbar^2}{2
m a^2}\frac{\Delta\sqrt{\rho}}{\sqrt{\rho}}
-m\Phi-eU-m h(\rho),
\label{class2}
\end{eqnarray}
\begin{eqnarray}
\frac{\partial {\bf v}}{\partial t}+H{\bf v}+\frac{1}{a}({\bf v}\cdot
\nabla){\bf v}=\frac{\hbar^2}{2m^2a^3}
\nabla\left ( \frac{\Delta\sqrt{\rho}}{\sqrt{\rho}}
\right )-\frac{1}{a}\nabla\Phi-\frac{
1}{\rho a}\nabla p
+\frac{e}{m a}\left ({\bf E}+{\bf v}\times {\bf B}\right ),
\label{class3}
\end{eqnarray}
\begin{eqnarray}
\frac{\Delta\Phi}{4\pi G a^2}=\rho
-\frac{3H^2}{8\pi G}.
\label{class4}
\end{eqnarray}
These equations have a clear physical
interpretation. Eq. (\ref{class1}), corresponding to the imaginary
part of the GP equation, is the continuity equation. It accounts for the
conservation of mass. Eq.
(\ref{class2}),
corresponding to the real part of the GP equation,
is the quantum Hamilton-Jacobi or  Bernoulli  equation. It involves the Madelung
quantum potential
\begin{eqnarray}
Q=-\frac{\hbar^2}{2ma^2}\frac{\Delta\sqrt{\rho}}{\sqrt{\rho}}.
\label{qpc}
\end{eqnarray}
Equation (\ref{class3}), obtained by
taking the gradient of  Eq. (\ref{class2}), is the quantum Euler-Lorentz
equation. It includes
the  quantum force
$-(1/a)\nabla Q$, the pressure force
$-(m/\rho a)\nabla p$ arising from the self-interaction of the bosons, the
Lorentz force $(e/m)({\bf E}+{\bf v}\times
{\bf B})$ and the gravitational force $-\nabla\Phi$.
Equation (\ref{class4}) is the Poisson equation. These equations
are coupled to the Maxwell equations (\ref{s20}) and (\ref{s21}) in
which the charge density and the current of charge are
expressed in terms of hydrodynamic variables according to Eq.
(\ref{ghydro4}). We stress
that the hydrodynamic equations
(\ref{class1})-(\ref{class4}) are equivalent to
the GPMP equations (\ref{nr1}) and (\ref{nr2}). Equations
(\ref{class1})-(\ref{class4}),  (\ref{s20}) and (\ref{s21}) form the quantum
barotropic Euler-Lorentz-Maxwell-Poisson (ELMP) equations. For $\hbar=0$, we
recover the barotropic ELMP equations.

{\it Remark:} Introducing the energy
\begin{eqnarray}
E=-\frac { \partial S}{\partial
t}-e U,
\label{paris2}
\end{eqnarray}
the quantum
Hamilton-Jacobi equation (\ref{class2}) takes the form
\begin{eqnarray}
E=\frac{1}{2}m{\bf v}^2-\frac{\hbar^2}{2
m a^2}\frac{\Delta\sqrt{\rho}}{\sqrt{\rho}}
+m\Phi+m h(\rho).
\label{enec2}
\end{eqnarray}
This is the sum of the kinetic energy, the quantum potential energy, the
gravitational energy, and the enthalpy. Equation (\ref{enec2}) can be
interpreted as
the quantum generalization of the classical equation of mechanics
$E={\bf p}^2/2m+m\Phi$ where $m{\bf v}$ represents the impulse ${\bf p}$.

\subsection{Hydrodynamic representation of the current}

Combining Eqs. (\ref{dwp}), (\ref{chc1b}) and (\ref{whydro2}), or taking the
nonrelativistic limit $c\rightarrow +\infty$ in Eq. (\ref{chc6}), we find that
the
density and current of mass and charge defined by Eqs. (\ref{nr5}) and
(\ref{nr7}) can be expressed in terms of  hydrodynamic
variables as
\begin{eqnarray}
{\bf J}=a\rho{\bf v},\qquad \rho_{e} =\frac {e}{m} \rho,\qquad {\bf
J}_e=a\frac{e}{m}\rho{\bf v}=a\rho_e {\bf v}.
\label{ghydro4}
\end{eqnarray}
Using these relations, we can check that the equation of continuity
(\ref{class1}) is equivalent to the local charge conservation equation
(\ref{nr9c}).

\section{Conclusion}
\label{sec_conclusion}

In this paper, we have developed a hydrodynamic
representation of the KGME equations describing the evolution of a complex
(charged) SF. The KG equation can be viewed as a
relativistic generalization of the Schr\"odinger and GP equations for spin-$0$
bosons forming a BEC at $T=0$. These equations may find applications in the
context of  dark matter, boson stars, and neutron stars with a superfluid core.
Most
authors work in terms of the SF $\varphi$ and solve the KG
wave equation. However, the hydrodynamic equations derived in the
present paper and in previous works may be easier to implement numerically and
interpret physically.
When applied to cosmology, they
can be viewed as a quantum generalization of the
standard hydrodynamic
equations of the CDM model.  The fundamental difference with the
CDM model is the existence of a
quantum force arising from the Heisenberg uncertainty principle that
acts at small scales in order to prevent singularities. Even if this term is
difficult to calculate numerically with a good accuracy as it involves 
third order derivatives, it provides a small-scale regularization that
stabilizes the system at the scale of DM halos. Our hydrodynamic
equations also involve a pressure force
arising from the self-interaction of the bosons. When the self-interaction
is repulsive (scattering), this force has a stabilizing role. On a physical
point of view, these quantum forces (uncertainty principle and
scattering) may solve the small-scale
problems of the CDM model such
as the cusp problem and the missing satellite problem. On a practical point of
view, they can stabilize the numerical algorithm used to solve the
hydrodynamic equations. In the present paper, we have developed a general
formalism associated with the KGME equations. Astrophysical
applications of this formalism will be given in future contributions. For
example, in cosmology, a charged SF could explain the existence of the
magnetic fields in galaxies. On the other hand, if the centers of neutron
stars condensate and form BECs, they could become superfluid
and/or superconducting. This could explain the high values of the
magnetic fields observed in neutron stars because, under these conditions, the
remanent electrons are very energetic and could generate such high fields. In a
different context, boson
stars have been proposed as black hole mimickers. Finally, some people have
discussed the possibility that the Higgs field has led some imprints in the
physics of the early universe. These are some of the applications of charged SF
in cosmology and astrophysics that could be developed in future works.

\appendix

\section{Useful identities}
\label{sec_id}

In this Appendix, we list various identities that are useful in our
calculations:
\begin{eqnarray}
D_{\mu}V^{\mu}=\frac{1}{\sqrt{-g}}\partial_{\mu}(\sqrt{-g}\,  V^{\mu}),
\label{id0}
\end{eqnarray}
\begin{equation}
D_{\mu}\varphi=\partial_{\mu}\varphi,
\label{id4c}
\end{equation}
\begin{equation}
D_{\mu}V_{\nu}=\partial_{\mu}V_{\nu}-\Gamma_{\mu\nu}^{\sigma}V_{\sigma},
\label{id1}
\end{equation}
\begin{equation}
\Gamma_{\mu\nu}^{\sigma}=\Gamma_{\nu\mu}^{\sigma},
\label{id2}
\end{equation}
\begin{equation}
D_{\mu}(\varphi V_{\nu})=\varphi D_{\mu}V_{\nu}+V_{\nu}\partial_{\mu}\varphi,
\label{id4b}
\end{equation}
\begin{equation}
D_{\mu}V^{\mu}=D_{\mu}(g^{\mu\nu}V_{\nu})=g^{\mu\nu}
D_{\mu}V_{\nu}=g^{\mu\nu}\partial_{\mu}V_{\nu}-g^{\mu\nu}
\Gamma_{\mu\nu}^{\sigma}V_{\sigma},
\label{id3}
\end{equation}
\begin{equation}
D_{\mu}(\varphi V^{\mu})=\varphi D_{\mu}V^{\mu}+V^{\mu}\partial_{\mu}\varphi,
\label{id4}
\end{equation}
\begin{eqnarray}
D_{\mu}(g^{\mu\sigma}F_{\sigma\nu})=g^{\mu\sigma}D_{\mu}F_{\sigma\nu},
\label{id5}
\end{eqnarray}
\begin{eqnarray}
D_{\mu}F_{\sigma\nu}=\partial_{\mu}F_{\sigma\nu}-\Gamma_{\sigma\mu}^{\rho}
F_{\rho\nu}
-\Gamma_{\nu\mu}^{\rho}F_{\sigma\rho},
\label{id6}
\end{eqnarray}
\begin{equation}
D_{\nu}({V}_{\mu}{V}^{\mu})=2{V}^{\mu}D_{\nu}{V}_{\mu},
\label{comp10}
\end{equation}
\begin{equation}
D_{\mu}D_{\nu}V_{\alpha}-D_{\nu}D_{\mu}V_{\alpha}=-R_{\alpha\mu\nu}^{\beta}V_{
\beta},
\label{comp10b}
\end{equation}
\begin{equation}
g^{\mu\alpha}R_{\alpha\mu\nu}^{\beta}V_{\beta}=-R_{\beta\nu}V^{\beta},
\label{comp10c}
\end{equation}
\begin{equation}
D_{\mu}D_{\nu}\varphi=D_{\nu}D_{\mu}\varphi.
\label{comp10d}
\end{equation}

\section{The generalized Klein-Gordon-Maxwell-Poisson equations in an expanding
background}
\label{sec_kgp}

In this Appendix, we consider a simplified model in which we introduce the
gravitational potential $\Phi({\bf r},t)$ in the ordinary KG
equation by hand, as an external potential, and assume that this potential
is produced
by the SF itself via a generalized Poisson equation in which the source is the
energy density $\epsilon$. This leads to
the generalized KGMP equations. We then show that these equations can be
rigorously justified from the KGME
equations in the weak field limit $\Phi/c^2\rightarrow 0$.

\subsection{The Klein-Gordon equation}

We consider the FLRW metric that describes an isotropic and
homogeneous expanding background. We assume that the Universe is flat in
agreement with the observations of the CMB. The line
element in the comoving frame is
\begin{equation}
ds^2=g_{\mu\nu}dx^{\mu}dx^{\nu}=c^2dt^2-a(t)^2\delta_{ij}dx^idx^j.
\label{s1}
\end{equation}
In order to take the self-gravity of the SF into account, we
introduce a Lagrangian of interaction that couples the gravitational potential
$\Phi({\bf r},t)$ to the SF $\varphi({\bf r},t)$ according to
\begin{equation}
\mathcal{L}_{\rm int}=-\frac{m^2}{\hbar^2}\Phi|\varphi|^2.
\label{s2}
\end{equation}
The total Lagrangian of the system (SF $+$ gravity) is
given by $\mathcal{L}=\mathcal{L}_\varphi+\mathcal{L}_{\rm int}$. The equation
of motion resulting from the stationarity of the
total action $S=S_{\varphi}+S_{\rm int}$, obtained by writing $\delta
S=0$, is the KG equation
\begin{equation}
\square_e\varphi+\frac{m^2c^2}{\hbar^2}\varphi+\frac{2m^2}{\hbar^2}
\Phi\varphi+2\frac { dV } { d|\varphi|^2 }
\varphi=0.
\label{s2b}
\end{equation}
In Eq. (\ref{s2b}) the gravitational potential  $\Phi({\bf
r},t)$ acts as an external
potential. This amounts to introducing a potential of interaction of the form
\begin{equation}
V_{\rm int}(|\varphi|^2)=\frac{m^2}{\hbar^2}\Phi|\varphi|^2
\label{s4}
\end{equation}
in the KG equation (\ref{kg2}). The electromagnetic d'Alembertian operator in a
FLRW  spacetime can be written as
\begin{equation}
\square_e\varphi=\frac{1}{c^2}\left (\frac{\partial}{\partial
t}+i\frac{e}{\hbar}U\right
)^2\varphi-\frac{1}{a^2}\left
(\nabla-i\frac{e}{\hbar}{\bf A}\right
)^2\varphi+3\frac{H}{c^2}\left (\frac{\partial}{\partial
t}+i\frac{e}{\hbar}U\right
)\varphi
\label{fkg4b}
\end{equation}
or, equivalently, as
\begin{eqnarray}
\square_e\varphi=\frac{1}{c^2}\frac{\partial^2\varphi}{\partial
t^2}+\frac{3H}{c^2}\frac{\partial\varphi}{\partial
t}-\frac{1}{a^2}\Delta\varphi
+i \frac{e}{\hbar} \left
(\frac{1}{a^2}\nabla\cdot
{\bf A}+\frac{1}{c^2}\frac{\partial
U}{\partial t}+\frac{3H}{c^2}U\right )\varphi
+\frac{e^2}{\hbar^2}\left (
\frac{1}{a^2}{\bf A}^2-\frac{1}{c^2}U^2\right )
\varphi\nonumber\\
+2i \frac{e}{\hbar} \left (\frac{1}{a^2}{\bf
A}\cdot\nabla\varphi+\frac{U}{c^2}\frac{\partial \varphi}{\partial
t}\right ).
\label{s3q}
\end{eqnarray}
The first three terms in Eq. (\ref{s3q})
correspond to the d'Alembertian in a FLRW spacetime
\begin{equation}
\Box=D_{\mu}\partial^{\mu}=\frac{1}{c^2}\frac{\partial^2}{\partial
t^2}+\frac{3H}{c^2}\frac{\partial}{\partial t}-\frac{1}{a^2}\Delta.
\label{dalflrw}
\end{equation}
The KG equation (\ref{s2b}) takes the form
\begin{eqnarray}
\frac{1}{c^2}\frac{\partial^2\varphi}{\partial
t^2}+\frac{3H}{c^2}\frac{\partial\varphi}{\partial
t}-\frac{1}{a^2}\Delta\varphi
+i \frac{e}{\hbar} \left
(\frac{1}{a^2}\nabla\cdot
{\bf A}+\frac{1}{c^2}\frac{\partial
U}{\partial t}\right )\varphi
+i \frac{3e}{\hbar c^2} H U
\varphi+\frac{e^2}{\hbar^2}\left (
\frac{1}{a^2}{\bf A}^2-\frac{1}{c^2}U^2\right )
\varphi\nonumber\\
+2i \frac{e}{\hbar} \left (\frac{1}{a^2}{\bf
A}\cdot\nabla\varphi+\frac{U}{c^2}\frac{\partial \varphi}{\partial
t}\right )
+\left (1+\frac{2\Phi}{c^2}\right) \frac{m^2
c^2}{\hbar^2}\varphi
+2\frac{dV}{d|\varphi|^2}\varphi=0.\label{s3qb}
\end{eqnarray}
The energy density and the pressure defined from the diagonal part of the
energy-momentum tensor (\ref{emt2}) are given by
\begin{eqnarray}
\epsilon=\frac{1}{2c^2}\left
|\frac{\partial\varphi}{\partial t}\right |^2
+\frac{1}{2a^2}|\nabla\varphi|^2+i\frac{e}{2\hbar}\frac{U}{c^2}\left
(\varphi\frac{\partial\varphi^*}{\partial
t}-\varphi^*\frac{\partial\varphi}{\partial
t}\right )\nonumber\\
+\frac{e^2}{2\hbar^2}\frac{U^2}{c^2}|\varphi|^2
-\frac{1}{2a^2}\frac{ie}{\hbar}{\bf A}\cdot
(\varphi\nabla\varphi^*-\varphi^*\nabla\varphi)
+\frac{1}{2a^2}\frac{e^2}{\hbar^2}{\bf
A}^2|\varphi|^2\nonumber\\
+\frac{m^2c^2}{2\hbar^2}|\varphi|^2+V(|\varphi|^2)+\frac{\epsilon_0}{2a^2}
{
\bf
E}^2+\frac{1}{2\mu_0a^2}{\bf
B}^2,
\label{s5}
\end{eqnarray}
\begin{eqnarray}
P=\frac{1}{2c^2}\left
|\frac{\partial\varphi}{\partial t}\right |^2
-\frac{1}{6a^2}|\nabla\varphi|^2+i\frac{e}{2\hbar}\frac{U}{c^2}\left
(\varphi\frac{\partial\varphi^*}{\partial
t}-\varphi^*\frac{\partial\varphi}{\partial
t}\right )\nonumber\\
+\frac{e^2}{2\hbar^2}\frac{U^2}{c^2}|\varphi|^2
+\frac{1}{6a^2}\frac{ie}{\hbar}{\bf A}\cdot
(\varphi\nabla\varphi^*-\varphi^*\nabla\varphi)
-\frac{1}{6a^2}\frac{e^2}{\hbar^2}{\bf
A}^2|\varphi|^2\nonumber\\
-\frac{m^2c^2}{2\hbar^2}|\varphi|^2-V(|\varphi|^2)+\frac{\epsilon_0}{6a^2}{\bf
E}^2+\frac{1}{6\mu_0 a^2}{\bf
B}^2.
\label{s6}
\end{eqnarray}

\subsection{The local conservation of charge}

The charge conservation equation can be written as 
\begin{eqnarray}
\frac{\partial \rho_e}{\partial t}+\frac{1}{a^2}\nabla\cdot {\bf
J}_e+3H\rho_e=0.
\label{s22}
\end{eqnarray}
The total charge is $Q=a^3\int\rho_e\, d{\bf r}$.

\subsection{The Maxwell equations}

The Maxwell equations are given by
\begin{eqnarray}
\frac{1}{a}\nabla\times {\bf E}=-\frac{\partial {\bf B}}{\partial t}-H{\bf
B},\qquad \nabla\cdot {\bf B}=0,
\label{s20}
\end{eqnarray}
\begin{eqnarray}
\frac{1}{a^2}\nabla\cdot {\bf
E}=\frac{\rho_e}{\epsilon_0},\qquad \frac{1}{a} \nabla\times {\bf B}=\mu_0{\bf
J}_e+\frac{1}{c^2}\frac{\partial
{\bf
E}}{\partial t}+\frac{H}{c^2}{\bf
E}.
\label{s21}
\end{eqnarray}
The Lorentz
gauge takes the form
\begin{eqnarray}
\frac{1}{c^2}\frac{\partial U}{\partial
t}+\frac{3H}{c^2}U+\frac{1}{a^2}\nabla\cdot {\bf A}=0.
\label{sa1}
\end{eqnarray}
Within the Lorentz gauge, the field equations satisfied by the potentials  $U$
and ${\bf A}$ are given by
\begin{eqnarray}
\square U-\frac{2}{a^2}H\nabla\cdot {\bf
A}+\frac{3\dot
H}{c^2}U=\frac{\rho_e}{\epsilon_0},
\label{sa2}
\end{eqnarray}
\begin{eqnarray}
\square {\bf A}+\frac{2H}{c^2}{\bf
E}=\mu_0 {\bf
J}_{e}.
\label{sa3}
\end{eqnarray}
We also note that the fourth term in
Eq. (\ref{s3q}) disappears if we make
the choice of the Lorentz gauge.

\subsection{The generalized Poisson equation}
\label{sec_poiss}

Equation (\ref{s3qb}) is the
electromagnetic KG equation for a SF in an
external potential $\Phi({\bf r},t)$ in an expanding background. We now state
that $\Phi({\bf r},t)$ is actually the gravitational potential produced by the
SF itself. We phenomenologically assume that the gravitational
potential is determined by a generalized Poisson equation of the
form
\begin{eqnarray}
\frac{\Delta\Phi}{4\pi
Ga^2}=\frac{1}{c^2}(\epsilon-\epsilon_b)
\label{s7}
\end{eqnarray}
in which the source of the gravitational potential is the energy density
$\epsilon$ of the SF (more precisely, its deviation from the
homogeneous background density $\epsilon_b(t)$). Using Eq. (\ref{s5}) for the
energy
density of a SF, and recalling the Friedmann equation \cite{weinberg}:
\begin{eqnarray}
H^2=\frac{8\pi G}{3c^2}\epsilon_b,
\label{s8}
\end{eqnarray}
the generalized Poisson equation can be written
as
\begin{eqnarray}
\frac{\Delta\Phi}{4\pi G a^2}=\frac{1}{2c^4}\left
|\frac{\partial\varphi}{\partial t}\right |^2
+\frac{1}{2a^2c^2}|\nabla\varphi|^2+\frac{m^2}{2\hbar^2}|\varphi|^2
+\frac{1}{c^2}V(|\varphi|^2)-\frac{3H^2}{8\pi
G}
+i\frac{e}{2\hbar}\frac{U}{c^4}\left
(\varphi\frac{\partial\varphi^*}{\partial
t}-\varphi^*\frac{\partial\varphi}{\partial
t}\right )\nonumber\\
+\frac{e^2}{2\hbar^2}\frac{U^2}{c^4}|\varphi|^2
-\frac{1}{2a^2}\frac{ie}{\hbar c^2}{\bf A}\cdot
(\varphi\nabla\varphi^*-\varphi^*\nabla\varphi)
+\frac{e^2}{2a^2\hbar^2c^2}{\bf
A}^2|\varphi|^2
+\frac{\epsilon_0}{2a^2}\frac{{\bf
E}^2}{c^2}+\frac{1}{2\mu_0a^2}\frac{{\bf
B}^2}{c^2}.
\label{s9}
\end{eqnarray}
Equations  (\ref{s3qb}), (\ref{s20}), (\ref{s21}) and (\ref{s9})
form the
generalized KGMP equations. They
have been introduced in an {\it ad hoc} manner by introducing the Newtonian
potential $\Phi$ by hand, as an external potential, but they
can be rigorously justified from the  KGME equations (\ref{em1}), (\ref{wkg1}),
(\ref{wmax1}), (\ref{wmax2}) and (\ref{wein4})
in the weak field limit $\Phi/c^2\rightarrow 0$ (which, of course, is different
from the
nonrelativistic limit $c\rightarrow
+\infty$). We note that we cannot neglect the term $2\Phi/c^2$ in the
last but one term of Eq. (\ref{wkg1}) because it is multiplied by
$c^2$. This is
how the Newtonian potential $\Phi$ arises  in the KG
equation (\ref{s3qb}). Therefore, we do not have to
introduce $\Phi$ by hand: the generalized KGMP equations can be
obtained from the KGME equations  by simply neglecting terms of order $\Phi/c^2$
(when they are not multiplied by $c^2$) in these equations. The
equations derived from the generalized KGMP equations can be obtained from the
ones derived from the KGME equations in the same manner.

{\it Remark:} We could consider a simplified model in which the source of the
Newtonian potential is the pseudo rest mass density $\rho$ instead of the
energy density $\epsilon/c^2$. In that case, the generalized Poisson equation
(\ref{s7}) would be replaced by the ordinary Poisson equation 
\begin{eqnarray}
\frac{\Delta\Phi}{4\pi
Ga^2}=\rho-\frac{\epsilon_b}{c^2}.
\label{s7bb}
\end{eqnarray}
However, there is no rigorous justification of this equation in the
relativistic regime.

\subsection{The generalized Gross-Pitaevskii-Maxwell-Poisson equations}

Making the Klein transformation (\ref{gpe1}) in the generalized KGMP equations
(\ref{s3qb}) and (\ref{s9}), or taking the weak
field limit $\Phi/c^2\rightarrow 0$ in the GPME equations (\ref{gpe2}) and
(\ref{gpe3}), we obtain the generalized GPMP equations
\begin{eqnarray}
i\hbar\frac{\partial\psi}{\partial t}-\frac{\hbar^2}{2m
c^2}\frac{\partial^2\psi}{\partial t^2}-\frac{3}{2}H\frac{\hbar^2}{m
c^2}\frac{\partial\psi}{\partial t}
+\frac{\hbar^2}{2 m a^2}\Delta\psi-m\Phi \psi
-m\frac{dV}{d|\psi|^2}\psi+\frac{3}{2}i\hbar
H\psi
-i \frac{e\hbar}{2m} \left (\frac{1}{a^2}\nabla\cdot
{\bf A}+\frac{1}{c^2}\frac{\partial
U}{\partial t}\right )\psi\nonumber\\
-i \frac{3e\hbar}{2mc^2}H U\psi
-\frac{e^2}{2m}\left (\frac{1}{a^2}{\bf A}^2-\frac{1}{c^2}U^2\right )
\psi
-i \frac{e\hbar}{m} \biggl \lbrack\frac{1}{a^2}{\bf A}\cdot\nabla\psi
+\frac{U}{c^2}\left
(\frac{\partial\psi}{\partial t}-\frac{imc^2}{\hbar}\psi\right )\biggr
\rbrack=0,
\label{s10}
\end{eqnarray}
\begin{eqnarray}
\frac{\Delta\Phi}{4\pi G a^2}=|\psi|^2
+\frac{\hbar^2}{2m^2c^4}\left
|\frac{\partial\psi}{\partial t}\right
|^2+\frac{\hbar^2}{2a^2c^2m^2}|\nabla\psi|^2
+\frac{1}{c^2}V(|\psi|^2)
-\frac{\hbar}{m c^2}{\rm Im} \left (\frac{\partial\psi}{\partial
t}\psi^*\right ) -\frac{3H^2}{8\pi G}\nonumber\\
+i\frac{e\hbar}{2m^2}\frac{U}{c^4}\left
(\psi\frac{\partial\psi^*}{\partial
t}-\psi^*\frac{\partial\psi}{\partial
t}\right )
-\frac{e}{mc^2}U|\psi|^2+\frac{e^2}{2m^2}\frac{U^2}{c^4}|\psi|^2
-\frac{1}{2a^2}\frac{ie\hbar}{m^2c^2}{\bf
A}\cdot
(\psi\nabla\psi^*-\psi^*\nabla\psi)\nonumber\\
+\frac{1}{2a^2}\frac{e^2}{m^2c^2}{\bf
A}^2|\psi|^2
+\frac{\epsilon_0}{2a^2}\frac{{\bf
E}^2}{c^2}+\frac{1}{2\mu_0a^2}\frac{{\bf
B}^2}{c^2}.
\label{s11}
\end{eqnarray}

The energy density and the pressure are given by
\begin{eqnarray}
\epsilon=c^2|\psi|^2
+\frac{\hbar^2}{2m^2c^2}\left
|\frac{\partial\psi}{\partial t}\right
|^2+\frac{\hbar^2}{2a^2m^2}|\nabla\psi|^2
+V(|\psi|^2)
-\frac{\hbar}{m}{\rm Im} \left (\frac{\partial\psi}{\partial
t}\psi^*\right )
+i\frac{e\hbar}{2m^2}\frac{U}{c^2}\left
(\psi\frac{\partial\psi^*}{\partial
t}-\psi^*\frac{\partial\psi}{\partial
t}\right )\nonumber\\
-\frac{e}{m}U|\psi|^2+\frac{e^2}{2m^2}\frac{U^2}{c^2}|\psi|^2
-\frac{1}{2a^2}\frac{ie\hbar}{m^2}{\bf A}\cdot
(\psi\nabla\psi^*-\psi^*\nabla\psi)
+\frac{1}{2a^2}\frac{e^2}{m^2}{\bf A}^2|\psi|^2
+\frac{\epsilon_0}{2a^2}{\bf
E}^2+\frac{1}{2\mu_0a^2}{\bf
B}^2,
\label{s12}
\end{eqnarray}
\begin{eqnarray}
P=\frac{\hbar^2}{2m^2c^2}\left
|\frac{\partial\psi}{\partial t}\right
|^2-\frac{\hbar^2}{6a^2m^2}|\nabla\psi|^2
-V(|\psi|^2)
-\frac{\hbar}{m}{\rm Im} \left (\frac{\partial\psi}{\partial
t}\psi^*\right )
+i\frac{e\hbar}{2m^2}\frac{U}{c^2}\left
(\psi\frac{\partial\psi^*}{\partial
t}-\psi^*\frac{\partial\psi}{\partial
t}\right )\nonumber\\
-\frac{e}{m}U|\psi|^2+\frac{e^2}{2m^2}\frac{U^2}{c^2}|\psi|^2
+\frac{1}{6a^2}\frac{ie\hbar}{m^2}{\bf A}\cdot
(\psi\nabla\psi^*-\psi^*\nabla\psi)
-\frac{1}{6a^2}\frac{e^2}{m^2}{\bf A}^2|\psi|^2
+\frac{\epsilon_0}{6a^2}{\bf
E}^2+\frac{1}{6\mu_0a^2}{\bf
B}^2.
\label{s13}
\end{eqnarray}

{\it Remark:} by introducing the electromagnetic d'Alembertian operator 
(\ref{s3q}), the generalized GP equation (\ref{s10}) can be written in the more
compact form
\begin{eqnarray}
i\hbar\frac{\partial\psi}{\partial
t}+\frac{3}{2}i\hbar
H\psi-\frac{\hbar^2}{2m}\square_e\psi-eU\psi-m\Phi \psi
-m \frac{dV}{d|\psi|^2}\psi=0.
\label{gpe2w}
\end{eqnarray}
In the nonrelativistic limit $c\rightarrow +\infty$, Eqs. (\ref{s11}) and
(\ref{gpe2w}) return the GPP
equations (\ref{nr1}) and (\ref{nr2}).

\subsection{The hydrodynamic equations}

Making the Madelung transformation defined by Eqs. (\ref{hggp5}) and
(\ref{whydro2})
in the generalized GPMP equations (\ref{s10}) and (\ref{s11}), or taking the
weak field limit $\Phi/c^2\rightarrow 0$ in the quantum barotropic ELME
equations (\ref{whydro5})-(\ref{whydro8}), we obtain
the hydrodynamic equations
\begin{eqnarray}
\frac{\partial\rho}{\partial t}+3H\left
(1+\frac{E}{mc^2}\right )\rho+\frac{1}{a}\nabla\cdot (\rho
{\bf
v})=-\frac{1}{mc^2}\frac{\partial}{\partial
t}(\rho E),
\label{s14}
\end{eqnarray}
\begin{eqnarray}
\frac{\partial S}{\partial t}+\frac{1}{2 m a^2}(\nabla S-e{\bf
A})^2=-\frac{\hbar^2}{2 m
c^2}\frac{\frac{\partial^2\sqrt{\rho}}{\partial
t^2}}{\sqrt{\rho}}
+\frac{\hbar^2}{2
m a^2}\frac{\Delta\sqrt{\rho}}{\sqrt{\rho}}\nonumber\\
-m\Phi-eU-m h(\rho)
+\frac{E^2}{2mc^2}-\frac{3H\hbar^2}{4 m c^2
\rho}\frac{\partial\rho}{\partial t},
\label{s15}
\end{eqnarray}
\begin{eqnarray}
\frac{\partial {\bf v}}{\partial t}+H{\bf v}+\frac{1}{a}({\bf v}\cdot
\nabla){\bf v}=-\frac{\hbar^2}{2m^2ac^2}\nabla \left
(\frac{\frac{\partial^2\sqrt{\rho}}{\partial t^2}}{\sqrt{\rho}}\right
)
+\frac{\hbar^2}{2m^2a^3}\nabla\left (\frac{\Delta\sqrt{\rho}}{\sqrt{\rho}}
\right )-\frac{1}{a}\nabla\Phi-\frac{
1}{\rho a}\nabla p
+\frac{e}{m a}\left ({\bf E}+{\bf v}\times {\bf
B}\right )\nonumber\\
+\frac{1}{2am^2c^2}\nabla
(E^2)
-\frac{3}{4}\frac{\hbar^2}{a m^2 c^2} H\nabla \left
(\frac{1}{\rho}\frac{\partial\rho}{\partial t}\right ),
\label{s16}
\end{eqnarray}
\begin{eqnarray}
\frac{\Delta\Phi}{4\pi G a^2}=\rho
+\frac{\hbar^2}{2m^2c^4}\left\lbrack
\frac{1}{4\rho}\left (\frac{\partial\rho}{\partial t}\right
)^2+\frac{\rho}{\hbar^2}\left (\frac{\partial S}{\partial t}\right
)^2\right\rbrack
+\frac{\hbar^2}{2a^2c^2m^2}\left\lbrack
\frac{1}{4\rho}(\nabla\rho)^2+\frac{\rho}{\hbar^2}(\nabla S)^2\right\rbrack
\nonumber\\
+\frac{1}{c^2}V(\rho)-\frac{1}{m c^2}\rho\frac{\partial
S}{\partial t}
-\frac{3H^2}{8\pi G}
+\frac{e}{m^2}\frac{U}{c^4}\rho\frac{\partial
S}{\partial t}
-\frac{e}{mc^2}U\rho\nonumber\\
+\frac{e^2}{2m^2}\frac{U^2}{c^4}\rho
-\frac{1}{a^2}\frac{e}{m^2c^2}\rho {\bf A}\cdot
\nabla S
+\frac{1}{2a^2}\frac{e^2}{m^2c^2}{\bf A}^2\rho
+\frac{\epsilon_0}{2a^2}\frac{{\bf
E}^2}{c^2}+\frac{1}{2\mu_0a^2}\frac{{\bf
B}^2}{c^2}.
\label{s17}
\end{eqnarray}
The energy density and the
pressure can be written in terms of hydrodynamic
variables as
\begin{eqnarray}
\epsilon=\rho c^2
+\frac{\hbar^2}{2m^2c^2}\left\lbrack
\frac{1}{4\rho}\left (\frac{\partial\rho}{\partial t}\right
)^2+\frac{\rho}{\hbar^2}\left (\frac{\partial S}{\partial t}\right
)^2\right\rbrack
+\frac{\hbar^2}{2a^2m^2}\left\lbrack
\frac{1}{4\rho}(\nabla\rho)^2+\frac{\rho}{\hbar^2}(\nabla S)^2\right\rbrack
\nonumber\\
+V(\rho)-\frac{1}{m}\rho\frac{\partial
S}{\partial t}
+\frac{e}{m^2}\frac{U}{c^2}\rho\frac{\partial S}{\partial t}
-\frac{e}{m}U\rho\nonumber\\
+\frac{e^2}{2m^2}\frac{U^2}{c^2}\rho
-\frac{1}{a^2}\frac{e}{m^2}\rho {\bf A}\cdot
\nabla S
+\frac{1}{2a^2}\frac{e^2}{m^2}{\bf A}^2\rho
+\frac{\epsilon_0}{2a^2}{\bf
E}^2+\frac{1}{2\mu_0a^2}{\bf
B}^2,
\label{s18}
\end{eqnarray}
\begin{eqnarray}
P=\frac{\hbar^2}{2m^2c^2}\left\lbrack
\frac{1}{4\rho}\left (\frac{\partial\rho}{\partial t}\right
)^2+\frac{\rho}{\hbar^2}\left (\frac{\partial S}{\partial t}\right
)^2\right\rbrack
-\frac{\hbar^2}{6a^2m^2}\left\lbrack
\frac{1}{4\rho}(\nabla\rho)^2+\frac{\rho}{\hbar^2}(\nabla S)^2\right\rbrack
\nonumber\\
-V(\rho)-\frac{1}{m}\rho\frac{\partial
S}{\partial t}
+\frac{e}{m^2}\frac{U}{c^2}\rho\frac{\partial S}{\partial t}
-\frac{e}{m}U\rho\nonumber\\
+\frac{e^2}{2m^2}\frac{U^2}{c^2}\rho
+\frac{1}{3a^2}\frac{e}{m^2}\rho {\bf A}\cdot
\nabla S
-\frac{1}{6a^2}\frac{e^2}{m^2}{\bf A}^2\rho
+\frac{\epsilon_0}{6a^2}{\bf
E}^2+\frac{1}{6\mu_0a^2}{\bf
B}^2.
\label{s19}
\end{eqnarray}
Equations (\ref{s20}), (\ref{s21}), and (\ref{s14})-(\ref{s17}) form the
generalized quantum barotropic ELMP equations. In the
nonrelativistic limit $c\rightarrow +\infty$, we recover the quantum barotropic
ELMP equations (\ref{class1})-(\ref{class4}),  (\ref{s20}) and (\ref{s21}).

{\it Remark:} By introducing the d'Alembertian operator
(\ref{dalflrw}), the hydrodynamic equations (\ref{s14})-(\ref{s16}) can be
written in the more
compact form
\begin{eqnarray}
\frac{\partial}{\partial
t}\left \lbrack \left (1+\frac{E}{mc^2}\right
)\rho\right\rbrack+3H  \left (1+\frac{E}{mc^2}\right
) \rho+\frac{1}{a}\nabla\cdot
(\rho
{\bf
v})=0,
\label{s23}
\end{eqnarray}
\begin{eqnarray}
\frac{\partial S}{\partial t}+\frac{1}{2 m a^2}(\nabla S-e{\bf
A})^2=-\frac{\hbar^2}{2 m}\frac{\square\sqrt{\rho}}{\sqrt{\rho}}
-m\Phi-eU-m h(\rho)
+\frac{E^2}{2mc^2},
\label{s24}
\end{eqnarray}
\begin{eqnarray}
\frac{\partial {\bf v}}{\partial t}+H{\bf v}+\frac{1}{a}({\bf v}\cdot
\nabla){\bf v}=-\frac{\hbar^2}{2m^2a}\nabla \left
(\frac{\square\sqrt{\rho}}{\sqrt{\rho}}\right
)-\frac{1}{a}\nabla\Phi-\frac{
1}{\rho a}\nabla p
+\frac{e}{m a}\left ({\bf E}+{\bf v}\times {\bf
B}\right )
+\frac{1}{2am^2c^2}\nabla
(E^2).
\label{s25}
\end{eqnarray}
Using Eq. (\ref{chc6}),  we can check that the equation of
continuity (\ref{s23}) is equivalent to the local charge conservation equation
(\ref{s22}). Introducing the energy (\ref{whydro4}), we can write Eq.
(\ref{s24}) as
\begin{eqnarray}
E+\frac{E^2}{2mc^2}=\frac{1}{2}m{\bf v}^2+\frac{\hbar^2}{2
m}\frac{\square\sqrt{\rho}}{\sqrt{\rho}}
+m\Phi+m h(\rho).
\label{whydro6q}
\end{eqnarray}
This is a second degree equation for $E$ whose  solutions are
\begin{eqnarray}
E=mc^2\left\lbrack -1\pm \sqrt{1+\frac{{\bf v}^2}{c^2}+\frac{\hbar^2}{
m^2c^2}\frac{\square\sqrt{\rho}}{\sqrt{\rho}}
+2\frac{\Phi}{c^2}+2\frac{1}{c^2}
h(\rho)}\right\rbrack.
\label{kr5b}
\end{eqnarray}
This expression can be substituted into  Eqs.  (\ref{s23}) and (\ref{s25}) to
obtain a closed system of hydrodynamic equations.

\subsection{The London equations}

The London equations are given by
\begin{eqnarray}
\square U-\frac{2}{a^2}H\nabla\cdot {\bf
A}+\frac{3\dot
H}{c^2}U=-\frac{e^2}{m^2c^2\epsilon_0}\rho U,
\label{sa4}
\end{eqnarray}
\begin{eqnarray}
\square {\bf A}+\frac{2H}{c^2}{\bf
E}=-\frac{e^2\mu_0}{m^2}\rho {\bf A}.
\label{sa5}
\end{eqnarray}
We note that these equations are coupled through the terms arising
from the expansion of the universe.

\subsection{The magnetohydrodynamic  equations}

The magnetohydrodynamic (MHD) equations can be formally obtained by taking
${\bf E}={\bf 0}$ in the hydrodynamic equations of the previous subsections.
Using Eq. (\ref{hydro17b}), the
Lorentz force in Eq. (\ref{s16}) can be written as $\frac{1}{\rho a}{\bf
J}_e\times {\bf
B}$. On the other hand, the Maxwell equation (\ref{s21}) reduces to
$\frac{1}{a}\nabla\times {\bf B}=\mu_0{\bf
J}_e$.  Combining these two relations and using the identity  $({\bf
B}\cdot\nabla){\bf
B}=\nabla
({{\bf B}^2}/{2})-{\bf B}\times
(\nabla\times {\bf B})$, we obtain
\begin{eqnarray}
\frac{1}{\rho a}{\bf J}_e\times {\bf
B}=-\frac{1}{\rho a^2}\nabla \left (\frac{{\bf B}^2}{2\mu_0}\right
)+\frac{1}{\rho\mu_0 a^2}({\bf B}\cdot \nabla){\bf B}.
\label{mhd1}
\end{eqnarray}
The first term on the r.h.s. is the magnetic pressure force and
the second term is the magnetic tension force. We can define a  magnetic
pressure as $p_m={\bf B}^2/2\mu_0a^2$. We note that the magnetic pressure in Eq.
(\ref{s19}) is  $P_m={\bf B}^2/6\mu_0a^2$.
Therefore, $P_m=p_m/3$. The difference
between $P_m$ and $p_m$ basically comes from the fact that the energy-momentum
tensor $T_{i}^{j}$ is not diagonal (see Sec. 12.10 of \cite{jackson}).

\section{The generalized Klein-Gordon-Maxwell-Poisson equations in a static
background}
\label{sec_kgps}

In this Appendix, we consider a flat and static spacetime with the Minkowskian
metric
\begin{equation}
ds^2=g_{\mu\nu}dx^{\mu}dx^{\nu}=c^2dt^2-\delta_{ij}dx^idx^j,
\label{s1flat}
\end{equation}
and we introduce the Newtonian potential $\Phi$ by hand as an external
potential (see Appendix \ref{sec_kgp}).  This is equivalent to taking the weak
field limit $\Phi/c^2\rightarrow 0$ in the equations of Sec. \ref{sec_wfa} and
setting $a=1$ and $H=0$. This particular case will allow  us to simplify
certain equations of Secs. \ref{sec_kgme} and \ref{sec_gpme}.

\subsection{The Klein-Gordon equation}

The KG equation is given by
\begin{equation}
\square_e\varphi+\frac{m^2c^2}{\hbar^2}\varphi+\frac{2m^2}{\hbar^2}
\Phi\varphi+2\frac { dV } { d|\varphi|^2 }
\varphi=0.
\label{fkg1}
\end{equation}
The electromagnetic d'Alembertian operator can
be written under the equivalent forms
\begin{equation}
\square_e\varphi=\left (\partial_{\mu}+i\frac{e}{\hbar}A_{\mu}\right
)\left (\partial^{\mu}+i\frac{e}{\hbar}A^{\mu}\right
)\varphi,
\label{fkg3}
\end{equation}
\begin{equation}
\square_e\varphi=\partial_{\mu}\partial^{\mu}
\varphi+i\frac{e}{\hbar}
(\partial_{\mu}A^{ \mu })\varphi+2i\frac{e} {
\hbar}A_{\mu}\partial^{\mu}\varphi-\frac{e^2}{\hbar^2}A_{\mu}A^{\mu}
\varphi,
\label{fkg3b}
\end{equation}
\begin{equation}
\square_e\varphi=\frac{1}{c^2}\left (\frac{\partial}{\partial
t}+i\frac{e}{\hbar}U\right
)^2\varphi-\left
(\nabla-i\frac{e}{\hbar}{\bf A}\right
)^2\varphi,
\label{fkg4c}
\end{equation}
\begin{eqnarray}
\square_e\varphi=\frac{1}{c^2}\frac{\partial^2\varphi}{\partial
t^2}-\Delta\varphi+i\frac{e}{\hbar}\left (\frac{1}{c^2}\frac{\partial
U}{\partial t}+\nabla\cdot {\bf A}\right )\varphi+2i \frac{e}{\hbar} \left
(\frac{U}{c^2}\frac{\partial \varphi}{\partial
t}+{\bf
A}\cdot\nabla\varphi \right )
-\frac{e^2}{\hbar^2}\left (\frac{U^2}{c^2}-{\bf A}^2\right )
\varphi.
\label{s3}
\end{eqnarray}
The first term in Eq. (\ref{fkg3b}) and the first two terms in Eq. (\ref{s3})
correspond to the d'Alembertian in a Minkowskian
spacetime
\begin{equation}
\Box=\partial_{\mu}\partial^{\mu}=\frac{1}{c^2}\frac{\partial^2}{\partial
t^2}-\Delta.
\label{dalflat}
\end{equation}

\subsection{The local conservation of charge}

The local charge 
conservation equation writes
\begin{eqnarray}
\partial_{\mu}J_e^{\mu}=0 \qquad \Leftrightarrow
\qquad \frac{\partial\rho_e}{\partial t}+\nabla\cdot {\bf J}_e=0
\label{floc}
\end{eqnarray}
and the total charge is given by
\begin{eqnarray}
Q=\frac{1}{c}\int J_e^{0}\, d^3x=\int \rho_e \,
d^3x.
\label{fglob}
\end{eqnarray}

\subsection{The Maxwell equations}

The Maxwell equations are given by
\begin{eqnarray}
{\bf E}=-\frac{\partial {\bf A}}{\partial t}-\nabla U,\qquad {\bf
B}=\nabla\times {\bf A},
\label{ds1}
\end{eqnarray}
\begin{eqnarray}
\nabla\times {\bf E}=-\frac{\partial {\bf B}}{\partial t},\qquad
\nabla\cdot {\bf B}=0,
\label{ds2}
\end{eqnarray}
\begin{eqnarray}
\nabla\cdot {\bf
E}=\frac{\rho_e}{\epsilon_0},\qquad  \nabla\times {\bf B}=\mu_0{\bf
J}_e+\frac{1}{c^2}\frac{\partial
{\bf
E}}{\partial t}.
\label{ds3}
\end{eqnarray}
The Lorentz
gauge takes the form
\begin{eqnarray}
\partial_{\mu}A^{ \mu }=0\qquad \Leftrightarrow\qquad
\frac{1}{c^2}\frac{\partial
U}{\partial t}+\nabla\cdot {\bf A}=0.
\label{s3gauge}
\end{eqnarray}
Within the Lorentz gauge, the field equations satisfied by the potentials $U$
and ${\bf A}$ are given by
\begin{eqnarray}
\square A_{\nu}=\mu_0 (J_e)_{\nu},\qquad \square
U=\frac{\rho_e}{\epsilon_0},\qquad \square
{\bf A}=\mu_0 {\bf J}_e.
\label{lw1bb}
\end{eqnarray}
We also note that the second term in 
Eq. (\ref{fkg3b})  and the third term in
Eq. (\ref{s3}) disappear if we make
the choice of the Lorentz gauge.

\subsection{The generalized Poisson equation}

The generalized Poisson equation writes (see Appendix \ref{sec_poiss}):
\begin{eqnarray}
\frac{\Delta\Phi}{4\pi
G}=\frac{\epsilon}{c^2}=\frac{1}{2c^4}\left
|\frac{\partial\varphi}{\partial t}\right |^2
+\frac{1}{2c^2}|\nabla\varphi|^2+\frac{m^2}{2\hbar^2}|\varphi|^2
+\frac{1}{c^2}V(|\varphi|^2)-\frac{3H^2}{8\pi
G}
+i\frac{e}{2\hbar}\frac{U}{c^4}\left
(\varphi\frac{\partial\varphi^*}{\partial
t}-\varphi^*\frac{\partial\varphi}{\partial
t}\right )\nonumber\\
+\frac{e^2}{2\hbar^2}\frac{U^2}{c^4}|\varphi|^2
-\frac{1}{2}\frac{ie}{\hbar c^2}{\bf A}\cdot
(\varphi\nabla\varphi^*-\varphi^*\nabla\varphi)
+\frac{e^2}{2\hbar^2c^2}{\bf
A}^2|\varphi|^2
+\frac{\epsilon_0}{2}\frac{{\bf
E}^2}{c^2}+\frac{1}{2\mu_0}\frac{{\bf
B}^2}{c^2}.
\label{ein2}
\end{eqnarray}
Equations (\ref{fkg1}), (\ref{ds2}), (\ref{ds3}) and (\ref{ein2}) form the
generalized KGMP equations.

\subsection{The complex hydrodynamic equations}
\label{sec_gr1}

Making the WKB transformation (\ref{comp2}) in the KG
equation (\ref{fkg1}),
we obtain the
complex quantum
relativistic Hamilton-Jacobi equation (we assume $V=0$):
\begin{equation}
\frac{1}{c^2}\left (\frac{\partial {\cal S}_{\rm
tot}}{\partial t}+e U\right )^2-\left ({\nabla{\cal S}_{\rm
tot}}-e {\bf A}\right )^2 -m^2c^2-2m^2\Phi=i\hbar
\square {\cal S}_{\rm tot}+ie\hbar \left (\frac{1}{c^2}\frac{\partial
U}{\partial t}+\nabla\cdot {\bf A}\right ).
\label{comp8}
\end{equation}
Introducing the complex
quadrivelocity (\ref{comp5}), it can be
rewritten as
\begin{equation}
{U}_{\mu}{U}^{\mu}-c^2-2\Phi=-i\frac{\hbar}{m}\partial_{\mu}{U}^{\mu}.
\label{comp9}
\end{equation}
Taking the gradient of this equation and proceeding as in Sec. \ref{sec_che},
we obtain the complex quantum relativistic Euler-Lorentz
equation 
\begin{equation}
\frac{d{U}_{\nu}}{d\tau}\equiv
{U}^{\mu}\partial_{\mu}{U}_{\nu}=-i\frac{\hbar}{2m } \square
{U}_{\nu}-\frac{e}{m}{U}^{\mu}F_{\mu\nu}-i\frac{\hbar
e}{2m^2}\partial^{\mu}F_{\mu\nu}+\partial_{\nu}\Phi.
\label{comp14ann}
\end{equation}
The first
term on the r.h.s. is a
relativistic viscous term with a complex viscosity given by Eq.
(\ref{comp14vis}), the second term is a complex Lorentz force, the third term is
a peculiar complex electromagnetic quantum force and the fourth term is the
gravitational force. The KG
equation (\ref{fkg1}) is equivalent to the complex hydrodynamic equation
(\ref{comp14ann}). For $\hbar=0$, we recover the
relativistic Euler-Lorentz
equation (in that case ${u}_{\mu}$ is real):
\begin{equation}
\frac{d{u}_{\nu}}{d\tau}\equiv
{u}^{\mu}\partial_{\mu}{u}_{\nu}=-\frac{e}{m}{u}^{\mu}F_{\mu\nu
}+\partial_{\nu}\Phi.
\label{comp14b}
\end{equation}
Introducing the complex energy (\ref{etot}), the complex quantum
relativistic Hamilton-Jacobi equation (\ref{comp8}) can be
written as
\begin{eqnarray}
{\cal E}_{\rm tot}^2=m^2c^4+m^2c^2{\bf U}^2+2m^2c^2\Phi-i\hbar\frac{\partial
{\cal E}_{\rm tot}}{\partial t}-i\hbar mc^2\nabla\cdot {\bf U}.
\label{kr6e}
\end{eqnarray}

\subsection{The real hydrodynamic equations}
\label{sec_gr2}

Making the de Broglie transformation (\ref{hydro1}) in the KG
equation (\ref{fkg1}), separating real and imaginary parts, and introducing the
quadrivelocity (\ref{hydro4}), we
obtain
the continuity equation
\begin{eqnarray}
\partial_{\mu}\left (\rho u^{\mu}\right )=0
\label{hydro7b}
\end{eqnarray}
and the quantum relativistic Hamilton-Jacobi equation
\begin{eqnarray}
u_{\mu}u^{\mu}=\frac{\hbar^2}{m^2}\frac{\square\sqrt{\rho}}{\sqrt{\rho}}
+c^2+2\Phi+2V'(\rho).
\label{hydro8b}
\end{eqnarray}
Taking the gradient of
Eq. (\ref{hydro8b}) and proceeding as in Sec. \ref{sec_rhe}, we obtain the
quantum relativistic Euler-Lorentz equation
\begin{eqnarray}
\frac{du_{\nu}}{d\tau}\equiv u^{\mu}\partial_{\mu}u_{\nu}=\frac{\hbar^2}{2m^2}
\partial_ { \nu } \left (\frac{
\square\sqrt {\rho} } { \sqrt {\rho} }\right
)-\frac{e}{m}u^{\mu}F_{ \mu\nu }+\partial_{\nu}\Phi+\partial_{\nu}V'(\rho).
\label{hydro12}
\end{eqnarray}
The first term on the r.h.s. is the relativistic quantum force, the second
term is the Lorentz force, the third term is the gravitational force and the
fourth term is a pressure force arising from the self-interaction of the SF.
Therefore, the KG equation (\ref{fkg1}) is equivalent to the hydrodynamic
equations (\ref{hydro7b})-(\ref{hydro12}). For $\hbar=0$, we recover the
relativistic Euler-Lorentz
equation (\ref{comp14b}) with the additional term $+\partial_{\nu}V'(\rho)$.
Introducing the energy 
(\ref{hydro5c}), the real quantum relativistic
Hamilton-Jacobi equation (\ref{hydro8b}) can be written as
\begin{eqnarray}
E_{\rm tot}^2=m^2c^4+m^2c^2{\bf
u}^2+\hbar^2c^2\frac{\square\sqrt{\rho}}{\sqrt{\rho}}
+2m^2c^2\Phi+2m^2c^2 V'(\rho).
\label{kr6d}
\end{eqnarray}
This equation can be interpreted as the quantum version of the relativistic
equation of
mechanics $E_{\rm tot}^2=p^2c^2+m^2c^4$ 
where $m u_{\mu}$ represents the impulse
$p_{\mu}$.

\subsection{The generalized Gross-Pitaeveskii-Maxwell-Poisson equations}
\label{sec_gpstat}

Making the Klein transformation (\ref{gpe1}) in the generalized KGMP equations
(\ref{fkg1}) and (\ref{ein2}), we obtain 
\begin{eqnarray}
i\hbar\frac{\partial\psi}{\partial t}-\frac{\hbar^2}{2m
c^2}\frac{\partial^2\psi}{\partial
t^2}+\frac{\hbar^2}{2 m}\Delta\psi-m\Phi \psi
-m\frac{dV}{d|\psi|^2}\psi
-i \frac{e\hbar}{2m} \left (\nabla\cdot
{\bf A}+\frac{1}{c^2}\frac{\partial
U}{\partial t}\right )\psi\nonumber\\
-\frac{e^2}{2m}\left ({\bf
A}^2-\frac{1}{c^2}U^2\right )
\psi
-i \frac{e\hbar}{m} \biggl \lbrack {\bf A}\cdot\nabla\psi
+\frac{U}{c^2}\left
(\frac{\partial\psi}{\partial t}-\frac{imc^2}{\hbar}\psi\right )\biggr
\rbrack=0,
\label{aps10}
\end{eqnarray}
\begin{eqnarray}
\frac{\Delta\Phi}{4\pi G}=|\psi|^2
+\frac{\hbar^2}{2m^2c^4}\left
|\frac{\partial\psi}{\partial t}\right
|^2+\frac{\hbar^2}{2c^2m^2}|\nabla\psi|^2
+\frac{1}{c^2}V(|\psi|^2)
-\frac{\hbar}{m c^2}{\rm Im} \left (\frac{\partial\psi}{\partial
t}\psi^*\right )\nonumber\\
+i\frac{e\hbar}{2m^2}\frac{U}{c^4}\left
(\psi\frac{\partial\psi^*}{\partial
t}-\psi^*\frac{\partial\psi}{\partial
t}\right )
-\frac{e}{mc^2}U|\psi|^2+\frac{e^2}{2m^2}\frac{U^2}{c^4}|\psi|^2
-\frac{1}{2}\frac{ie\hbar}{m^2c^2}{\bf
A}\cdot
(\psi\nabla\psi^*-\psi^*\nabla\psi)\nonumber\\
+\frac{1}{2}\frac{e^2}{m^2c^2}{\bf
A}^2|\psi|^2
+\frac{\epsilon_0}{2}\frac{{\bf
E}^2}{c^2}+\frac{1}{2\mu_0}\frac{{\bf
B}^2}{c^2}.
\label{aps11}
\end{eqnarray}
Equations  (\ref{ds2}), (\ref{ds3}), (\ref{aps10}) and (\ref{aps11}) form the
generalized GPMP equations. The relativistic GP equation (\ref{aps10}) can be
written more compactly as
\begin{eqnarray}
i\hbar\frac{\partial\psi}{\partial t}-\frac{\hbar^2}{2m}\left
(\partial_{\mu}+i\frac{e}{\hbar}A_{\mu}\right
)\left (\partial^{\mu}+i\frac{e}{\hbar}A^{\mu}\right
)\psi-e U
\psi-m\Phi\psi-m \frac{dV}{d|\psi|^2}\psi=0.
\label{ggp3}
\end{eqnarray}
This equation can be directly obtained from Eq. (\ref{ggp1})
by considering that the metric is approximately flat. Since the
term $g^{00}-1$ in Eq. (\ref{ggp1}) is multiplied by $c^2$, we have to use the
expression of the metric $g^{00}\simeq 1-2\Phi/c^2$ valid at first order in
$\Phi/c^2$ to evaluate this term. This is how the Newtonian potential
$\Phi$ enters
into 
Eq. (\ref{ggp3}).

In the nonrelativistic
limit $c\rightarrow +\infty$, Eqs. (\ref{aps10}) and (\ref{aps11})
reduce to
\begin{eqnarray}
i\hbar\frac{\partial\psi}{\partial
t}=-\frac{\hbar^2}{2 m}\left (\nabla-i\frac{e}{\hbar}{\bf
A}\right )^2\psi+m\Phi \psi+eU\psi
+m\frac{dV}{d|\psi|^2}\psi,
\label{ggp4}
\end{eqnarray}
\begin{eqnarray}
\Delta\Phi=4\pi G |\psi|^2.
\label{ggp5}
\end{eqnarray}
Equations (\ref{ds2}), (\ref{ds3}), (\ref{ggp4}) and (\ref{ggp5}) form the
GPMP equations. The GPMP equations are usually derived in a
different manner, from the mean
field Schr\"odinger equation (see, e.g., Sec. II.A of \cite{prd1}). Here, we
have rigorously derived these equations from the KGME equations in the
nonrelativistic limit $c\rightarrow +\infty$. This derivation shows that the
nonlinearity in the GP equation (\ref{ggp4}) is related to the 
potential of self-interaction $V$ in the KG equation (\ref{fkg1}).
For the quartic potential (\ref{lag4}), in the absence of the electromagnetic
field, we
recover the usual GP equation
\begin{eqnarray}
i\hbar\frac{\partial\psi}{\partial
t}=-\frac{\hbar^2}{2 m }\Delta\psi+m\Phi \psi
+\frac{4\pi a_s\hbar^2}{m^2}|\psi|^2\psi
\label{nr4}
\end{eqnarray}
with a cubic nonlinearity. We also
note that in the GP
equation (\ref{ggp4}) the
gravitational (external) potential $\Phi$ appears on the same footing as the
electric potential $U$ while in the KG equation (\ref{fkg1}) or in the
relativistic GP equation (\ref{aps10}) they appear at different places.

\subsection{The complex hydrodynamic equations}

The complex hydrodynamic equations corresponding to the
relativistic GP equation (\ref{aps10}) can be obtained by making the WKB
transformation (\ref{hggp1}) and proceed as in Appendix \ref{sec_gr1}.
Alternatively, they  can be directly obtained from the complex hydrodynamic
equations (\ref{comp8})-(\ref{kr6e}) corresponding
to the KG equation (\ref{fkg1}) by using the relations
(\ref{hggp3})-(\ref{hggp8c}) of Sec. \ref{sec_wa}.
The complex
quantum relativistic Hamilton-Jacobi equation and the complex Euler-Lorentz
equation write (we assume $V=0$):
\begin{eqnarray}
\frac{1}{c^2}\left (\frac{\partial {\cal S}}{\partial t}+eU\right
)^2-2m\left (\frac{\partial {\cal S}}{\partial t}+eU\right
)-\left (\nabla {\cal S}-e{\bf A}\right
)^2-2m^2\Phi=i\hbar
\Box{\cal S}+ie\hbar \left (\frac{1}{c^2}\frac{\partial U}{\partial
t}+\nabla\cdot {\bf A}\right ),
\label{comp17}
\end{eqnarray}
\begin{eqnarray}
V_{\mu}V^{\mu}=2\Phi-2cV^0-i\frac{\hbar}{m}\partial_{\mu}V^{\mu},
\label{qhggp4b}
\end{eqnarray}
\begin{equation}
\frac{d{V}_{\nu}}{d\tau}\equiv
{V}^{\mu}\partial_{\mu}{V}_{\nu}=-c\partial^0V_{\nu}-i\frac{\hbar}{2m } \square
{V}_{\nu}-\frac{e}{m}{V}^{\mu}F_{\mu\nu}-\frac{e}{m}cF_{\nu}^0-i\frac{\hbar
e}{2m^2}\partial^{\mu}F_{\mu\nu}+\partial_{\nu}\Phi,
\label{hcomp14as}
\end{equation}
\begin{eqnarray}
\frac{{\cal E}^2}{2mc^2}+{\cal E}=\frac{1}{2}m{\bf
V}^2+m\Phi-\frac{i\hbar}{2mc^2}\frac{\partial {\cal E}}{\partial
t}-i\frac{\hbar}{2}\nabla\cdot {\bf V}.
\label{kr6b}
\end{eqnarray}

\subsection{The real hydrodynamic equations}

The real hydrodynamic equations corresponding to the
relativistic GP equation (\ref{aps10})
can be obtained by making the Madelung transformation (\ref{hggp5}) and
proceeding as in Appendix \ref{sec_gr2}. Alternatively, they  can be directly
obtained from
the real hydrodynamic equations
(\ref{hydro7b})-(\ref{kr6d}) corresponding to the KG
equation (\ref{fkg1}) by using the relations (\ref{hggp7})-(\ref{hggp8b}) of
Sec. \ref{sec_rheb}. 
The
continuity equation, the quantum relativistic Hamilton-Jacobi equation, and the
quantum relativistic  Euler-Lorentz equation write
\begin{eqnarray}
\frac{\partial\rho}{\partial t}+\partial_{\mu}\left (\rho v^{\mu}\right
)=0,
\label{aot4}
\end{eqnarray}
\begin{eqnarray}
v_{\mu}v^{\mu}=-2 c
v^0+2\Phi+\frac{\hbar^2}{m^2}\frac{\square\sqrt{\rho}}{\sqrt{ \rho} }+2
V'(\rho),
\label{aot5}
\end{eqnarray}
\begin{eqnarray}
\frac{dv_{\nu}}{d\tau}\equiv
v^{\mu}\partial_{\mu}v_{\nu}=-c\partial^0v_{\nu}+\frac{\hbar^2}{2m^2}
\partial_ { \nu } \left (\frac{
\square\sqrt {\rho} } { \sqrt {\rho} }\right
)-\frac{e}{m}v^{\mu}F_{ \mu\nu }-\frac{e}{m}c F_{\nu
}^0+\partial_{\nu}\Phi+\partial_{\nu}V'(\rho),
\label{hydro12ash}
\end{eqnarray}
\begin{eqnarray}
\frac{E^2}{2mc^2}+E=\frac{1}{2}m{\bf
v}^2+\frac{\hbar^2}{2m}\frac{\square\sqrt{\rho}}{\sqrt{\rho}}
+m\Phi+m V'(\rho).
\label{kr6das}
\end{eqnarray}

\subsection{The hydrodynamic equations}

Making the Madelung transformation defined by Eqs. (\ref{hggp5}) and
(\ref{whydro2}) in the generalized GPMP equations (\ref{aps10}) and
(\ref{aps11}), or taking the
weak field limit $\Phi/c^2\rightarrow 0$ in the quantum barotropic ELME
equations (\ref{whydro8}) and (\ref{kr1})-(\ref{kr3}) and setting $a=1$ and
$H=0$, we obtain
the hydrodynamic equations 
\begin{eqnarray}
\frac{\partial}{\partial t}\left\lbrack \left (1+\frac{E}{mc^2}\right
)\rho\right\rbrack+\nabla\cdot (\rho {\bf v})=0,
\label{cann}
\end{eqnarray}
\begin{eqnarray}
\frac{\partial S}{\partial t}+\frac{1}{2 m }(\nabla S-e{\bf
A})^2=-\frac{\hbar^2}{2 m}\frac{\square\sqrt{\rho}}{\sqrt{\rho}}
-m\Phi-eU-m h(\rho)
+\frac{E^2}{2mc^2},
\label{conn}
\end{eqnarray}
\begin{eqnarray}
\frac{\partial {\bf v}}{\partial t}+({\bf v}\cdot
\nabla){\bf v}=-\frac{\hbar^2}{2m^2}\nabla \left
(\frac{\square\sqrt{\rho}}{\sqrt{\rho}}\right
)-\nabla\Phi-\frac{
1}{\rho}\nabla p
+\frac{e}{m}\left ({\bf E}+{\bf v}\times {\bf
B}\right )
+\frac{1}{2m^2c^2}\nabla
(E^2),
\label{s25b}
\end{eqnarray}
\begin{eqnarray}
\frac{\Delta\Phi}{4\pi G}=\rho
+\frac{\hbar^2}{2m^2c^4}\left\lbrack
\frac{1}{4\rho}\left (\frac{\partial\rho}{\partial t}\right
)^2+\frac{\rho}{\hbar^2}\left (\frac{\partial S}{\partial t}\right
)^2\right\rbrack
+\frac{\hbar^2}{2c^2m^2}\left\lbrack
\frac{1}{4\rho}(\nabla\rho)^2+\frac{\rho}{\hbar^2}(\nabla S)^2\right\rbrack
\nonumber\\
+\frac{1}{c^2}V(\rho)-\frac{1}{m c^2}\rho\frac{\partial
S}{\partial t}
+\frac{e}{m^2}\frac{U}{c^4}\rho\frac{\partial
S}{\partial t}
-\frac{e}{mc^2}U\rho\nonumber\\
+\frac{e^2}{2m^2}\frac{U^2}{c^4}\rho
-\frac{e}{m^2c^2}\rho {\bf A}\cdot
\nabla S
+\frac{1}{2}\frac{e^2}{m^2c^2}{\bf A}^2\rho
+\frac{\epsilon_0}{2}\frac{{\bf
E}^2}{c^2}+\frac{1}{2\mu_0}\frac{{\bf
B}^2}{c^2}.
\label{as17}
\end{eqnarray}
Equations (\ref{ds2}), (\ref{ds3}), and (\ref{cann})-(\ref{as17}) form the
generalized quantum barotropic ELMP equations. Introducing the energy
(\ref{whydro4}), the quantum relativistic Hamilton-Jacobi equation (\ref{conn})
can be written as 
\begin{eqnarray}
E+\frac{E^2}{2mc^2}=\frac{1}{2}m {\bf v}^2+\frac{\hbar^2}{2
m}\frac{\square\sqrt{\rho}}{\sqrt{\rho}}
+m\Phi+m V'(\rho).
\label{connaz}
\end{eqnarray}
Solving for $E$ in Eq. (\ref{connaz}), we obtain
\begin{eqnarray}
E=mc^2\left\lbrack -1\pm \sqrt{1+\frac{{\bf v}^2}{c^2}+\frac{\hbar^2}{
m^2c^2}\frac{\square\sqrt{\rho}}{\sqrt{\rho}}
+2\frac{\Phi}{c^2}+2\frac{1}{c^2}
h(\rho)}\right\rbrack.
\label{kr5c}
\end{eqnarray}
This expression can be substituted into Eqs. (\ref{cann}) and (\ref{s25b}) to
obtain a closed system of hydrodynamic equations.

In the
nonrelativistic limit $c\rightarrow +\infty$, the hydrodynamic equations
(\ref{cann})-(\ref{as17}) reduce to
\begin{eqnarray}
\frac{\partial\rho}{\partial t}+\nabla\cdot (\rho {\bf v})=0,
\label{el}
\end{eqnarray}
\begin{eqnarray}
\frac{\partial S}{\partial t}+\frac{1}{2 m }(\nabla S-e{\bf
A})^2=\frac{\hbar^2}{2 m}\frac{\Delta\sqrt{\rho}}{\sqrt{\rho}}
-m\Phi-eU-m h(\rho),
\label{qconn}
\end{eqnarray}
\begin{eqnarray}
\frac{\partial {\bf v}}{\partial t}+({\bf v}\cdot
\nabla){\bf v}=\frac{\hbar^2}{2m^2}
\nabla\left ( \frac{\Delta\sqrt{\rho}}{\sqrt{\rho}}
\right )-\nabla\Phi-\frac{
1}{\rho}\nabla p
+\frac{e}{m}\left ({\bf E}+{\bf v}\times {\bf B}\right ),
\label{class3b}
\end{eqnarray}
\begin{eqnarray}
\Delta\Phi=4\pi G\rho.
\label{class3bq}
\end{eqnarray}
Equations (\ref{ds2}), (\ref{ds3}), and (\ref{el})-(\ref{class3bq}) form
the quantum barotropic ELMP equations. Introducing the energy (\ref{whydro4}),
the quantum Hamilton-Jacobi equation (\ref{qconn}) can be written as
\begin{eqnarray}
E=\frac{1}{2}m {\bf v}^2-\frac{\hbar^2}{2
m}\frac{\Delta\sqrt{\rho}}{\sqrt{\rho}}
+m\Phi+m h(\rho).
\label{hjn}
\end{eqnarray}
This equation can be interpreted as the quantum generalization of the classical
equation of
mechanics $E={\bf p}^2/2m+m\Phi$ where $m{\bf v}$ represents the impulse ${\bf
p}$.

\subsection{The London equations}

The London equations are given by
\begin{eqnarray}
\square A_{\nu}=-\mu_0\frac{e^2}{m^2}
\rho A_{\nu},\qquad \square U=-\frac{e^2}{m^2c^2\epsilon_0}\rho U,\qquad \square
{\bf
A}=-\frac{e^2\mu_0}{m^2}\rho {\bf A}.
\label{lw1bc}
\end{eqnarray}
In a static spacetime, the equations for $U$ and ${\bf A}$ are decoupled.
In the
absence of electric field, ${\bf A}$ is stationary and Eq. (\ref{lw1bc})
reduces to
\begin{eqnarray}
\Delta {\bf A}=\frac{e^2\mu_0}{m^2}\rho {\bf A}.
\label{lw1bca}
\end{eqnarray}
This is the standard London equation which displays a characteristic length
scale $\lambda=(m^2/\mu_0 e^2)^{1/2}$ over which the potential vector, hence
the magnetic field, is exponentially suppressed. This accounts for the Meissner
effect \cite{meissner} in the theory of superconductivity wherein a material
exponentially expels all internal magnetic fields as it crosses the
superconducting threshold. The length scale $\lambda$ is the London penetration
depth.

\section{The  relativistic and nonrelativistic eigenvalue equations}
\label{sec_kgm}

In the absence of magnetic field (${\bf A}={\bf 0}$) and self-interaction
($V=0$), the KG equation (\ref{fkg1}) can be written as
\begin{eqnarray}
\frac{1}{c^2}\left (\frac{\partial}{\partial
t}+i\frac{e}{\hbar}U\right
)^2\varphi-\Delta\varphi+\frac{m^2
c^2}{\hbar^2}\varphi+\frac{2m^2}{\hbar^2}\Phi\varphi
=0,\label{st1}
\end{eqnarray}
or, equivalently, as
\begin{eqnarray}
\frac{1}{c^2}\frac{\partial^2\varphi}{\partial
t^2}-\Delta\varphi+i\frac{e}{\hbar c^2}\frac{\partial
U}{\partial t}\varphi+2i\frac{e}{\hbar c^2}U\frac{\partial
\varphi}{\partial
t}-\frac{e^2}{\hbar^2c^2}U^2
\varphi+\frac{m^2
c^2}{\hbar^2}\varphi+\frac{2m^2}{\hbar^2}\Phi\varphi
=0.\label{st2}
\end{eqnarray}
Looking for a stationary solution of the form
\begin{eqnarray}
\varphi=\frac{\hbar}{m}\sqrt{\rho} \, e^{-iE_{\rm tot}t/\hbar},
\label{st3}
\end{eqnarray}
where $E_{\rm tot}$ is a constant, we obtain the eigenvalue
equation
\begin{eqnarray}
(E_{\rm
tot}-eU)^2\varphi+c^2\hbar^2\Delta\varphi-2m^2c^2\Phi\varphi-m^2c^4\varphi=0.
\label{st4}
\end{eqnarray}
This
equation can also be obtained from the
Hamilton-Jacobi equation (\ref{hydro8b})  with
${\bf A}={\bf 0}$, $V=0$ and $S_{\rm tot}=-E_{\rm tot}t$.\footnote{We note that
the eigenenergy $E_{\rm tot}$ defined by Eq. (\ref{st3}) is different from the
energy defined by Eq. (\ref{hydro5c}) because of the
presence of the term $eU$. However, in order to avoid the proliferation of
notations, we use the same symbol. The same remark applies to the eigenenergy
$E$ defined by Eq. (\ref{st7}) below.}

In the absence of magnetic field (${\bf A}={\bf 0}$) and self-interaction
($V=0$), the GP equation (\ref{ggp4}) reduces to the Schr\"odinger equation
\begin{eqnarray}
i\hbar\frac{\partial\psi}{\partial
t}=-\frac{\hbar^2}{2 m}\Delta\psi+m\Phi \psi+eU\psi.
\label{st6}
\end{eqnarray}
Looking for a stationary solution  of the form
\begin{eqnarray}
\psi= \sqrt{\rho} \, e^{-iE t/\hbar},
\label{st7}
\end{eqnarray}
where $E$ is a constant, we obtain the eigenvalue equation
\begin{eqnarray}
-\frac{\hbar^2}{2m}\Delta\psi+(eU+m\Phi)\psi=E\psi.
\label{st8}
\end{eqnarray}
This equation can also be obtained from the
Hamilton-Jacobi equation (\ref{qconn})  with
${\bf A}={\bf 0}$, $V=0$ and $S=-E t$. Similarly to the remark made in Sec.
\ref{sec_gpstat}, the
gravitational (external) potential
$\Phi$ and the electric potential $U$ appear on the same footing in the
nonrelativistic
eigenvalue equation (\ref{st8}) while they appear at different places in the
relativistic eigenvalue equation (\ref{st4}). On the other hand, if we set
$E_{\rm tot}=E+mc^2$ in Eq. (\ref{st4}) and take the nonrelativistic
limit $c\rightarrow +\infty$, we recover Eq. (\ref{st8}).

If we consider a free particle ($U=\Phi=0$), expand the
wave functions $\varphi$ and $\psi$ as plane waves of the form $e^{i({\bf
k}\cdot {\bf r}-\omega t)}$, and use the de Broglie relations
$E_{(\rm tot)}=\hbar\omega$ and ${\bf p}=\hbar {\bf k}$, we obtain from the KG
equation (\ref{st1}) the relativistic
dispersion relation 
\begin{eqnarray}
\omega^2=c^2 k^2+\frac{m^2c^4}{\hbar^2} \qquad \Leftrightarrow \qquad
E_{\rm tot}^2=p^2c^2+m^2c^4,
\label{dr2}
\end{eqnarray}
and from the Schr\"odinger equation (\ref{st6}) the
classical
dispersion relation
\begin{eqnarray}
\omega=\frac{\hbar k^2}{2m}\qquad \Leftrightarrow \qquad E=\frac{p^2}{2m}.
\label{dr1}
\end{eqnarray}
In the relativistic case, we find that $E_{\rm tot}=\pm \sqrt{p^2c^2+m^2c^4}$.
There exist solutions both for positive and negative energies. The
solutions yielding negative energies led Dirac \cite{dirac2} to postulate the
existence of antiparticles.

\section{The pilot wave theory of de Broglie}
\label{sec_pilot}

De Broglie \cite{broglie1927a,broglie1927b} performed the
transformation
(\ref{hydro1}) on the KG equation
(\ref{kg2}) with $V=0$ and derived the quantum relativistic Hamilton-Jacobi
equation (\ref{hydro3}) which contains the  relativistic covariant quantum
potential (\ref{hydro9}). He also proposed to formally rewrite Eq.
(\ref{hydro3}) in the form of a relativistic Hamilton-Jacobi equation
\begin{eqnarray}
\left (\partial_{\mu}S_{\rm tot}+e A_{\mu}\right
)\left (\partial^{\mu} S_{\rm tot}+e A^{\mu}\right
)-M^2c^2=0
\label{db1}
\end{eqnarray}
by introducing an effective mass
\begin{eqnarray}
M=\sqrt{m^2+\frac{\hbar^2}{c^2}\frac{\square\sqrt{\rho}}{\sqrt{\rho}}}
\label{db2}
\end{eqnarray}
which includes the relativistic quantum potential. Equation
(\ref{db1}) can be rewritten as  $u_{\mu}u^{\mu}=c^2$ where $u_{\mu}$ is defined
by Eq. (\ref{hydro4}) with $M$ instead of $m$.

De Broglie also developed a pilot wave theory that is closely related to the
results obtained in Sec. \ref{sec_kgme}. We mention here some connections with
his work. As shown in Sec. \ref{sec_hrc}, the quadricurrent of charge is
$(J_e)_{\mu}=(e/m)\rho
u_{\mu}$, the density of charge is
$\rho_e=(e/m)\rho u_0/c$, and the current of charge is ${\bf
J}_e=(e/m)\rho{\bf u}$. We note that, in the relativistic regime,
$(J_e)_{\mu}\neq \rho_e
u_{\mu}$, $\rho_e\neq (e/m)\rho$, and ${\bf
J}_e\neq
\rho_e{\bf
u}$ as we could naively expect (these relations are only valid in the
nonrelativistic limit $c\rightarrow +\infty$). We have instead $(J_e)_{\mu}=
\rho_e (u_{\mu}/u_0)c$, $(J_e)_0=
\rho_e c$, and ${\bf J}_e= \rho_e ({\bf u}/u_0)c$.

We can write the quadricurrent of charge as
\begin{eqnarray}
(J_e)_{\mu}=\rho_e (u_e)_{\mu}
\label{db4}
\end{eqnarray}
by introducing a charge quadrivelocity
\begin{eqnarray}
(u_e)_{\mu}=\frac{u_{\mu}}{u_0}c=\frac{\partial_{\mu}S_{\rm
tot}+eA_{\mu}}{\partial_{0}S_{\rm tot}+eA_{0}}c.
\label{db5}
\end{eqnarray}
We note that $(u_e)_0=c$. The components of the quadricurrent of charge satisfy
the
relations
\begin{eqnarray}
(J_e)_0= \rho_e c,\qquad
{\bf J}_e=\rho_e {\bf u}_e.
\label{db10}
\end{eqnarray}
The charge density and the charge velocity are given by
\begin{eqnarray}
\rho_e=-\frac{e}{m}\rho \frac{\frac{\partial S_{\rm tot}}{\partial
t}+eU}{mc^2}, \qquad {\bf u}_e=\frac{{\bf u}}{u_0}c=-\frac{\nabla S_{\rm
tot}-e{\bf A}}{\frac{\partial S_{\rm tot}}{\partial t}+eU}c^2.
\label{db6}
\end{eqnarray}
Using these relations, we find that the local charge conservation equation
(\ref{charge6}), which is equivalent to the equation of continuity
(\ref{hydro7}), can be written as
\begin{eqnarray}
D_{\mu}(\rho_e u_e^{\mu})=0.
\label{db8}
\end{eqnarray}
We found these equations by ourselves. However, by studying the early
history of quantum mechanics to prepare the Introduction, we discovered
with surprise that these equations were first obtained by de Broglie
\cite{broglie1927a} a very
long time ago in relation to his  pilot wave theory  [see
his Eqs.
(I) and (II)] and presented as {\it formules fondamentales} of his theory.

\section{The historical derivation of the Schr\"odinger equation}
\label{sec_ori}

In most textbooks of quantum mechanics, the Schr\"odinger equation is derived
from the correspondence principle. This is not, however, how Schr\"odinger
initially derived it. In this Appendix, we give a short account of the manner
how
Schr\"odinger obtained his famous equation in his historical papers since his
original approach is not well-known. 

\subsection{The pre-historic derivation of the Schr\"odinger equation}

We first begin by the manner how Schr\"odinger obtained his equation according
to Dirac \cite{dirachistoric}. However, we could not find any trace of
this derivation (or even a mention of it) in any published paper. Therefore,
the claim of Dirac must be considered  with circumspection. According to
Dirac \cite{dirachistoric}, de Broglie introduced the
relativistic wave equation
\begin{equation}
\frac{1}{c^2}\frac{\partial^2 \varphi}{\partial
t^2}-\Delta\varphi+\frac{m^2c^2}{\hbar^2}\varphi=0.
\label{fkg6}
\end{equation}
This is what we now call the KG equation.\footnote{We could not find any
published paper of de Broglie containing the KG equation prior to 1926.
There is an equation similar to the KG equation in \cite{broglie25} but it has
the opposite sign: $\square\varphi-(m^2c^2/\hbar^2)\varphi=0$ instead of
$\square\varphi+(m^2c^2/\hbar^2)\varphi=0$. On this point again, the claim of
Dirac (the attribution of the KG equation to de Broglie) must be considered with
circumspection.} Then, still according to
Dirac
\cite{dirachistoric}, Schr\"odinger extended this equation in order to include
the 
electromagnetic field and ``guessed'' the equation
\begin{equation}
\frac{1}{c^2}\left (\frac{\partial}{\partial
t}+i\frac{e}{\hbar}U\right
)^2\varphi-\left
(\nabla-i\frac{e}{\hbar}{\bf A}\right
)^2\varphi+\frac{m^2c^2}{\hbar^2}
\varphi=0.
\label{fkg7}
\end{equation}
He then considered the nonrelativistic limit of this equation (Dirac
\cite{dirachistoric} does not explain how he did) and
obtained, in the absence of magnetic field,
the wave equation
\begin{eqnarray}
i\hbar\frac{\partial\psi}{\partial
t}=-\frac{\hbar^2}{2 m}\Delta\psi+eU\psi.
\label{fkg8}
\end{eqnarray}
This is the Schr\"odinger equation. However, in his published
papers
\cite{schrodinger1,schrodinger2,schrodinger3,schrodinger4},
Schr\"odinger introduced this equation in
a completely different manner, adopting from the start a nonrelativistic
approach. He introduced the KG equation (\ref{fkg7}) in his fourth
paper \cite{schrodinger4}, but did not explicitly take the nonrelativistic limit
of this equation to make the connection with Eq. (\ref{fkg8}). To our
knowledge, this connection was first made by Klein \cite{klein1,klein2}.

\subsection{The first derivation of the Schr\"odinger equation}
\label{sec_fi}

In his first paper \cite{schrodinger1} (see also \cite{schrodingerCORR}),
Schr\"odinger derives the eigenvalue
equation (\ref{st8}) from a variational principle. He starts from the classical
Hamilton-Jacobi equation
\begin{equation}
\label{fsch1}
E=\frac{1}{2m}(\nabla S)^2+m\Phi({\bf r}),
\end{equation}
where $E$ is the classical energy (which is a constant) and $S=S({\bf r})$ is
the classical action which is related to the classical impulse by ${\bf
p}=\nabla S$. He introduces a (real) wave function $\psi({\bf r})$ through the
substitution\footnote{This can be written as $\psi=e^{S/\hbar}$ which
corresponds to the WKB transformation (\ref{hggp1}) with a purely
imaginary action
${\cal S}=-iS$.}
\begin{equation}
\label{fsch2}
S=\hbar\ln\psi.
\end{equation}
Equation (\ref{fsch1}) is then rewritten in terms of $\psi$ as
\begin{equation}
\label{fsch3}
(\nabla\psi)^2-\frac{2m}{\hbar^2}(E-m\Phi)\psi^2=0.
\end{equation}
At that point, Schr\"odinger introduces the functional
\begin{equation}
\label{fsch4}
J=\int \left\lbrack (\nabla\psi)^2-\frac{2m}{\hbar^2}(E-m\Phi)\psi^2\right\rbrack\, d{\bf r}
\end{equation}
and considers its minimization with respect to variations on $\psi$. The
condition $\delta J=0$ gives
\begin{equation}
\label{fsch5}
\Delta\psi+\frac{2m}{\hbar^2}(E-m\Phi)\psi=0,
\end{equation}
which is the time-independent Schr\"odinger equation.\footnote{
Schr\"odinger puts much emphasis on this equation in his papers, so this
equation really is {\it the} Schr\"odinger equation (more than
the time-dependent equation (\ref{sch10}) introduced later).} This is an
eigenvalue
equation for the energy $E$. This is how the energy becomes quantized. Indeed,
for a given potential $\Phi$, this equation has physical solutions only
for some particular values of the energy. Therefore, this equation determines
the quantification of the energy through the boundary conditions. In the
{\it Zusatz bei der Korrektur}, Schr\"odinger
remarks that
his variational principle can be written as the minimization of an
energy functional
\begin{equation}
\label{fsch6}
E_{\rm tot}=\int \left\lbrack \frac{\hbar^2}{2m}(\nabla\psi)^2+m\Phi\psi^2\right\rbrack\, d{\bf r}
\end{equation}
under the constraint
\begin{equation}
\label{fsch7}
\int \psi^2\, d{\bf r}=1.
\end{equation}
Indeed, if we write the variational principle as
\begin{equation}
\label{fsch8}
\delta E_{\rm tot}-E\delta \int \psi^2\, d{\bf r}=0,
\end{equation}
we obtain Eq. (\ref{fsch5}) where the eigenenergy $E$ appears as the Lagrange multiplier associated with the constraint (\ref{fsch7}).

{\it Remark:} Nowadays, we derive the time-independent Schr\"odinger equation
as follows. We write the Hamiltonian (total energy) as
\begin{eqnarray}
E_{\rm tot}=\int\frac{\hbar^2}{2m}|\nabla\psi|^2\, d{\bf
r}+\int m\Phi|\psi|^2\, d{\bf r},
\label{lh4}
\end{eqnarray}
where the first term is the kinetic energy and  the second term
is the potential energy. One can show that the Hamiltonian (\ref{lh4}) and the
integral $\int |\psi|^2\, d{\bf r}$ (normalization) are conserved by the
Schr\"odinger equation (\ref{sch10}). As a result, a minimum of energy
$E_{\rm tot}$ under the normalization condition  $\int |\psi|^2\, d{\bf r}=1$ is
a stationary solution of the Schr\"odinger equation that is formally nonlinearly
dynamically stable.\footnote{This is how we now  justify the
minimization problem $\min \lbrace E_{\rm tot}  | \int |\psi|^2\,
d{\bf r} \, {\rm fixed}\rbrace$. Schr\"odinger introduced this
minimization problem as a postulate, without any justification. This remarkably
led him to the Schr\"odinger equation. However, this cannot be considered as a
{\it
derivation} of this equation.} Writing the variational principle as
\begin{eqnarray}
\label{lh5b}
\delta E_{\rm tot}-\mu\delta\int |\psi|^2\, d{\bf r}=0,
\end{eqnarray}
where $\mu$ is a Lagrange multiplier (chemical potential), we obtain the
time-independent Schr\"odinger equation
(\ref{fsch5}) provided that $\mu=E$. This shows that the chemical potential
$\mu$
can be
identified with the eigenenergy $E$ and {\it vice versa}. 

\subsection{The second derivation of the
Schr\"odinger equation}

In his second paper \cite{schrodinger2}, 
Schr\"odinger proposes another, more physical, derivation of his equation by
developing an analogy between geometric and undulatory mechanics. His derivation
combines arguments coming from mechanics, optics and dispersive waves
using the principles of Hamilton, Huygens, and Fermat. We
very succintely recall the main steps of his derivation. 

Schr\"odinger was
strongly inspired by the fundamental researches of
de Broglie
on matter waves (see his Introduction in \cite{schrodingerPR}). Following de
Broglie,
he assumes that each particle (like an electron or a proton) is described by a
wave of the form
\begin{equation}
\label{sch1}
\psi({\bf r},t)=Ae^{i({\bf k}\cdot {\bf r}-\omega t)}.
\end{equation}
Using the de Broglie relations
\begin{equation}
\label{sch2}
E=\hbar \omega,\qquad {\bf p}=\hbar {\bf k},
\end{equation}
and the classical
expression of the energy
\begin{equation}
\label{sch5}
E=\frac{p^2}{2m}+m\Phi({\bf r})
\end{equation}
of a particle of mass $m$ moving in a potential $\Phi({\bf r})$, Schr\"odinger
finds that the phase velocity of the matter wave is given
by\footnote{Note that the phase velocity (\ref{sch6}) differs from the group
velocity
$v_g=d\omega/dk=dE/dp=p/m$ which coincides with the velocity $v=p/m$ of the
particle.}
\begin{equation}
\label{sch6}
v_{\phi}=\frac{\omega}{k}=\frac{E}{p}=\frac{E}{\sqrt{2m\lbrack E-m\Phi({\bf
r})\rbrack}}.
\end{equation}
Therefore, the phase velocity of the wave depends on the position
${\bf r}$.
This is at
variance with the electromagnetic wave for which $v_{\phi}=c$ is a constant.
Schr\"odinger then assumes that the evolution of the wave function $\psi({\bf
r},t)$ is given by the ordinary (second order in time) wave equation in a
dispersive medium
\begin{equation}
\label{sch4}
\frac{1}{v_{\phi}^2({\bf r})}\frac{\partial^2\psi}{\partial t^2}=\Delta\psi.
\end{equation}
He also argues that the wave function must be of the form
\begin{equation}
\label{sch3b}
\psi({\bf r},t)\propto e^{-iEt/\hbar},
\end{equation}
which corresponds to the temporal term in Eq. (\ref{sch1}) when we use the de
Broglie relation (\ref{sch2}). From Eq. (\ref{sch3b}), we get
\begin{equation}
\label{sch7}
\frac{\partial^2\psi}{\partial t^2}=-\frac{E^2}{\hbar^2}\psi.
\end{equation}
Combining Eqs. (\ref{sch6}), (\ref{sch4}) and (\ref{sch7}), Schr\"odinger
recovers the fundamental eigenvalue equation
\begin{equation}
\label{sch8}
-\frac{\hbar^2}{2m}\Delta\psi+m\Phi\psi=E\psi
\end{equation}
that he previously obtained from a variational principle \cite{schrodinger1}.
This is the stationary
Schr\"odinger equation. In his fourth paper \cite{schrodinger4}, using the
identity
\begin{equation}
\label{sch9}
\frac{\partial\psi}{\partial t}=-i\frac{E}{\hbar}\psi,
\end{equation}
he eliminates the energy from Eq. (\ref{sch8}) and rewrites this
equation under the form
\begin{equation}
\label{sch10}
i\hbar \frac{\partial\psi}{\partial t}=-\frac{\hbar^2}{2m}\Delta\psi+m\Phi\psi.
\end{equation}
This is the time-dependent Schr\"odinger equation.

{\it Remark:} Before arriving at Eq. (\ref{sch10}),
Schr\"odinger derives
from Eq. (\ref{sch8}) the equation 
\begin{equation}
\label{sch11}
\left (-\frac{\hbar^2}{2m}\Delta+m\Phi\right )^2\psi=E^2\psi.
\end{equation}
Using Eq. (\ref{sch7}) he eliminates the energy from Eq. (\ref{sch11}) and
obtains an
equation the form
\begin{equation}
\label{sch12}
\hbar^2\frac{\partial^2\psi}{\partial t^2}+\left
(-\frac{\hbar^2}{2m}\Delta+m\Phi\right )^2\psi=0
\end{equation}
in which there is no complex $i$. Of course, Eq. (\ref{sch12}) can also
be obtained from Eq. (\ref{sch10})
by time derivation.

\subsection{The correspondence principle}

The nowadays standard procedure to derive the Schr\"odinger equation  is based
on the correspondence principle (this procedure can also be extended to the KG
equation). The correspondence principle first
appeared in the works of Schr\"odinger \cite{schrodingerCORR} and de
Broglie \cite{broglie1,broglie2}, and was rapidly adopted
by many other authors. The quantum equations (KG and Schr\"odinger) can be
obtained
from the non-quantum ones by using the correspondence principle 
\begin{equation}
p_{\mu}\leftrightarrow i\hbar\partial_{\mu},\qquad E_{(\rm
tot)}\leftrightarrow i\hbar \frac{\partial}{\partial t},\qquad
{\bf p}\leftrightarrow -i\hbar \nabla,
\label{corrp}
\end{equation}
where $p_0=E_{\rm tot}/c$ and ${\bf p}=(-p_1,-p_2,-p_3)$ are the components
of the quadri-impulse $p_{\mu}$. In this way, the
classical equation of mechanics $E=p^2/2m+m\Phi$ yields the
Schr\"odinger equation (\ref{sch10}) and the relativistic equation of mechanics
$E_{\rm tot}^2=p^2c^2+m^2c^4$, equivalent to $p_{\mu}p^{\mu}=m^2c^2$, yields the
KG
equation (\ref{fkg6}). The electromagnetic field can be taken into account
through the substitutions 
$\partial_{\mu}\rightarrow \partial_{\mu}+\frac{ie}{\hbar}A_{\mu}$, 
$\partial_{t}\rightarrow
\partial_{t}+\frac{ie}{\hbar}U$ and $\nabla\rightarrow
\nabla-\frac{ie}{\hbar}{\bf A}$, leading to the electromagnetic Schr\"odinger
equation (\ref{ggp4}) and
to the electromagnetic KG equation (\ref{fkg7}). An external potential $\Phi$
can be introduced in the KG
equation by writing the relativistic equation of mechanics  as
$E_{\rm tot}^2=p^2c^2+m^2c^4+2m^2c^2\Phi$. This  leads to Eq. (\ref{fkg1}).
However, the rigorous way to take gravity into account in the
KG equation is to consider a curved spacetime according to Einstein's theory of
general relativity as we have done in the present paper. In the weak field limit
$\Phi/c^2\rightarrow 0$, it is found that the KG
equation reduces to Eq. (\ref{fkg1}).

\subsection{The relativistic generalization Schr\"odinger's variational
principle}

Schr\"odinger's variational principle described in Sec.
\ref{sec_fi} was generalized to relativistic particles by De Donder
\cite{ddd,dd0,dd1,dd3} and Gordon \cite{gordon}. They started from the
relativistic Hamilton-Jacobi equation 
\begin{equation}
\partial_{\mu}S_{\rm tot}\partial^{\mu}S_{\rm tot}-m^2c^2=0
\label{rg1}
\end{equation}
and performed the transformation
\begin{equation}
S_{\rm tot}=\hbar\ln\varphi,
\label{rg2}
\end{equation}
thereby obtaining
\begin{equation}
\partial_{\mu}\varphi \partial^{\mu}\varphi-\frac{m^2c^2}{\hbar^2}\varphi^2=0.
\label{rg3}
\end{equation}
Integrating this equation over the four dimensional spacetime, they obtained the
functional
\begin{equation}
S_{\varphi}=\int \left (\frac{1}{2}\partial_{\mu}\varphi
\partial^{\mu}\varphi-\frac{m^2c^2}{2\hbar^2}\varphi^2\right )\, d^4x
\label{rg4}
\end{equation}
which can be interpreted as the action of a real SF (Gordon
\cite{gordon} used an extension of this method to deal with a complex
SF). The principle of
least action $\delta S_{\varphi}=0$ then leads to the KG equation (\ref{fkg6}).

\section{Detail of the calculations leading to Eq. (\ref{ggp1})}
\label{sec_det}

From Eq. (\ref{gpe1}) we obtain
\begin{eqnarray}
\partial_{\nu}\varphi=\frac{\hbar}{m}\left
(\partial_{\nu}\psi-i\frac{mc}{\hbar}\delta_{\nu}^{0}\psi\right )e^{-i m c^2
t/\hbar}.
\label{det1}
\end{eqnarray}
Using
\begin{eqnarray}
D_{\mu}\partial^{\mu}\varphi=D_{\mu}(g^{\mu\nu}\partial_{\nu}\varphi)=g^{\mu\nu}
D_{\mu}\partial_{\nu}\varphi
\label{det2}
\end{eqnarray}
and Eq. (\ref{det1}), we get (see Appendix \ref{sec_id}):
\begin{eqnarray}
D_{\mu}\partial^{\mu}\varphi=g^{\mu\nu}\frac{\hbar}{m}e^{-i m
c^2
t/\hbar}\left\lbrack
D_{\mu}\left (\partial_{\nu}\psi-i\frac{mc}{\hbar}\delta_{\nu}^{0}\psi\right )
-\frac{imc}{\hbar}\delta_{\mu}^0\left
(\partial_{\nu}\psi-i\frac{mc}{\hbar}\delta_{\nu}^{0}\psi\right )\right \rbrack.
\label{det3}
\end{eqnarray}
Performing the tensorial product of the term in bracket with $g^{\mu\nu}$,
and using the identity 
\begin{eqnarray}
D_{\mu}\left (\delta_{\nu}^{0}\psi\right
)=\partial_{\mu}\left (\delta_{\nu}^{0}\psi\right
)-\Gamma_{\mu\nu}^{\sigma}\delta_{\sigma}^{0}\psi=\delta_{\nu}^0\partial_{\mu}
\psi-\Gamma_{\mu\nu}^0\psi,
\label{det4}
\end{eqnarray}
we obtain
\begin{eqnarray}
D_{\mu}\partial^{\mu}\varphi=\frac{\hbar}{m}e^{-i m
c^2
t/\hbar}\left\lbrack
D_{\mu}\partial^{\mu}\psi-2i\frac{mc}{\hbar}\partial^{0}\psi-\frac{m^2c^2}{
\hbar^2}g^{00}\psi+\frac{imc}{\hbar}\Gamma_{\mu\nu}^0g^{\mu\nu}\psi\right
\rbrack.
\label{det5}
\end{eqnarray}
Substituting Eqs. (\ref{det1}) and (\ref{det5}) into Eq. (\ref{kg7}) and
rearranging terms, we get
\begin{eqnarray}
\square_e\varphi=\frac{\hbar}{m}e^{-i m
c^2
t/\hbar}\left\lbrack
\square_e\psi-2i\frac{mc}{\hbar}\partial^{0}\psi-\frac{m^2c^2}{
\hbar^2}g^{00}\psi+2\frac{mce}{\hbar^2}A^0\psi+\frac{imc}{\hbar}\Gamma_{\mu\nu}
^0g^ { \mu\nu } \psi\right
\rbrack.
\label{det6}
\end{eqnarray}
Finally, substituting Eq. (\ref{det6}) into the KG equation (\ref{kg2}), we
obtain the general relativistic GP equation (\ref{ggp1}). We can also proceed
slightly differently. Substituting Eq. (\ref{det1}) into the identity
\begin{eqnarray}
D_{\mu}\partial_{\nu}\varphi=\partial_{\mu}\partial_{\nu}
\varphi-\Gamma_{\mu\nu}^{\sigma}\partial_{\sigma}\varphi
\label{det7}
\end{eqnarray}
and expanding the terms, we get
\begin{eqnarray}
D_{\mu}\partial_{\nu}\varphi=\frac{\hbar}{m}e^{-i m
c^2
t/\hbar}\left\lbrack
D_{\mu}\partial_{\nu}\psi-i\frac{mc}{\hbar}(\delta_{\nu}^0\partial_{\mu}
\psi+\delta_{\mu}^0\partial_{\nu}\psi)-\frac{m^2c^2}{\hbar^2}\delta_{\nu}
^0\delta_{\mu}^0\psi+i\frac{mc}{\hbar}\Gamma_{\mu\nu}^0\psi\right\rbrack.
\label{det8}
\end{eqnarray}
Performing the tensorial product of Eq. (\ref{det8}) with $g^{\mu\nu}$ and
using Eq. (\ref{det2}), we recover Eq. (\ref{det5}), then Eq. (\ref{ggp1}).

\acknowledgements

This work was partially supported by CONACyT M\'exico under grants 
CB-2011 no. 166212, CB-2014-01 no. 240512,
Xiuhcoatl and Abacus clusters at Cinvestav 
and I0101/131/07
C-234/07 of the Instituto Avanzado de Cosmolog\'ia (IAC)
collaboration (http://www.iac.edu.mx/).


\section*{References}
\begin{thebibliography}{99}



\bibitem{dirachistoric}{\small P.A.M. Dirac, Sov. Phys. Usp. {\bf 22}, 648
(1979)}
\bibitem{brogliethese}{\small L. de Broglie, Ann. de Physique {\bf 3}, 22
(1925)}
\bibitem{einstein}{\small A. Einstein, Ann. Phys. {\bf 322}, 132 (1905)}
\bibitem{bhaumik}{\small M. Bhaumik, {\it Was Einstein wrong on quantum
physics?} [arXiv:1511.05098]}
\bibitem{dg}{\small C. Davisson, L. Germer, Phys. Rev. {\bf
30}, 707 (1927)}
\bibitem{schrodinger1}{\small E. Schr\"odinger, Ann. Phys. {\bf
384}, 361 (1926)}
\bibitem{schrodinger2}{\small E. Schr\"odinger, Ann. Phys. {\bf
384}, 489 (1926)}
\bibitem{schrodinger3}{\small E. Schr\"odinger, Ann. Phys. {\bf
385}, 437 (1926)}
\bibitem{schrodinger4}{\small E. Schr\"odinger, Ann. Phys. {\bf
386}, 109 (1926)}
\bibitem{bohr1}{\small N. Bohr, Phil. Mag. {\bf
26}, 1 (1913)}
\bibitem{bohr2}{\small N. Bohr, Phil. Mag. {\bf
26}, 476 (1913)}
\bibitem{heisenberg}{\small W. Heisenberg, Z. Phys. {\bf 33}, 879 (1925)}
\bibitem{bornjordan}{\small M. Born, P. Jordan, Z. Phys. {\bf 34}, 858 (1925)}
\bibitem{bhj}{\small M. Born,  W. Heisenberg, P. Jordan, Z. Phys. {\bf 35}, 557
(1926)}
\bibitem{schrodingerCORR}{\small E. Schr\"odinger, Ann. Phys.  {\bf
384}, 734 (1926)}
\bibitem{bornf}{\small M. Born, Zeitschrift f\"ur
Physik {\bf 38}, 803 (1926)}
\bibitem{born}{\small M. Born, Zeitschrift f\"ur
Physik {\bf 40}, 167 (1927)}
\bibitem{broglie1}{\small L. de Broglie,  Compt. Rend. Acad.
Sci. Paris {\bf 183}, 272 (1926)}
\bibitem{broglie2}{\small L. de Broglie, J. Physique {\bf 7}, 321 (1926)}
\bibitem{wentzel}{\small G. Wentzel, Z. Phys. {\bf 38}, 518 (1926)}
\bibitem{brillouin}{\small L. Brillouin, Compt. Rend. Acad.
Sci. Paris {\bf 183}, 24 (1926)}
\bibitem{kramers}{\small H.A. Kramers, Z. Phys. {\bf 39}, 828 (1926)}
\bibitem{schrodingerPR}{\small E. Schr\"odinger, Phys. Rev. {\bf
28}, 1049 (1926)}
\bibitem{klein1}{\small O. Klein, Z. Phys. {\bf 37}, 895
(1926)}
\bibitem{fock1}{\small V. Fock, Z. Phys. {\bf 38}, 242
(1926)}
\bibitem{ddd}{\small T. De Donder, F.H. van den Dungen, Compt. Rend. Acad.
Sci. Paris {\bf 183}, 22 (1926)}
\bibitem{fock2}{\small V. Fock, Z. Phys. {\bf 39}, 226
(1926)}
\bibitem{kudar}{\small J. Kudar, Ann. d. Phys. {\bf 386}, 632
(1926)}
\bibitem{eu}{\small P. Ehrenfest, G.E. Uhlenbeck,  Z. Phys. {\bf 39}, 495
(1926)}
\bibitem{gi}{\small G. Gamow, D. Iwanenko,  Z. Phys. {\bf 39}, 865
(1926)}
\bibitem{gordon}{\small W. Gordon, Z. Phys. {\bf 40}, 117
(1926)}
\bibitem{dd0}{\small T. De Donder, Compt. Rend. Acad.
Sci. Paris {\bf 183}, 594 (1926)}
\bibitem{schrodingerKG12}{\small E. Schr\"odinger, Ann. Phys.  {\bf
387}, 257 (1927)}
\bibitem{klein2}{\small O. Klein, Z. Phys. {\bf 41}, 407
(1927)}
\bibitem{schrodingerKG2}{\small E. Schr\"odinger, Ann. Phys.
{\bf
387}, 265 (1927)}
\bibitem{kaluza}{\small T. Kaluza, Berliner Sitzungsberichte, 966  (1921)}
\bibitem{dd1}{\small T. De Donder, Bull. Acad. Roy. Bel. {\bf
13}, 79 (1927)}
\bibitem{dd2}{\small T. De Donder, Bull. Acad. Roy. Bel. {\bf
13}, 103 (1927)}
\bibitem{donderp504}{\small T. De Donder, Bull. Acad. Roy. Bel. {\bf
13}, 504 (1927)}
\bibitem{dd3}{\small T. De Donder, Bull. Acad. Roy. Bel. {\bf
13}, 756 (1927)}
\bibitem{rosenfeld1}{\small L. Rosenfeld, Bull. Acad. Roy. Bel. {\bf
13}, 304 (1927)}
\bibitem{rosenfeld2}{\small L. Rosenfeld, Bull. Acad. Roy. Bel. {\bf
13}, 447 (1927)}
\bibitem{rosenfeld3}{\small L. Rosenfeld, Bull. Acad. Roy. Bel. {\bf
13}, 573 (1927)}
\bibitem{rosenfeld4}{\small L. Rosenfeld, Bull. Acad. Roy. Bel. {\bf
13}, 661 (1927)}
\bibitem{broglie1927b}{\small L. de Broglie,  Compt. Rend. Acad.
Sci. Paris {\bf 185}, 380 (1927)}
\bibitem{broglie1927a}{\small L. de Broglie, J. Physique {\bf 8}, 225 (1927)}
\bibitem{dirac1}{\small P.A.M. Dirac, Proc. Royal Soc. A {\bf 117}, 610
(1928)}
\bibitem{spin}{\small G.E. Uhlenbeck, S. Goudsmit, Nature {\bf 117}, 264 (1926)}
\bibitem{compton}{\small A.K. Compton, Journ. Frankl. Inst. {\bf 192}, 145
(1921)}
\bibitem{pauli}{\small W. Pauli, Z. Phys. {\bf 43}, 601
(1927)}
\bibitem{darwin}{\small C.G. Darwin, Roy. Soc. Proc. A  {\bf 116}, 227 (1927)}
\bibitem{dirac2}{\small P.A.M. Dirac, Proc. Royal Soc. A {\bf 126}, 360
(1930)}
\bibitem{weyl}{\small H. Weyl, Gruppentheorie und Quantenmechanik, 2nd ed. p.
234 (1931)}
\bibitem{anderson}{\small C. Anderson, Phys. Rev. {\bf 43}, 491
(1933)}
\bibitem{pw}{\small W. Pauli, V. Weisskopf, Helv. Phys. Act. {\bf 7}, 709
(1934)}
\bibitem{bose}{\small S.N. Bose, Z. Phys. {\bf  26}, 178 (1924)}
\bibitem{einsteinb}{\small A. Einstein, Sitzber. Kgl. Preuss. Akad. Wiss. {\bf
1}, 3  (1925)}
\bibitem{fermi}{\small E. Fermi, Rend. Acc. Lincei {\bf 3}, 145
(1926)}
\bibitem{dirac}{\small P. Dirac, Proc. R. Soc. A {\bf 112}, 661 (1926)}
\bibitem{bogoliubov}{\small N. Bogoliubov, J. Phys. {\bf 11}, 23
(1947)}
\bibitem{hy}{\small K. Huang, C.N. Yang, Phys. Rev. {\bf  105}, 767 (1957)}
\bibitem{lhy}{\small T.D. Lee, K. Huang, C.N. Yang, Phys. Rev. {\bf  106}, 1135
(1957)}
\bibitem{gross1}{\small E.P. Gross, Ann. of Phys. {\bf 4}, 57 (1958)}
\bibitem{gross2}{\small E.P. Gross,  Nuovo
Cimento {\bf 20}, 454 (1961)}
\bibitem{gross3}{\small E.P. Gross, J. Math. Phys. {\bf 4}, 195 (1963)}
\bibitem{pitaevskii1}{\small L.P. Pitaevskii, Sov. Phys. JETP {\bf 9}, 830
(1959)}
\bibitem{pitaevskii2}{\small L.P. Pitaevskii, Sov. Phys. JETP {\bf
13}, 451 (1961)}
\bibitem{kaup}{\small D.J. Kaup, Phys. Rev. {\bf  172}, 1331 (1968)}
\bibitem{rb}{\small R. Ruffini, S. Bonazzola, Phys. Rev. {\bf  187}, 1767
(1969)}
\bibitem{wheeler}{\small  J.A. Wheeler, Phys. Rev. {\bf 97}, 511 (1955)}
\bibitem{colpi}{\small M. Colpi, S.L. Shapiro, I. Wasserman, Phys. Rev. Lett.
{\bf  57}, 2485 (1986)}
\bibitem{prd1}{\small P.H. Chavanis, Phys. Rev. D {\bf 84}, 043531 (2011)}
\bibitem{revueabril}{\small  A. Su\'arez, V.H. Robles, T. Matos, Astrophys.
Space Sci. Proc. {\bf 38}, 107 (2014)}
\bibitem{revueshapiro}{\small T. Rindler-Daller, P.R. Shapiro, Astrophys.
Space Sci. Proc. {\bf 38}, 163 (2014)}
\bibitem{bookspringer}{\small P.H. Chavanis, {\it Self-gravitating Bose-Einstein
condensates}, in Quantum Aspects of Black Holes, edited by X.
Calmet (Springer, 2015)}
\bibitem{madelung}{\small E. Madelung, Zeit. F. Phys. {\bf 40}, 322 (1927)}
\bibitem{kennard}{\small E.H. Kennard, Phys. Rev. {\bf 31}, 876 (1928)}
\bibitem{spiegel}{\small E.A. Spiegel, Physica D {\bf 1}, 236 (1980)}
\bibitem{broglie1927c}{\small L. de Broglie,  Compt. Rend. Acad.
Sci. Paris {\bf 185}, 1118 (1927)}
\bibitem{london}{\small F. London, Zeit. F. Phys. {\bf 42}, 375 (1927)}
\bibitem{solvay}{\small L. de Broglie, {\it The Interpretation
of Wave Mechanics with the help of Waves with Singular Regions}
[arXiv:1005.4534]}
\bibitem{bohm1}{\small D. Bohm, Phys. Rev. {\bf 85}, 166 (1952)}
\bibitem{bohm2}{\small D. Bohm, Phys. Rev. {\bf 85}, 180 (1952)}
\bibitem{takabayasi1}{\small  T. Takabayasi, Prog. Theor. Phys. {\bf 8}, 143
(1952)}
\bibitem{takabayasi2}{\small  T. Takabayasi, Prog. Theor. Phys. {\bf 9}, 187
(1953)}
\bibitem{ndb1}{\small L. de Broglie,  Compt. Rend. Acad.
Sci. Paris {\bf 233}, 641 (1951)}
\bibitem{ndb2}{\small L. de Broglie,  Compt. Rend. Acad.
Sci. Paris {\bf 234}, 265 (1952)}
\bibitem{ndb3}{\small L. de Broglie,  Compt. Rend. Acad.
Sci. Paris {\bf 235}, 557 (1952)}
\bibitem{ndb4}{\small L. de Broglie, Found. Phys. {\bf 1}, 5 (1970)}
\bibitem{onsager}{\small L. Onsager, Nuovo Cimento {\bf 6}, 279 (1949)}
\bibitem{feynman}{\small R.P. Feynman, {\it Progress in Low Temperature
Physics, Vol. 1.} (North-Holland, Amsterdam, 1955)}
\bibitem{diracmm}{\small P.A.M. Dirac, Proc. Royal Soc. A {\bf 133}, 60
(1931)}
\bibitem{landauS}{\small L. Landau, J. Phys. {\bf 5}, 71 (1941)}
\bibitem{londonS}{\small F. London, Rev. Mod. Phys. {\bf 17}, 310 (1945)}
\bibitem{bohmer}{\small C.G. B\"ohmer, T. Harko, J. Cosmol. Astropart. Phys.
{\bf 06}, 025 (2007)}
\bibitem{prd2}{\small P.H. Chavanis, L. Delfini, Phys. Rev. D {\bf 84}, 043532
(2011)}
\bibitem{chavaniskpz}{\small P.H. Chavanis, Phys. Rev. D {\bf 84}, 063518
(2011)}
\bibitem{chavaniscosmo}{\small P.H. Chavanis, Astron. Astrophys. {\bf 537}, A127
(2012)}
\bibitem{tanja}{\small T. Rindler-Daller, P.R. Shapiro, Mon. Not. R. Astron.
Soc.
{\bf 422}, 135 (2012)}
\bibitem{chavharko}{\small P.H. Chavanis, T. Harko, Phys. Rev. D {\bf 86},
064011 (2012)}
\bibitem{abrilMNRAS}{\small A. Su\'arez, T. Matos, Mon. Not. R. Astron. Soc.
{\bf 416}, 87 (2011)}
\bibitem{smz}{\small  A. Su\'arez, T. Matos, Class. Quantum Grav. {\bf
31}, 045015 (2014)}
\bibitem{abrilph1}{\small  A. Su\'arez, P.H. Chavanis, Phys. Rev. D {\bf 92},
023510 (2015)}
\bibitem{abrilph2}{\small  A. Su\'arez, P.H. Chavanis, J. Phys.: Conf. Series
{\bf 654}, 012088 (2015)}
\bibitem{mrm}{\small  T. Matos, M.A. Rodr\'iguez-Meza, J. Phys.: Conf. Series
{\bf 545}, 012009 (2014)}
\bibitem{mc}{\small  T. Matos, E. Castellanos, AIP Conf. Proc.
{\bf 1577}, 181 (2014)}
\bibitem{weinberg}{\small  S. Weinberg, {\it Gravitation and Cosmology} (John
Wiley, 1972)}
\bibitem{brother}{\small  F. London, H. London, Proc. Royal Soc. A {\bf
149}, 71 (1935)}
\bibitem{ma}{\small C.-P. Ma, E. Bertschinger, Astrophys. J. {\bf
455}, 7
(1995)}
\bibitem{li}{\small B. Li, T. Rindler-Daller, P.R. Shapiro, Phys. Rev. D
{\bf 89}, 083536 (2014)}
\bibitem{revuebec}{\small F. Dalfovo, S. Giorgini, L.P. Pitaevskii, S.
Stringari, Rev. Mod. Phys. {\bf 71}, 463 (1999)}
\bibitem{abrilphprep}{\small  A. Su\'arez, P.H. Chavanis, in preparation}
\bibitem{meissner}{\small W. Meissner, R. Ochsenfeld, Naturwissenschaften {\bf
21}, 787 (1933)}
\bibitem{jackson}{\small J.D. Jackson, {\it Classical Electrodynamics} (John
Wiley, 1975)}
\bibitem{broglie25}{\small L. de Broglie,  Compt. Rend. Acad.
Sci. Paris {\bf 180}, 498 (1925)}

\end{thebibliography}
\end{document}